\documentclass[9pt,journal]{IEEEtran}
\pdfoutput=1

\usepackage{times}   
\usepackage{color}
\usepackage{amsfonts}
\usepackage{amssymb}  
\usepackage[cmex10]{amsmath}  
\usepackage{epsfig}  
\usepackage{graphics}  
\usepackage{array}  
\usepackage{rotating}  
\usepackage{cite}  
\usepackage{hhline}  
\usepackage{supertabular}  
\usepackage{multirow}  
\usepackage{fancyvrb} 
\usepackage{cite} 

\newcommand{\lp}{l_p}
\newcommand{\bi}{\begin{itemize}}
\newcommand{\ei}{\end{itemize}}

\newcommand{\bs}{\backslash}
\newcommand{\real}{\mathbb{R}}

\newcommand{\vha}{\hat{\boldsymbol a}}
\newcommand{\vhb}{\hat{\boldsymbol b}}
\newcommand{\vhh}{\hat{\boldsymbol h}}
\newcommand{\vta}{\tilde{\boldsymbol a}}

\newcommand{\va}{\boldsymbol a}
\newcommand{\vb}{\boldsymbol b}

\newcommand{\ve}{\boldsymbol e}
\newcommand{\vf}{\boldsymbol f}

\newcommand{\vh}{\boldsymbol h}
\newcommand{\vq}{\boldsymbol q}
\newcommand{\vr}{\boldsymbol r}
\newcommand{\vs}{\boldsymbol s}

\newcommand{\vw}{\boldsymbol w}
\newcommand{\vx}{\boldsymbol x}
\newcommand{\vy}{\boldsymbol y}

\newcommand{\vA}{\boldsymbol A}
\newcommand{\vB}{\boldsymbol B}
\newcommand{\vC}{\boldsymbol C}
\newcommand{\vD}{\boldsymbol D}
\newcommand{\vE}{\boldsymbol E}
\newcommand{\vH}{\boldsymbol H}
\newcommand{\vI}{\boldsymbol I}

\newcommand{\vX}{\boldsymbol X}

\newcommand{\vphi}{\boldsymbol \phi}
\newcommand{\mhH}{\hat{\mathbf{H}}}

\newcommand{\mC}{\mathbf{C}}
\newcommand{\mD}{\mathbf{D}}
\newcommand{\mF}{\mathbf{F}}
\newcommand{\mG}{\mathbf{G}}
\newcommand{\mH}{\mathbf{H}}
\newcommand{\mJ}{\mathbf{J}}

\newcommand{\mV}{\mathbf{V}}
\newcommand{\mW}{\mathbf{W}}

\newcommand{\al}{\alpha} 
\newcommand{\ep}{\epsilon}
\newcommand{\lam}{\lambda}
\newcommand{\om}{\omega}
\newcommand{\sig}{\sigma}
\newcommand{\vep}{\varepsilon}
\newcommand{\Om}{\Omega}
\newcommand{\FT}{\mathcal F}              

\providecommand{\abs}[1]{\lvert#1\rvert}
\providecommand{\norm}[1]{\lVert#1\rVert}
\providecommand{\normp}[1]{\lVert#1\rVert_p}
\providecommand{\normsq}[1]{\lVert#1\rVert_2}

\newcommand{\opt}{^\star}
\newcommand{\be}{\begin{equation}}
\newcommand{\ee}{\end{equation}}
\newcommand{\ben}{\begin{equation*}}
\newcommand{\een}{\end{equation*}}
\newcommand{\bea}{\begin{eqnarray}}
\newcommand{\eea}{\end{eqnarray}}
\newcommand{\bean}{\begin{eqnarray*}}
\newcommand{\eean}{\end{eqnarray*}}
\newcommand{\circonv}{\mathbin{\bigcirc\mkern-14mu {\mbox{{\it \footnotesize L}}}\;}}
\newtheorem{thm}{Theorem}

\definecolor{med}{gray}{.85} 
\definecolor{dark}{gray}{.65}
\definecolor{light}{gray}{.85}
\newcommand{\Co}{\hbox{\sf I}\kern-0.40em \hbox{\sf C}}
\newcommand{\diag}{\mbox{diag}} 
\newcommand{\e}{\mathcal E}

\newcommand{\pa}{\partial} 

\begin{document}
\title{Iterative Design of $\lp$ Digital Filters}
\author{Ricardo A. Vargas and C. Sidney Burrus \\ Electrical and Computer Engrineering Dept. \\ Rice University \\ February 2009 }
\maketitle

\begin{abstract}
The design of digital filters is a fundamental process in the context of digital signal processing. The purpose of this paper is to study the use of $\lp$ norms (for $2\!<\!p\!<\!\infty$) as design criteria for digital filters, and to introduce a set of algorithms for the design of {\it Finite} (FIR) and {\it Infinite} (IIR) {\it Impulse Response} digital filters based on the {\it Iterative Reweighted Least Squares} (IRLS) algorithm. The proposed algorithms rely on the idea of breaking the $\lp$ filter design problem into a sequence of approximations rather than solving the original $\lp$ problem directly. It is shown that one can efficiently design filters that arbitrarily approximate a desired $\lp$ solution (for $2\!<\!p\!<\!\infty$) including the commonly used $l_\infty$ (or minimax) design problem. A method to design filters with different norms in different bands is presented (allowing the user for better control of the signal and noise behavior per band). Among the main contributions of this work is a method for the design of {\it magnitude} $\lp$ IIR filters. Experimental results show that the algorithms in this work are robust and efficient, improving over traditional off-the-shelf optimization tools. The group of proposed algorithms form a flexible collection that offers robustness and efficiency for a wide variety of digital filter design applications.
\end{abstract}

\section{Introduction}
\label{ch:int}
The design of digital filters has fundamental importance in digital signal processing. One can find applications of digital filters in many diverse areas of science and engineering including medical imaging, audio and video processing, oil exploration, and highly sophisticated military applications. Furthermore, each of these applications benefits from digital filters in particular ways, thus requiring different properties from the filters they employ. Therefore it is of critical importance to have efficient design methods that can shape filters according to the user's needs.

In this work we use the discrete $\lp$ norm as the criterion for designing efficient digital filters. We also introduce a set of algorithms, all based on the {\it Iterative Reweighted Least Squares} ({\bf IRLS}) method, to solve a variety of relevant digital filter design problems. The proposed family of algorithms has proven to be efficient in practice; these algorithms share theoretical justification for their use and implementation. Finally, the document makes a point about the relevance of the $\lp$ norm as a useful tool in filter design applications.

The rest of this chapter is devoted to motivating the problem. Section \ref{int:digfilt} introduces the general filter design problem and some of the signal processing concepts relevant to this work. Section \ref{int:irls} presents the {\it basic} Iterative Reweighted Least Squares method, one of the key concepts in this document. Section \ref{int:fir} introduces {\it Finite Impulse Response} ({\bf FIR}) filters and covers theoretical motivations for $\lp$ design, including previous knowledge in $\lp$ optimization (both from experiences in filter design as well as other fields of science and engineering). Similarly, Section \ref{int:iir} introduces {\it Infinite Impulse Response} ({\bf IIR}) filters.  These last two sections lay down the structure of the proposed algorithms, and provide an outline for the main contributions of this work. 

Chapters \ref{ch:fir} and \ref{ch:iir} formally introduce the different $l_p$ filter design problems considered in this work and discuss their IRLS-based algorithms and corresponding results. Each of these chapters provides a literary review of related previous work as well as a discussion on the proposed methods and their corresponding results. An important contribution of this work is the extension of known and well understood concepts in $\lp$ FIR filter design to the IIR case.

\subsection{Digital filter design}
\label{int:digfilt}

When designing digital filters for signal processing applications one is often interested in creating objects $\vh\in\real^N$ in order to alter some of the properties of a given vector $\vx\in\real^M$ (where $0\!<\!M,N\!<\!\infty$). Often the properties of $\vx$ that we are interested in changing lie in the frequency domain, with $\vX=\FT(\vx)$ being the {\it frequency domain} representation of $\vx$ given by
\[ \vx \stackrel{\FT}{\leftrightarrow} \vX=\vA_X e^{j\om\vphi_X} \]
where $\vA_X$ and $\vphi_X$ are the {\it amplitude} and {\it phase} components of $\vx$, and $\FT(\cdot):\real^N\mapsto\real^\infty$ is the {\it Fourier transform} operator defined by
\be \label{int:filtfft} \FT\{\vh\} = H(\om) \triangleq \sum_{n=0}^{N-1} h_n e^{-j\om n} \qquad \forall\; \om\in [-\pi,\pi] \ee
So the idea in filter design is to create filters $\vh$ such that the Fourier transform $\vH$ of $\vh$ posesses desirable {\it amplitude} and {\it phase} characteristics.

The {\it filtering} operator is the convolution operator (${\boldsymbol \ast}$) defined by
\[ (\vx\ast\vh)(n) = \sum_m x(m)h(n-m) \]
An important property of the convolution operator is the {\it Convolution Theorem} \cite{sigsys_cd} which states that
\be \vx\ast\vh \stackrel{\FT}{\leftrightarrow} \vX\cdot\vH=(\vA_X\cdot\vA_H) \, e^{j\om(\vphi_X+\vphi_H)} \ee
where $\left\{\vA_X,\vphi_X\right\}$ and $\left\{\vA_H,\vphi_H\right\}$ represent the amplitude and phase components of $\vX$ and $\vH$ respectively.
It can be seen that by {\it filtering} $\vx$ with $\vh$ one can apply a {\it scaling} operator to the amplitude of $\vx$ and a {\it biasing} operator to its phase.

A common use of digital filters is to remove a certain {\it band} of frequencies from the frequency spectra of $\vx$ (such as typical {\it lowpass} filters). Other types of filters include {\it band-pass}, {\it high-pass} or {\it band-reject} filters, depending on the range of frequencies that they alter.

\subsection{The notion of approximation in $\lp$ filter design}
\label{int:notion}

Once a filter design concept has been selected, the design problem becomes finding the optimal vector $\vh\in\real^n$ that most closely {\it approximates} our desired frequency response concept (we will denote such {\it optimal} vector by $\vh\opt$). This approximation problem will heavily depend on the {\it measure} by which we evaluate all vectors $\vh\in\real^N$ to choose $\vh\opt$.

In this document we consider the discrete $\lp$ norms defined by
\be \label{int:lpbasic} \normp{\va} = \sqrt[p]{\sum_k \abs{a_k}^p} \qquad \forall\; a\in\real^N \ee
as measures of optimality, and consider a number of filter design problems based upon this criterion.  The work explores the {\it Iterative Reweighted Least Squares} (IRLS) approach as a design tool, and provides a number of algorithms based on this method. Finally, this work considers critical theoretical aspects and evaluates the numerical properties of the proposed algorithms in comparison to existing {\it general purpose} methods commonly used. It is the belief of the author (as well as the author's advisor) that the IRLS approach offers a more tailored route to the $\lp$ filter design problems considered, and that it contributes an example of a {\it made-for-purpose} algorithm best suited to the characteristics of $\lp$ filter design.

\subsection{The IRLS algorithm}
\label{int:irls}

Iterative Reweighted Least Squares (IRLS) algorithms define a family of iterative methods that solve an otherwise complicated numerical optimization problem by breaking it into a series of {\it weighted least squares} (WLS) problems, each one easier in principle than the original problem. At iteration $i$ one must solve a weighted least squares problem of the form
\be \label{int:wls1} \min_{h_i} \;\norm{w(h_{i-1})f(h_i)}_2 \ee
where $w(\cdot)$ is a specific weighting function and $f(\cdot)$ is a function of the filter. Obviously a large class of problems could be written in this form (large in the sense that both $w( \cdot)$ and $f(\cdot)$ can be defined arbitrarily).  One case worth considering is the {\it linear approximation} problem defined by
\be \label{int:wls2} \min_{\vh} \: \norm{\vD-\mC\vh} \ee
where $\vD\in\real^M$ and $\mC\in\real^{M\times N}$ are given, and $\norm{\cdot}$ is an arbitrary measure. One could write $f(\cdot)$ in (\ref{int:wls1}) as
\[ f(\vh) = \vD - \mC\vh \]
and attempt to find a suitable function $w(\cdot)$ to minimize the arbitrary norm $\norm{\cdot}$ in (\ref{int:wls2}). In vector notation, at iteration $i$ one can write (\ref{int:wls1}) as follows,
\be \label{int:wls3} \min_{\vh_i} \;\norm{w(\vh_{i-1}) \left(\vD-\mC\vh_i\right) }_2 \ee
One can show that the solution of (\ref{int:wls3}) for any iteration is given by 
\[ \vh = (\mC^T \mW\mC)^{-1} \mC^T\mW\vD \]
with $\mW = \diag(\vw^2)$ (where $\vw$ is the weighting vector). To solve problem (\ref{int:wls3}) above, one could use the following algorithm:
\begin{enumerate}
\item Set initial weights $\vw_0$
\item At the $i$-th iteration find $\vh_i = (\mC^T \mW_{i-1} \mC)^{-1} \mC^T \mW_{i-1} \vD$
\item Update $\mW_i$ as a function of $\vh_i$ (i.e. $\mW_i=\mW(\vh_i)$ )
\item Iterate steps 2 and 3 until a certain stopping criterion is reached
\end{enumerate}
This method will be referred in this work as the {\it basic} IRLS algorithm.

An IRLS algorithm is said to {\it converge} if the algorithm produces a sequence of points {$h_i$} such that
\[ \lim_{i\rightarrow \infty} \vh_i = \vh^* \]
where $\vh^*$ is a {\it fixed point} defined by
\[ \vh^* = (\mC^T \mW^*\mC)^{-1} \mC^T \mW^*\vD \]
with $\mW^*=\mW(\vh^*)$. In principle one would want $\vh^*=\vh\opt$ (as defined in Section \ref{int:notion}).

IRLS algorithms have been used in different areas of science and engineering. Their atractiveness stem from the idea of simplifying a difficult problem as a sequence of weighted least squares problems that can be solved efficiently with programs such as Matlab or LAPACK. However (as it was mentioned above) success is determined by the existence of a weighting function that leads to a fixed point that happens to be at least a local solution of the problem in question. This might not be the case for any given problem. In the case of $\lp$ optimization one can justify the use of IRLS methods by means of the following theorem:
\begin{thm}[Weight Function Existence theorem]
\label{int:wexist} 
 Let $g_k(\omega)$ be a Chebyshev set and define
\[H(\vh;\omega)=\sum_{k=0}^M h_kg_k(\omega)\]
where $\vh=(h_0,h_1,\ldots,h_M)^T$. Then, given $D(\om)$ continuous on $[0,\pi]$ and $1\!<\!q\!<\!p\!\leq\!\infty$ the following are identical sets:
\begin{itemize}
\item \{$\vh\mid H(\vh;\omega)$ is a best weighted $L_p$ approximation to$D(\om)$ on $[0,\pi]$\}.
\item \{$\vh\mid H(\vh;\omega)$ is a best weighted $L_q$ approximation to $D(\om)$ on $[0,\pi]$\}.
\end{itemize}
Furthermore, the theorem above is valid if the interval $[0,\pi]$ is replaced by a finite point set $\Omega\subset[0,\pi]$ (this theorem is accredited to Motzkin and Walsh \cite{motwal,motwal2}).
\end{thm}

Theorem \ref{int:wexist} is fundamental since it establishes that weights exist so that the solution of an $L_p$ problem is indeed the solution of a weighted $L_q$ problem  (for arbitrary $p,q\!>\!1$). Furthermore the results of Theorem \ref{int:wexist} remain valid for $l_p$ and $l_q$. For our purposes, this theorem establishes the existence of a weighting function so that the solution of a weighted $l_2$ problem is indeed the solution of an $\lp$ problem; the challenge then is to find the corresponding {\it weighting function}. The remainder of this document explores this task for a number of relevant filter design problems and provides a consistent computational framework.

\subsection{Finite Impulse Response (FIR) $\lp$ design}
\label{int:fir}
A {\it Finite Impulse Response} ({\bf FIR}) filter is an ordered vector $\vh\in\real^N$ (where $0\!!<\!N\!<\!\infty$), with a complex polynomial form in the frequency domain given by
\[ H(\om) = \sum_{n=0}^{N-1} h_n e^{-j\om n} \]
The filter $H(\om)$ contains amplitude and phase components $\left\{A_H(\om),\phi_H(\om)\right\}$ that can be designed to suit the user's purpose. 

Given a desired frequency response $D(\om)$, the general $\lp$ approximation problem is given by 
\[ \min_{\vh} \;\normp{D(\om)-H(\vh;\om)} \]
In the most basic scenario $D(\om)$ would be a complex valued function, and the optimization algorithm would minimize the $\lp$ norm of the complex error function $\ep(\om)=D(\om)-H(\om)$; we refer to this case as the {\it complex} $\lp$ design problem (refer to Section \ref{fir:complex}). 

One of the caveats of solving complex approximation problems is that the user must provide desired magnitude and phase specifications. In many applications one is interested in removing or altering a range of frequencies from a signal; in such instances it might be more convenient to only provide the algorithm with a desired magnitude function while allowing the algorithm to find a phase that corresponds to the optimal magnitude design. The {\it magnitude $\lp$} design problem is given by
\[ \min_{\vh} \; \normp{D(\om) - \abs{H(\vh;\om)}\,} \]
where $D(\om)$ is a real, positive function. This problem is discussed in Section \ref{fir:magn}.

Another problem that uses no phase information is the {\it linear phase} $\lp$ problem. It will be shown in Section \ref{fir:linph} that this problem can be formulated so that only real functions are involved in the optimization problem (since the phase component of $H(\om)$ has a specific linear form).

An interesting case results from the idea of combining different norms in different frequency bands of a desired function $D(\om)$. One could assign different $p$-values for different bands (for example, minimizing the error energy ($\vep_2$) in the passband while using a minimax error ($\vep_{\infty}$) approach in the stopband to keep control of noise).  The {\it frequency-varying} $\lp$ problem is formulated as follows,
\[ \min_{\vh} \;\normp{(D-H)(\om_{pb})}+\norm{(D-H)(\om_{sb})}_q \]
where $\left\{\om_{pb},\om_{pb}\right\}$ are the passband and stopband frequency ranges respectively (and $2\!<\!p,q\!<\!\infty$).

Perhaps the most relevant problem addressed in this work is the {\it Constrained Least Squares} ({\bf CLS}) problem. In a continuous sense, a CLS problem is defined by 
\[ \begin{array}{ll}
\underset{\vh}{\min} & \normsq{d(\om)-H(\om)} \\
\mbox{subject to} & \abs{d(\om)-H(\om)} \!\leq\! \tau \end{array} \]
The idea is to minimize the error energy across all frequencies, but ensuring first that the error at each frequency does not exceed a given tolerance $\tau$. Section \ref{fir:cls} explains the details for this problem and shows that this type of formulation makes good sense in filter design and can efficiently be solved via IRLS methods.

\subsubsection{The IRLS algorithm and FIR literature review}
A common approach to dealing with highly structured approximation problems consists in breaking a complex problem into a series of simpler, smaller problems. Often, one can even prove important mathematical properties in this way.  Consider the $\lp$ approximation problem introduced in (\ref{int:lpbasic}),
\be \label{int:irls1} \min_{\vh} \;\normp{f(\vh)} \ee
For simplicity at this point we can assume that $f(\cdot):\real^N\mapsto\real^M$ is linear. It is relevant to mention that (\ref{int:irls1}) is equivalent to
\be \label{int:irls2} \min_{\vh} \;\normp{f(\vh)}^p \ee
In its most basic form the $\lp$ IRLS algorithm works by rewriting (\ref{int:irls2}) into a weighted least squares problem of the form
\be \label{int:irls3} \min_{\vh} \;\normsq{w(\vh)f(\vh)}^2 \ee
Since a linear weighted least squares problem like (\ref{int:irls3}) has a closed form solution, it can be solved in one step. Then the solution is used to update the weighting function, which is kept constant for the next closed form solution and so on (as discussed in Section \ref{int:irls}).

One of the earlier works on the use of IRLS methods for $\lp$ approximation was written by Charles Lawson \cite{lawphd,rul,appfunc}, in part motivated by problems that might not have a suitable $l_{\infty}$ algorithm. He looked at a basic form of the IRLS method to solve $l_{\infty}$ problems and extended it by proposing a multiplicative update of the weighting coefficients at each iteration (that is, $w_{k+1}(\om)=f(\om)\cdot w_k(\om)$). Lawson's method triggered a number of papers and ideas; however his method is sensitive to the weights becoming numerically zero; in this case the algorithm must restart. A number of ideas \cite{rul,appfunc} have been proposed (some from Lawson himself) to prevent or deal with these occurrences, and in general his method is considered somewhat slow.

John Rice and Karl Usow \cite{rul,irls} extended Lawson's method to the general $\lp$ problem ($2\!<\!p\!<\!\infty$) by developing an algorithm based on Lawson's that also updates the weights in a multiplicative form. They used the results from Theorem \ref{int:wexist} by Motzkin and Walsh \cite{motwal,motwal2} to guarantee that a solution indeed exists for the $\lp$ problem. They defined
\[ w_{k+1}(\om)=w_k^\alpha(\om)\abs{\ep_k(\om)}^\beta \]
where
\[\alpha=\frac{\gamma(p-2)}{\gamma(p-2)+1}\] 
and
\[\beta=\frac{\alpha}{2\gamma}=\frac{p-2}{2(\gamma(p-2)+1)} \]
with $\gamma$ being a convergence parameter and $\ep(\om)=d(\om)-H(\om)$. The rest of the algorithm works the same way as the basic IRLS method; however the proper selection of $\gamma$ could allow for strong convergence (note that for $\gamma=0$ we obtain the basic IRLS algorithm).

Another approach to solve (\ref{int:irls1}) consists in a {\it partial updating} strategy of the {\it filter coefficients} rather than the {\it weights}, by using a temporary coefficient vector defined by
\begin{equation} \label{int:irls4} \hat{\va}_{k+1}=[\mC^T\mW_k^T\mW_k\mC]^{-1}\mC^T\mW_k^T\mW_k\vA_d \end{equation}
The filter coefficients after each iteration are then calculated by 
\be \va_{k+1} = \lam\hat{\va}_{k+1}+(1-\lam)\va_k \ee
where $\lam$ is a {\it convergence parameter} (with $0\!<\!\lam\!<\!1$). This approach is known as the Karlovitz method \cite{karlov}, and it has been claimed that it converges to the global optimal solution for {\bf even} values of $p$ such that $4\!\leq\! p\!<\!\infty$. However, in practice several convergence problems have been found even under such assumptions.  One drawback is that the convergence parameter $\lam$ has to be optimized at each iteration via an expensive line search process. Therefore the overall execution time becomes rather large.

S. W. Kahng \cite{kahng} developed an algorithm based on Newton-Raphson's method that uses 
\be \label{int:irls5} \lam = \frac{1}{p-1} \ee
to get
\be \label{int:irls6} \va_{k+1}=\frac{\hat{\va}_{k+1}+(p-2)\va_k}{p-1} \ee
This selection for $\lam$ is based upon Newton's method to minimize $\ep$ (the same result was derived independently by Fletcher, Grant and Hebden\cite{fletlp}).  The rest of the algorithm follows Karlovitz's approach; however since $\lam$ is fixed there is no need to perform the linear search for its best value. Since Kahng's method is based on Newton's method, it converges quadratically to the optimal solution. Kahng proved that his method converges for all cases of $\lam$ and for any problem (at least in theory). It can be seen that Kahng's method is a particular case of Karlovitz's algorithm, with $\lam$ as defined in (\ref{int:irls5}). Newton-Raphson based algorithms are not warranted to converge to the optimal solution unless they are somewhat close to the solution since they require to know and invert the Hessian matrix of the objective function (which must be {\it positive definite} \cite{intropt}). However, their associated quadratic convergence makes them an appealing option.

Burrus, Barreto and Selesnick developed a method \cite{irls,barms,leastp} that combines the powerful quadratic convergence of Newton's methods with the robust initial convergence of the basic IRLS method, thus overcoming the initial sensitivity of Newton-based algorithms and the slow linear convergence of Lawson-based methods. To accelerate initial convergence, their approach to solve (\ref{int:irls1}) uses $p=\sig*2$, where $\sig$ is a convergence parameter (with $1\!<\! \sig \!\leq\! 2$). At any given iteration, $p$ increases its value by a factor of $\sig$. This is done at each iteration, so to satisfy 
\be \label{int:irls7} p_k=\min \;(p_{des},\sig\cdot p_{k-1}) \ee
where $p_{des}$ corresponds to the desired $\lp$ norm. The implementation of each iteration follows Karlovitz's method using the particular selection of $p$ given by (\ref{int:irls7}).
\begin{figure}[ht]
    \centerline{\psfig{figure=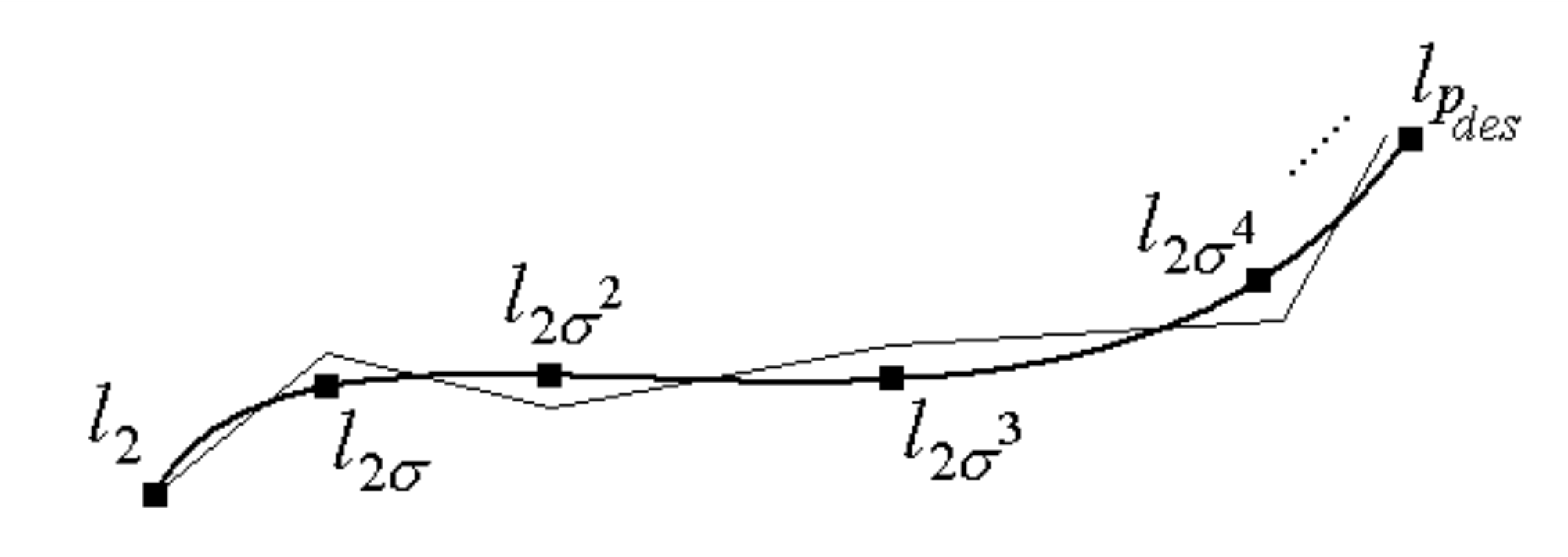,width=3.0in}}
    \caption{Homotopy approach for IRLS $\lp$ filter design.}
    \label{int:homotopy}
\end{figure}

It is worth noting that the method outlined above combines several ideas into a powerful approach. By not solving the desired $\lp$ problem from the first iteration, one avoids the potential issues of Newton-based methods where convergence is guaranteed within a radius of convergence. It is well known that for $2\!\leq\! p\!\leq\!\infty$ there exists a continuum of $\lp$ solutions (as shown in Figure \ref{int:homotopy}). By slowly increasing $p$ from iteration to iteration one hopes to follow the continuum of solutions from $l_2$ towards the desired $p$. By choosing a reasonable $\sig$ the method can only spend one iteration at any given $p$ and still remain close enough to the optimal path. Once the algorithm reaches a neighborhood of the desired $p$, it can be allowed to iterate at such $p$, in order to converge to the optimal solution. This process is analogous to {\it homotopy}, a commonly used family of optimization methods \cite{numopt}.

While $l_2$ and $l_{\infty}$ designs offer meaningful approaches to filter design, the Constrained Least Squares (CLS) problem offers an interesting tradeoff to both approaches \cite{adamspcsl}. In the context of filter design, the CLS problem seems to be first presented by John Adams \cite{adamsfircls} in 1991. The problem Adams posed is a {\it Quadratic Programming} (QP) problem, well suited for off-the-shelf QP tools like those based on Lagrange multiplier theory \cite{adamsfircls}. However, Adams posed the problem in such a way that a transition band is required. Burrus et al. presented a formulation \cite{constls,constls2D,constls_mult} where only a {\it transition frequency} is required; the transition band is {\it induced}; it does indeed exist but is not specified (it adjusts itself optimally according to the constraint specifications). The method by Burrus et al. is based on Lagrange multipliers and the Karush-Kuhn-Tucker (KKT) conditions. 

An alternative to the KKT-based method mentioned above is the use of IRLS methods where a suitable weighting function serves as the constraining function over frequencies that exceed the constraint tolerance. Otherwise no weights are used, effectively forcing a least-squares solution. While this idea has been suggested by Burrus et al., one of the main contributions of this work is a thorough investigation of this approach, as well as proper documentation of numerical results, theoretical findings and proper code.

\subsection{Infinite Impulse Response (IIR) $\lp$ design}
\label{int:iir}
In contrast to FIR filters, an {\it Infinite Impulse Response} ({\bf IIR}) filter is defined by two ordered vectors $\va\in\real^N$ and $\vb\in\real^{M+1}$ (where $0\!<\!M,N\!<\!\infty$), with frequency response given by 
\[ H(\om)=\frac{B(\om)}{A(\om)}=\frac{\sum\limits_{n=0}^M b_n e^{-j\om n}}{1+\sum\limits_{n=1}^N a_n e^{-j\om n}} \]
Hence the general $l_p$ approximation problem is
\be \label{int:iirlpp} \min_{a_n,b_n} \left\| \frac{\sum\limits_{n=0}^M b_n e^{-j\om n}}{1+\sum\limits_{n=1}^N a_n e^{-j\om n}} - D(\om) \right\|_p \ee
which can be  posed as a weighted least squares problem of the form
\be \label{int:lpiirp} \min_{a_n,b_n} \left\| w(\om)\cdot \left(\frac{\sum\limits_{n=0}^M b_n e^{-j\om n}}{1+\sum\limits_{n=1}^N a_n e^{-j\om n}} - D(\om)\right) \right\|_2^2 \ee
It is possible to design similar problems to the ones outlined in Section \ref{int:fir} for FIR filters. However, it is worth keeping in mind the additon al complication s that IIR design involves, including the nonlinear least squares problem presented in Section \ref{int:l2iirrev} below.  

\subsubsection{Least squares IIR literature review}
\label{int:l2iirrev}

The weighted nonlinear formulation presented in (\ref{int:lpiirp}) suggests the possibility of taking advantage of the flexibilities in design from the FIR problems. However this point comes at the expense of having to solve at each iteration a weighted nonlinear $l_2$ problem.  Solving least squares approximations with rational functions is a nontrivial problem that has been studied extensively in diverse areas including statistics, applied mathematics and electrical engineering. One of the contributions of this document is a presentation in Section \ref{iir:ls} on the subject of $l_2$ IIR filter design that captures and organizes previous relevant work. It also sets the framework for the proposed methods used in this document.

In the context of IIR digital filters there are three main groups of approaches to (\ref{int:lpiirp}). Section \ref{iir:ls_trad} presents relevant work in the form of traditional optimization techniques. These are methods derived mainly from the applied mathematics community and are in general efficient and well understood. However the generality of such methods occasionally comes at the expense of being inefficient for some particular problems. Among the methods found in literature, the Davidon-Flecther-Powell ({\bf DFP}) algorithm \cite{dfp}, the damped Gauss-Newton method \cite{num_met_unc,invfreqz}, the Levenberg-Marquardt algorithm \cite{newt_freq,soren_newt}, and the method of Kumaresan \cite{kumar_burrus,hilde} form the basis of a number of methods to solve (\ref{int:iirlpp}).

A different approach to (\ref{int:iirlpp}) from traditional optimization methods consists in {\it linearizing} (\ref{int:lpiirp}) by transforming the problem into a simpler, linear form. While in principle this proposition seems inadequate (as the original problem is being transformed), Section \ref{iir:ls_lin} presents some logical attemps at linearizing (\ref{int:lpiirp}) and how they connect with the original problem. The concept of {\it equation error} (a {\bf weighted} form of the {\it solution error} that one is actually interested in solving) has been introduced and employed by a number of authors. In the context of filter design, E. Levy \cite{levy} presented an equation error linearization formulation in 1959 applied to analog filters. An alternative equation error approach presented by C. S. Burrus \cite{dfd} in 1987 is based on the methods by Prony \cite{prony} and Pade \cite{pade}. The method by Burrus can be applied to frequency domain digital filter design, and is used in selected stages in some of the algorithms presented in this work.

An extension of the equation error methods is the group of {\it iterative prefiltering} algorithms presented in Section \ref{iir:iterfilt}. These methods build on equation error methods by weighting (or {\it prefiltering}) their equation error formulation iteratively, with the intention to converge to the minimum of the solution error.  Sanathanan and Koerner \cite{trans_ratio} presented in 1963 an algorithm ({\bf SK}) that builds on an extension of Levy's method by iterating on Levy's formulation.  Sid-Ahmed, Chottera and Jullien \cite{sidahmed} presented in 1978 a similar algorithm to the SK method but applied to the digital filter problem.

A popular and well understood method is the one by Steiglitz and McBride \cite{time_iter,smb} introduced in 1966.  The {\bf SMB} method is time-domain based, and has been extended to a number of applications, including the frequency domain filter design problem \cite{soe}. Steiglitz and McBride used a two-phase method based on linearization. Initially (in {\it Mode-1}) their algorithm is essentially that of Sanathanan and Koerner but in time. This approach often diverges when close to the solution; therefore their method can optionally switch to {\it Mode-2}, where a more traditional derivative-based approach is used.  

A more recent linearization algorithm was presented by L. Jackson \cite{jackson} in 2008. His approach is an iterative prefiltering method based directly in frequency domain, and uses diagonalization of certain matrices for efficiency.

While intuitive and relatively efficient, most linearization methods share a common problem: they often diverge close to the solution (this effect has been noted by a number of authors; a thorough review is presented in \cite{soe}).  Section \ref{iir:soe} presents the {\it quasilinearization} method derived by A. Soewito \cite{soe} in 1990.  This algorithm is robust, efficient and well-tailored for the least squares IIR problem, and is the method of choice for this work.

\section{Finite Impulse Response Filters}
\label{ch:fir}
 
This chapter discusses the problem of designing Finite Impulse Response (FIR) digital filters according to the $\lp$ error criterion using Iterative Reweighted Least Squares methods. Section \ref{fir:trad} gives an introduction to FIR filter design, including an overview of traditional FIR design methods. For the purposes of this work we are particularly interested in $l_2$ and $l_{\infty}$ design methods, and their relation to relevant $\lp$ design problems. Section \ref{fir:linph} formally introduces the linear phase problem and presents results that are common to most of the problems considered in this work. Finally, Sections \ref{fir:complex} through \ref{fir:freqp} present the application of the Iterative Reweighted Least Squares algorithm to other important problems in FIR digital filter design, including the relevant contributions of this work.

\subsection{Traditional design of FIR filters}
\label{fir:trad} 

Section \ref{int:digfilt} introduced the notion of digital filters and filter design. In a general sense, an FIR filter design problem has the form
\[ \min_{\vh} \; \norm{f(\vh)} \]
where $f(\cdot)$ defines an error function that depends on $\vh$, and $\norm{\cdot}$ is an abitrary norm. While one could come up with a number of error formulations for digital filters, this chapter elaborates on the most commonly used, namely the linear phase and complex problems (both satisfy the linear form $f(\vh)=\vD-\mC\vh$ as will be shown later in this chapter). As far as norms, typically the $l_2$ and $l_{\infty}$ norms are used. One of the contributions of this work is to demonstrate the usefulness of the more general $\lp$ norms and their feasibility by using efficient IRLS-based algorithms.

\subsubsection{Traditional design of least squares ($l_2$) FIR filters}

Typically, FIR filters are designed by {\it discretizing} a desired frequency response $H_d(\om)$ by taking $L$ frequency samples at $\{\om_0, \om_1,\ldots,\om_{L-1}\}$. One could simply take the inverse Fourier transform of these samples and obtain $L$ filter coefficients; this approach is known as the {\it Frequency Sampling design method} \cite{dfd}, which basically interpolates the frequency spectrum over the samples. However, it is often more desirable to take a large number of samples to design a small filter (large in the sense that $L\gg N$, where $L$ is the number of frequency samples and $N$ is the filter order). The weighted least-squares $(l_2)$ norm (which considers the error energy) is defined by 
\be \label{fir:l2cont} \vep_2\triangleq\normsq{\ep(\om)}=\left(\frac{1}{\pi}\int_0^\pi W(\om)\abs{D(\om)-H(\om)}^2d\om \right)^\frac{1}{2} \ee
where $D(\om)$ and $H(\om)=\FT(\vh)$ are the desired and designed amplitude responses respectively. By acknowledging the convexity of (\ref{fir:l2cont}), one can drop the root term; therefore a discretized form of (\ref{fir:l2cont}) is given by
\be \label{fir:l2disc} \vep_2=\sum_{k=0}^{L-1} W(\om_k)\abs{D(\om_k)-H(\om_k)}^2 \ee
The solution of Equation (\ref{fir:l2disc}) is given by
\be \label{fir:l2disc2} \vh=\left(\mC^T\mW^T\mW\mC\right)^{-1} \mC^T\mW^T\mW\vD \ee
where $\mW=\diag(\sqrt{\vw})$ contains the weighting vector $\vw$. By solving (\ref{fir:l2disc2}) one obtains an optimal $l_2$ approximation to the desired frequency response $D(\om)$. Further discussion and other variations on least squares FIR design can be found in \cite{dfd}.

\subsubsection{Traditional design of minimax ($l_{\infty}$) FIR filters}

In contrast to $l_2$ design, an $l_{\infty}$ filter minimizes the maximum error across the designed filter's frequency response. A formal formulation of the problem \cite{digfiltanton,digfiltcunning} is given by
\be \label{fir:licont} \min_{\vh} \max_{\om} \; \abs{D(\om)-H(\om;\vh)} \ee
A discrete version of (\ref{fir:licont}) is given by
\be \label{fir:lidisc} \min_{\vh} \max_k \; \abs{D(\om_k)-C_k \vh} \ee
Within the scope of filter design, the most commonly approach to solving (\ref{fir:lidisc}) is the use of the {\it Alternation Theorem} \cite{approxcheney}, in the context of linear phase filters (to be discussed in Section \ref{fir:linph}).  In a nutshell the alternation theorem states that for a length-$N$ FIR linear phase filter there are at least $N+1$ {\it extrema points} (or frequencies). The Remez exchange algorithm \cite{dfd,digfiltanton,digfiltcunning} aims at finding these extrema frequencies iteratively, and is the most commonly used method for the minimax linear phase FIR design problem. Other approaches use more standard linear programming methods including the Simplex algorithm \cite{linprogchvat,strangappmath} or interior point methods such as Karmarkar's algorithm \cite{karmarl1}.

The $l_{\infty}$ problem is fundamental in filter design. While this document is not aimed covering the $l_{\infty}$ problem in depth, portions of this work are devoted to the use of IRLS methods for standard problems as well as some innovative uses of minimax optimization.

\subsection{Linear phase $\lp$ filter design}
\label{fir:linph} 

Linear phase FIR filters are important tools in signal processing. As will be shown below, they do not require the user to specify a phase response in their design (since the assumption is that the desired phase response is indeed linear). Besides, they satisfy a number of symmetry properties that allow for the reduction of dimensions in the optimization process, making them easier to design computationally. Finally, there are applications where a linear phase is desired as such behavior is more physically meaningful.

\subsubsection{Four types of linear phase filters}
\label{fir:4linph} 
The frequency response of an FIR filter $h(n)$ is given by
\[ H(\om) = \sum_{n=0}^{N-1} h(n) e^{-j\om n} \]
In general, $H(\om)=R(\om)+jI(\om)$ is a periodic complex function of $\om$ (with period $2\pi$). Therefore it can be written as follows,
\begin{align}
H(\om) &= R(\om) +jI(\om) \notag \\
&= A(\om)e^{j\phi(\om)} \label{fir:complex1}
\end{align}
where the {\it magnitude} response is given by 
\be \label{fir:linphmag} A(\om) =\abs{H(\om)}=\sqrt{R(\om)^2 + I(\om)^2} \ee
and the {\it phase} response is
\[ \phi(\om)= \arctan\left(\frac{I(\om)}{R(\om)}\right) \]
However $A(\om)$ is not analytic and $\phi(\om)$ is not continuous. From a computational point of view (\ref{fir:complex1}) would have better properties if both $A(\om)$ and $\phi(\om)$ were continuous analytic functions of $\om$; an important class of filters for which this is true is the class of {\it linear phase} filters \cite{dfd}.

Linear phase filters have a frequency response of the form 
\be \label{fir:linphform} H(\om) = A(\om) e^{j\phi(\om)} \ee 
where $A(\om)$ is the real, continuous {\it amplitude} response of $H(\om)$ and 
\[ \phi(\om)= K_1  +K_2\om \]
is a {\bf linear} phase function in $\om$ (hence the name); $K_1$ and $K_2$ are constants. The jumps in the phase response correspond to sign reversals in the magnitude resulting as defined in (\ref{fir:linphmag}).

Consider a length-$N$ FIR filter (assume for the time being that $N$ is odd). Its frequency response is given by 
\begin{align}
H(\om) &= \sum_{n=0}^{N-1} h(n)e^{-j\om n} \notag \\
&= e^{-j\om M} \sum_{n=0}^{2M} h(n)e^{j\om(M-n)} \label{fir:b} 
\end{align}
where $M = \frac{N-1}{2}$. Equation (\ref{fir:b}) can be written as follows, 
\bea 
H(\om) & = & e^{-j\om M} [ h(0) e^{j\om M} + \ldots + h(M-1) e^{j\om} + h(M) \nonumber \\
 & & \qquad + h(M+1) e^{-j\om} + \ldots + h(2M) e^{-j\om M} \label{fir:braces} ] 
\eea
It is clear that for an odd-length FIR filter to have the linear phase form described in (\ref{fir:linphform}), the term inside braces in (\ref{fir:braces}) must be a real function (thus becoming $A(\om)$). By imposing even symmetry on the filter coefficients about the midpoint ($n=M$), that is 
\[ h(k) = h(2M-k) \]
equation (\ref{fir:braces}) becomes
\be H(\om) = e^{-j\om M} \left[h(M) + 2\sum_{n=0}^{M-1} h(n) \cos\om(M-n)\right] \ee
Similarly, with odd symmetry (i.e. $h(k)=h(2M-k)$) equation (\ref{fir:braces}) becomes 
\be H(\om) = e^{j(\frac{\pi}{2}-\om M)} 2\sum_{n=0}^{M-1} h(n) \sin\om(M-n) \ee
Note that the term $h(M)$ disappears as the symmetry condition requires that
\[h(M)=h(N-M-1)=-h(M)=0\]
Similar expressions can be obtained for an even-length FIR filter, 
\begin{eqnarray} H(\om) & = & \sum_{n=0}^{N-1} h(n)e^{-j\om n} \nonumber \\ & = & e^{-j\om M} \sum_{n=0}^{\frac{N}{2}-1} h(n)e^{j\om(M-n)} \label{fir:lpeven} \end{eqnarray}
It is clear that depending on the combinations of $N$ and the symmetry of $h(n)$, it is possible to obtain four types of filters \cite{dfd,dtsp,cbesp}. Table \ref{fir:lptable} shows the four possible linear phase FIR filters described by (\ref{fir:linphform}), where the second column refers to the type of filter symmetry.

\begin{table}[th]
\begin{center}
\begin{tabular}{|>{\setlength{\parindent}{.3cm}}p{.4cm}||l|l|}
\hline & Even & \begin{tabular}{l@{$\;$=$\;$}l}
$A(\om)$ & $\displaystyle h(M)+2\sum_{n=0}^{M-1}h(n) \cos\om(M-n)$ \\
$\phi(\om)$ & $-\om M$ \end{tabular}
\\\cline{2-3}
\begin{rotate}{90} \hspace{2mm} $N$ Odd \end{rotate} & Odd &
\begin{tabular}{l@{$\;$=$\;$}l}
$A(\om)$ & $\displaystyle 2\sum_{n=0}^{M-1}h(n) \sin\om(M-n)$
\\ $\phi(\om)$ & $\frac{\pi}{2}-\om M$ \end{tabular}
\\\hline \hline
& Even & \begin{tabular}{l@{$\;$=$\;$}l} $A(\om)$ &
$\displaystyle h(M)+2\sum_{n=0}^{\frac{N}{2}-1}h(n)
\cos\om(M-n)$ \\ $\phi(\om)$ & $-\om M$ \end{tabular}
\\\cline{2-3}
\begin{rotate}{90} \hspace{1mm} $N$ Even \end{rotate} & Odd
&
\begin{tabular}{l@{$\;$=$\;$}l}
$A(\om)$ & $\displaystyle 2\sum_{n=0}^{\frac{N}{2}-1}h(n)
\sin\om(M-n)$ \\ $\phi(\om)$ & $\frac{\pi}{2}-\om M$
\end{tabular}
\\\hline
\end{tabular}
\end{center}
\caption{The four types of linear phase FIR filters.}
\label{fir:lptable}
\end{table}

\subsubsection{IRLS-based methods}
\label{fir:irlsmeth}

Section \ref{fir:4linph} introduced linear phase filters in detail. In this section we cover the use of IRLS methods to design linear phase FIR filters according to the $\lp$ optimality criterion. Recall from Section \ref{fir:4linph} that for any of the four types of linear phase filters their frequency response can be expressed as
\[ H(\om)=A(\om)e^{j(K_1+K_2\om)} \]
Since $A(\om)$ is a real continuous function as defined by Table \ref{fir:lptable}, one can write the linear phase $\lp$ design problem as follows
\be \label{fir:linphcont1} \min_{\va} \; \normp{D(\om)-A(\om;\va)}^p \ee
where $\va$ relates to $\vh$ by considering the symmetry properties outlined in Table \ref{fir:lptable}. Note that the two objects from the objective function inside the $\lp$ norm are real. By sampling (\ref{fir:linphcont1}) one can write the design problem as follows
\[ \min_{\va} \; \sum_k \abs{D(\om_k)-A(\om_k;\va)}^p \]
or
\be \label{fir:linphdisc1} \min_{\va} \; \sum_k \abs{D_k - \vC_k\va}^p \ee
where $D_k$ is the $k$-th element of the vector $\vD$ representing the sampled desired frequency response $D(\om_k)$, and $\vC_k$ is the $k$-th row of the trigonometric kernel matrix as defined by Table \ref{fir:lptable}.

One can apply the basic IRLS approach described in Section \ref{int:irls} to solve (\ref{fir:linphdisc1}) by posing this problem as a weighted least squares one:
\be \label{fir:err1} \min_{\va} \; \sum_k w_k \abs{\vD_k - \vC_k\va}^2 \ee
The main issue becomes iteratively finding suitable weights $\vw$ for (\ref{fir:err1}) so that the algorithm converges to the optimal solution $\va\opt$ of the $\lp$ problem (\ref{fir:linphcont1}). Existence of adequate weights is guaranteed by Theorem \ref{int:wexist} as presented in Section \ref{int:irls}; finding these optimal weights is indeed the difficult part. Clearly a reasonable choice for $\vw$ is that which turns (\ref{fir:err1}) into (\ref{fir:linphdisc1}), namely
\[ \vw=\abs{\vD - \mC\va}^{p-2} \]
Therefore the basic IRLS algorithm for problem (\ref{fir:linphdisc1}) would be:
\begin{enumerate}
\item Initialize the weights $\vw_0$ (a reasonable choice is to make them all equal to one).
\item At the $i$-th iteration the solution is given by
\be \label{fir:basic1} \va_{i+1}=[\mC^T\mW_i^T\mW_i\mC]^{-1} \mC^T\mW_i^T\mW_i\vD \ee
\item Update the weights with
\[ \vw_{i+1} = \abs{\vD - \mC\va_{i+1}}^{p-2} \]
\item Repeat the last steps until convergence is reached.
\end{enumerate}
It is important to note that $\mW_i=\diag(\sqrt{\vw_i})$. In practice it has been found that this approach has practical defficiencies, since the inversion required by (\ref{fir:basic1}) often leads to an ill-posed problem and, in most cases, convergence is not achieved.

As mentioned before, the basic IRLS method has drawbacks that make it unsuitable for practical implementations. Charles Lawson considered a version of this algorithm applied to the solution of $l_{\infty}$ problems (for details refer to \cite{lawphd}). His method has linear convergence and is prone to problems with proportionately small residuals that could lead to zero weights and the need for restarting the algorithm. In the context of $\lp$ optimization, Rice and Usow \cite{rul} built upon Lawson's method by adapting it to $\lp$ problems. Like Lawson's methods, the algorithm by Rice and Usow updates the weights in a multiplicative manner; their method shares similar drawbacks with Lawson's. Rice and Usow defined
\[ w_{i+1}(\om) = w_i^\alpha(\om)\abs{\ep_i(\om)}^\beta \]
where
\[ \alpha = \frac{\gamma(p-2)}{\gamma(p-2)+1} \] 
and
\[ \beta = \frac{\alpha}{2\gamma}=\frac{p-2}{2\gamma(p-2)+2} \]
and follow the basic algorithm. 

L. A. Karlovitz realized the computational problems associated with the basic IRLS method and improved on it by partially updating the filter coefficient vector. He defines
\be \label{fir:karl1} \hat{\va}_{i+1}=[\mC^T\mW_i^T\mW_i\mC]^{-1}\mC^T\mW_i^T\mW_i\vD \ee
and uses $\hat{\va}$ in
\be \label{fir:karl2} \va_{i+1} = \lam\hat{\va}_{i+1}+(1-\lam)\va_i \ee
where $\lam\!\in\![0,1]$ is a partial step parameter that must be adjusted at each iteration. Karlovitz's method \cite{karlov} has been shown to converge globally for even values of $p$ (where $2\!\leq\! p\!<\!\infty$). In practice, convergence problems have been found even under such assumptions. Karlovitz proposed the use of line searches to find the {\it optimal} value of $\lambda$ at each iteration, which basically creates an independent optimization problem nested inside each iteration of the IRLS algorithm. While computationally this search process for the optimal $\lambda$ makes Karlovitz's method impractical, his work indicates the feasibility of IRLS methods and proves that partial updating indeed overcomes some of the problems in the basic IRLS method. Furthermore, Karlovitz's method is the first one to depart from a multiplicative updating of the weights in favor of an additive updating on the filter coefficients. In this way some of the problems in the Lawson-Rice-Usow approach are overcome, especially the need for restarting the algorithm.

S. W. Kahng built upon the findings by Karlovitz by considering the process of finding an adequate $\lambda$ for partial updating. He applied Newton-Raphson's method to this problem and proposed a closed form solution for $\lambda$, given by
\be \label{fir:kahng1} \lambda = \frac{1}{p-1}\ee
resulting in
\be \label{fir:kahng2} \va_{i+1} = \lambda\hat{\va}_{i+1}+(1-\lam)\va_i \ee 
The rest of Kahng's algorithm follows Karlovitz's approach. However, since $\lam$ is fixed, there is no need to perform the linear search at each iteration. Kahng's method has an added benefit: since it uses Newton's method to find $\lam$, the algorithm tends to converge much faster than previous approaches. It has indeed been shown to converge quadratically. However, Newton-Raphson-based algorithms are not guaranteed to converge globally unless at some point the existing solution lies close enough to the solution, within their radius of convergence \cite{intropt}. Fletcher, Grant and Hebden\cite{fletlp} derived the same results independently.

Burrus, Barreto and Selesnick \cite{irls,barms,leastp} modified Kahng's methods in several important ways in order to improve on their initial and final convergence rates and the method's stability (we refer to this method as BBS). The first improvement is analogous to a {\it homotopy} \cite{numopt}. Up to this point all efforts in $\lp$ filter design attempted to solve the {\it actual} $\lp$ problem from the first iteration. In general there is no reason to believe that an initial guess derived from an unweighted $l_2$ formulation (that is, the $l_2$ design that one would get by setting $\vw_0=\hat{1}$) will look in any way similar to the actual $\lp$ solution that one is interested in. However it is known that there exists a continuity of $\lp$ solutions for $1\!<\!p\!<\!\infty$. In other words, if $\va_2^{\star}$ is the optimal $l_2$ solution, there exists a $p$ for which the optimal $\lp$ solution $\va_p^{\star}$ is arbitrarily close to $\va_2^{\star}$; that is, for a given $\delta\!>\!0$
\[ \norm{\va_2^{\star}-\va_p^{\star}}\!\leq\! \delta \hspace{1cm} \mbox{for some } p\in(2,\infty) \]
This fact allows anyone to {\it gradually move} from an $\lp$ solution to an $l_q$ solution.

To accelerate initial convergence, the BBS method of Burrus et al. initially solves for $l_2$ by setting $p_0=2$ and then sets $p_i=\sig\cdot p_{i-1}$, where $\sig$ is a convergence parameter defined by $1\!\leq\! \sig\!\leq\! 2$. Therefore at the $i$-th iteration 
\be \label{fir:irls1} p_i=\min \; (p_{des},\sig p_{i-1}) \ee
where $p_{des}$ corresponds to the desired $\lp$ solution. The implementation of each iteration follows Karlovitz's method with Kahng's choice of $\lam$, using the particular selection of $p$ given by (\ref{fir:irls1}).

To summarize, define the class of IRLS algorithms as follows: after $i$ iterations, given a vector $\va_i$ the IRLS iteration requires two steps,
\begin{enumerate}
\item Find $\vw_i=f(\va_i)$
\item Find $\va_{i+1}=g(\vw_i,\va_i)$
\end{enumerate}

The following is a summary of the IRLS-based algorithms discussed so far and their corresponding updating functions:
\begin{enumerate}
\item Basic IRLS algorithm.
\begin{itemize}
\item $\vw_i = \abs{\vD-\mC\va_i}^{p-2}$
\item $\mW_i = \diag(\sqrt{\vw_i})$
\item $\va_{i+1} = \left[\mC^T\mW_i^T\mW_i\mC\right]^{-1}\mC^T\mW_i^T\mW_i\vD$
\end{itemize}
\item Rice-Usow-Lawson (RUL) method
\begin{itemize}
\item $\vw_i = \vw_{i-1}^\alpha \abs{\vD-\mC\va_i}^\frac{\alpha}{2\gamma}$
\item $\mW_i = \diag(\vw_i)$
\item $\va_{i+1} = \left[ \mC^T\mW_i^T\mW_i\mC\right]^{-1}\mC^T\mW_i^T\mW_i\vD$
\item $\alpha = \frac{\gamma(p-2)}{\gamma(p-2)+1}$
\item $\gamma$ constant
\end{itemize}
\item Karlovitz' method
\begin{itemize}
\item $\vw_i = \abs{\vD-\mC\va_i}^{p-2}$
\item $\mW_i = \diag(\sqrt{\vw_i})$
\item $\va_{i+1}  = \lam\left[ \mC^T\mW_i^T\mW_i\mC\right]^{-1}\mC^T\mW_i^T\mW_i\vD+(1-\lam) \va_i$
\item $\lam$ constant
\end{itemize}
\item Kahng's method
\begin{itemize}
\item $\vw_i = \abs{\vD-\mC\va_i}^{p-2}$
\item $\mW_i = \diag(\sqrt{\vw_i})$
\item $\va_{i+1} = \left(\frac{1}{p-1}\right) \left[ \mC^T\mW_i^T\mW_i\mC\right]^{-1}\mC^T\mW_i^T\mW_i\vD+ \left(\frac{p-2}{p-1}\right) \va_i$
\end{itemize}
\item BBS method
\begin{itemize}
\item $p_i = \min \; (p_{des},\sig\cdot p_{i-1})$
\item $\vw_i = \abs{\vD-\mC\va_i}^{p_i-2}$
\item $\mW_i = \diag(\sqrt{\vw_i})$
\item $\va_{i+1} = \left(\frac{1}{p_i-1}\right) \left[ \mC^T\mW_i^T\mW_i\mC\right]^{-1}\mC^T\mW_i^T\mW_i\vD + \left(\frac{p_i-2}{p_i-1}\right) a_i$
\item $\sig$ constant
\end{itemize}
\end{enumerate}

\subsubsection{Modified {\it adaptive} IRLS algorithm}
\label{fir:modirls}

Much of the performance of a method is based upon whether it can actually converge given a certain error measure.  In the case of the methods described above, both convergence rate and stability play an important role in their performance. Both Karlovitz and RUL methods are supposed to converge linearly, while Kahng's and the BBS methods converge quadratically, since they both use a Newton-based additive update of the weights.

\begin{figure}[ht]
    \centerline{\psfig{figure=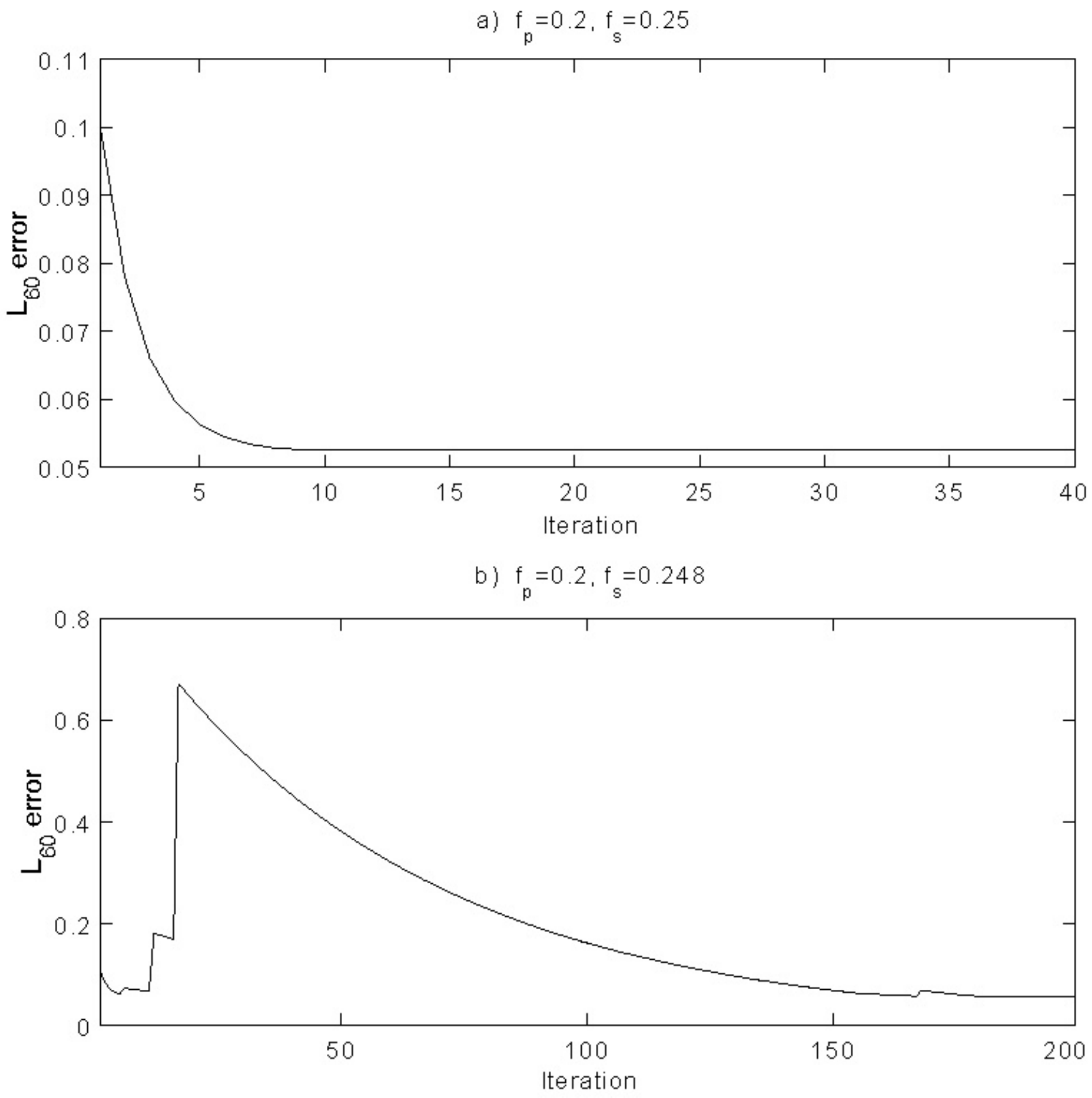,width=3.4in}}
    \caption{Error jumps on IRLS methods.}
    \label{fir:adap1}
\end{figure}

Barreto showed in \cite{barms} that the modified version of Kahng's method (or BBS) typically converges faster than the RUL algorithm. However, this approach presents some peculiar problems that depend on the transition bandwidth $\beta$. For some particular values of $\beta$, the BBS method will result in an ill-posed weight matrix that causes the $\lp$ error to increase dramatically after a few iterations as illustrated in Figure \ref{fir:adap1} (where $f=\om/2\pi$).

Two facts can be derived from the examples in Figure \ref{fir:adap1}: for this particular bandwidth the error increased slightly after the fifth and eleventh iterations, and increased dramatically after the sixteenth. Also, it is worth to notice that after such increase, the error started to decrease quadratically and that, at a certain point, the error became flat (thus reaching the numerical accuracy limits of the digital system).

\begin{figure}[ht]
    \centerline{\psfig{figure=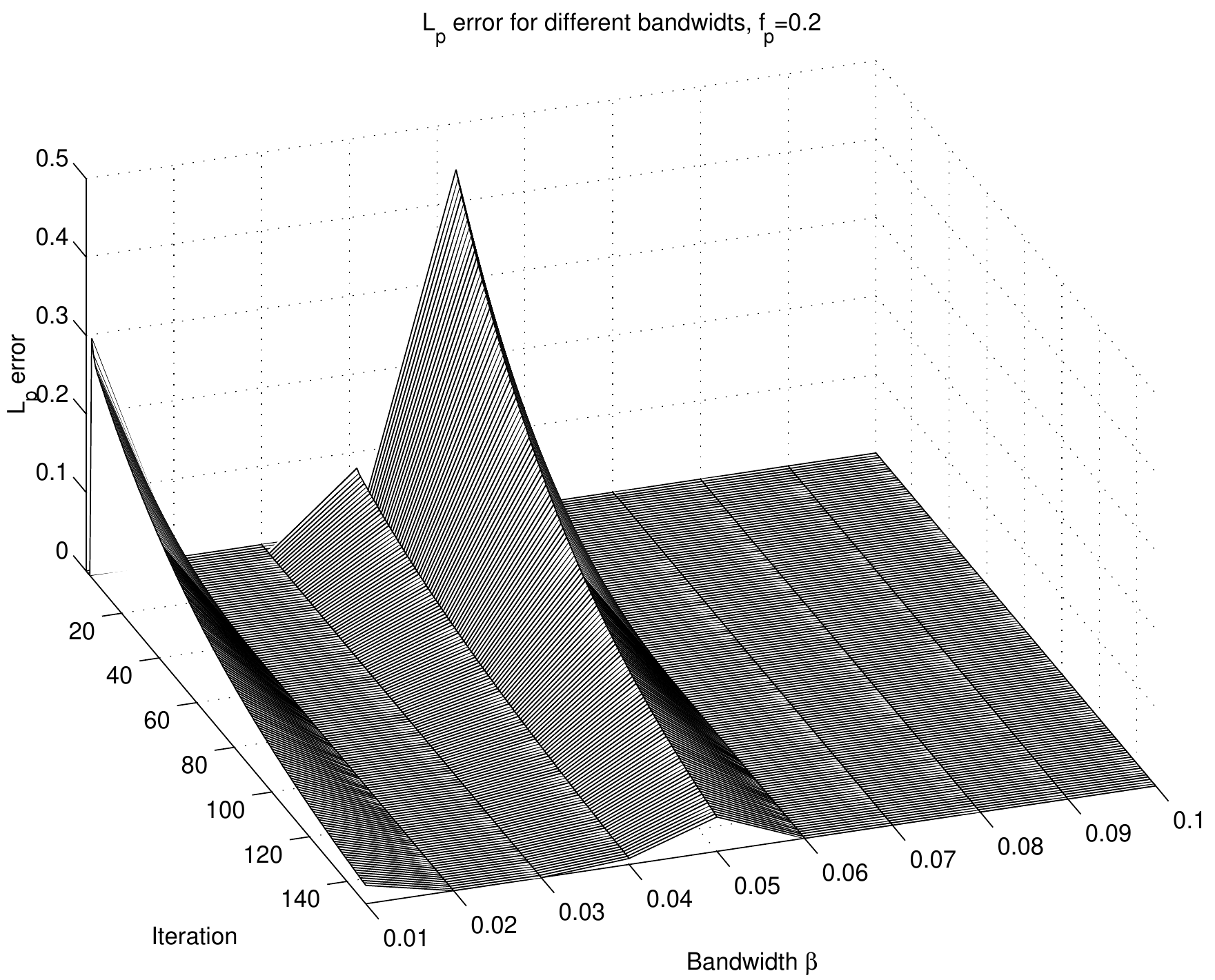,width=3.4in}}
    \caption{Relationship between bandwidth and error jumps.}
    \label{fir:adap2}
\end{figure}

The effects of different values of $\sig$ were studied to find out if a relationship between $\sig$ and the error increase could be determined. Figure \ref{fir:adap2} shows the $\lp$ error for different values of $\beta$ and for $\sig=1.7$. It can be seen that some particular bandwidths cause the algorithm to produce a very large error.

\begin{figure}[ht]
    \centerline{\psfig{figure=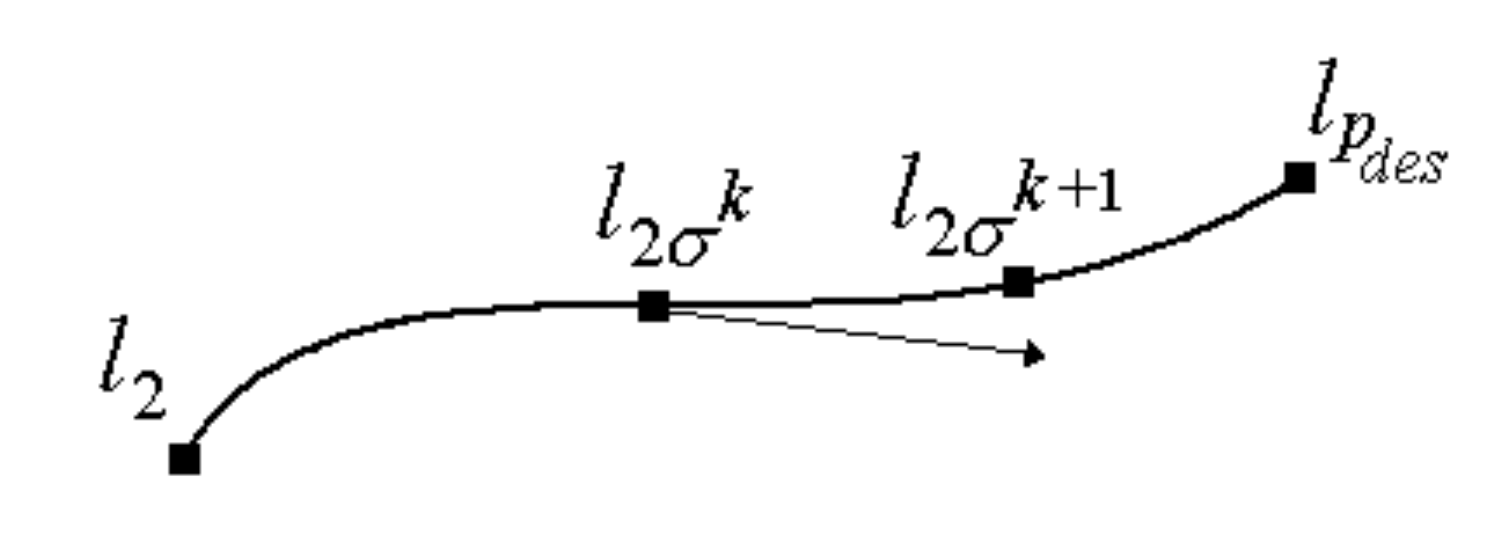,width=2.4in}}
    \caption{A step too long for IRLS methods.}
    \label{fir:bbslong}
\end{figure}

Our studies (as well as previous work from J. A. Barreto \cite{barms}) demonstrate that this error explosion occurs only for a small range of bandwidth specifications. Under most circumstances the BBS method exhibits fast convergence properties to the desired solution. However at this point it is not understood what causes the error increase and therefore this event cannot be anticipated. In order to avoid such problem, we propose the use of an adaptive scheme that modifies the BBS step. As $p$ increases the step from a current $\lp$ guess to the next also increases, as described in (\ref{fir:irls1}). In other words, at the $i$-th iteration one approximates the $l_{2\sig^i}$ solution (as long as the algorithm has not yet reached the desired $p$); the next iteration one approximates $l_{2\sig^{i+1}}$. There is always a possibility that these two solutions lie far apart enough that the algorithm takes a descent step so that the $l_{2\sig^{i+1}}$ {\it guess} is too far away from the actual $l_{2\sig^{i+1}}$ {\it solution}. This is better illustrated in Figure \ref{fir:bbslong}.

The conclusions derived above suggest the possibility to use an adaptive algorithm \cite{adap_irls_lp} that changes the value of $\sig$ so that the error always decreases. This idea was implemented by calculating temporary new weight and filter coefficient vectors that will not become the updated versions unless their resulting error is smaller than the previous one. If this is not the case, the algorithm "tries" two values of $\sig$, namely 
\be \label{fir:new1} \sig_L=\sig*(1-\delta) \qquad \mbox{and} \qquad \sig_H=\sig*(1+\delta) \ee
(where $\delta$ is an updating variable). The resulting errors for each attempt are calculated, and $\sig$ is updated according to the value that produced the smallest error. The error of this new $\sig$ is compared to the error of the nonupdated weights and coefficients, and if the new $\sig$ produces a smaller error, then such vectors are updated; otherwise another update of $\sig$ is performed. The {\it modified adaptive IRLS algorithm} can be summarized as follows,

\begin{enumerate}
\item Find the unweighted approximation $\va_0=\left[\mC^T\mC\right]^{-1}\mC^T\vD$ and use $p_0=2\sig$ (with $1\!\leq\! \sig\!\leq\! 2$)
\item Iteratively solve (\ref{fir:karl1}) and (\ref{fir:karl2}) using $\lam_i=\frac{1}{p_i-1}$ and find the resulting error $\vep_i$ for the $i$-th iteration
\item If $\vep_i \gg \vep_{i-1}$,
\begin{itemize}
\item Calculate (\ref{fir:new1})
\item Select the smallest of $\vep_{\sig_L}$ and $\vep_{\sig_H}$ to compare it with $\vep_i$ until a value is found that results in a decreasing error
\end{itemize}
Otherwise iterate as in the BBS algorithm.
\end{enumerate}

\begin{figure}
  \centerline{\psfig{figure=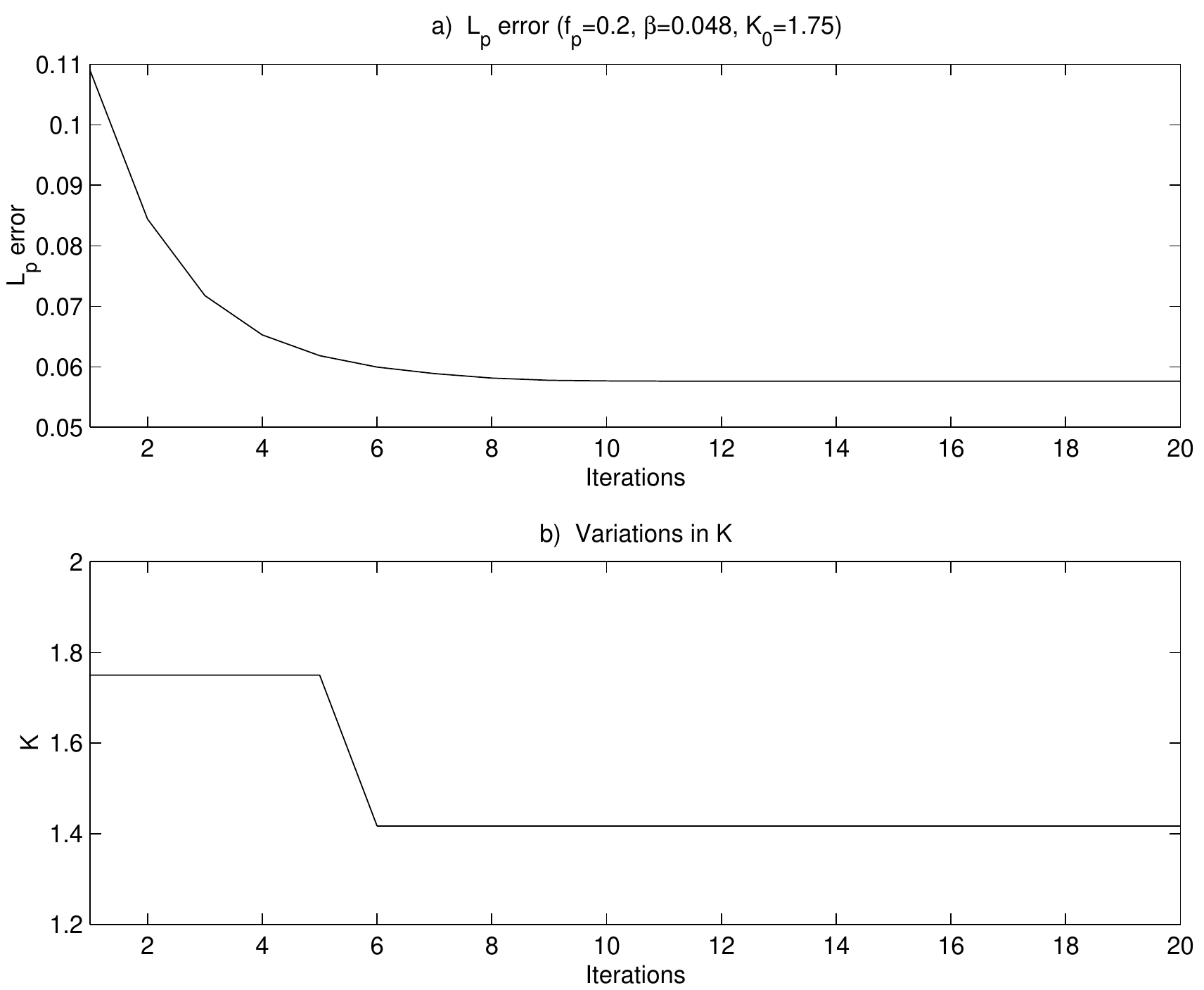,height=2.9in}}
  \caption{FIR design example using adaptive method.  a) $\lp$ error obtained with the adaptive method; b) Change of $\sig$.} 
  \label{fir:adapresults} 
\end{figure}
The algorithm described above changes the value of $\sig$ that causes the algorithm to produce a large error.  The value of $\sig$ is updated as many times as necessary without changing the values of the weights, the filter coefficients, or $p$. If an optimal value of $\sig$ exists, the algorithm will find it and continue with this new value until another update in $\sig$ becomes necessary. 

The algorithm described above was implemented for several combinations of $\sig$ and $\beta$; for all cases the new algorithm converged faster than the BBS algorithm (unless the values of $\sig$ and $\beta$ are such that the error never increases). The results are shown in Figure \ref{fir:adapresults}.a for the specifications from Figure \ref{fir:adap1}. Whereas using the BBS method for this particular case results in a large error after the sixteenth iteration, the adaptive method converged before ten iterations.

Figure \ref{fir:adapresults}.b illustrates the change of $\sig$ per iteration in the adaptive method, using an update factor of $\delta=0.1$. The $\lp$ error stops decreasing after the fifth iteration (where the BBS method introduces the large error); however, the adaptive algorithm adjusts the value of $\sig$ so that the $\lp$ error continues decreasing. The algorithm decreased the initial value of $\sig$ from 1.75 to its final value of 1.4175 (at the expense of only one additional iteration with $\sig=1.575$).

\begin{figure}[ht]
    \centerline{\psfig{figure=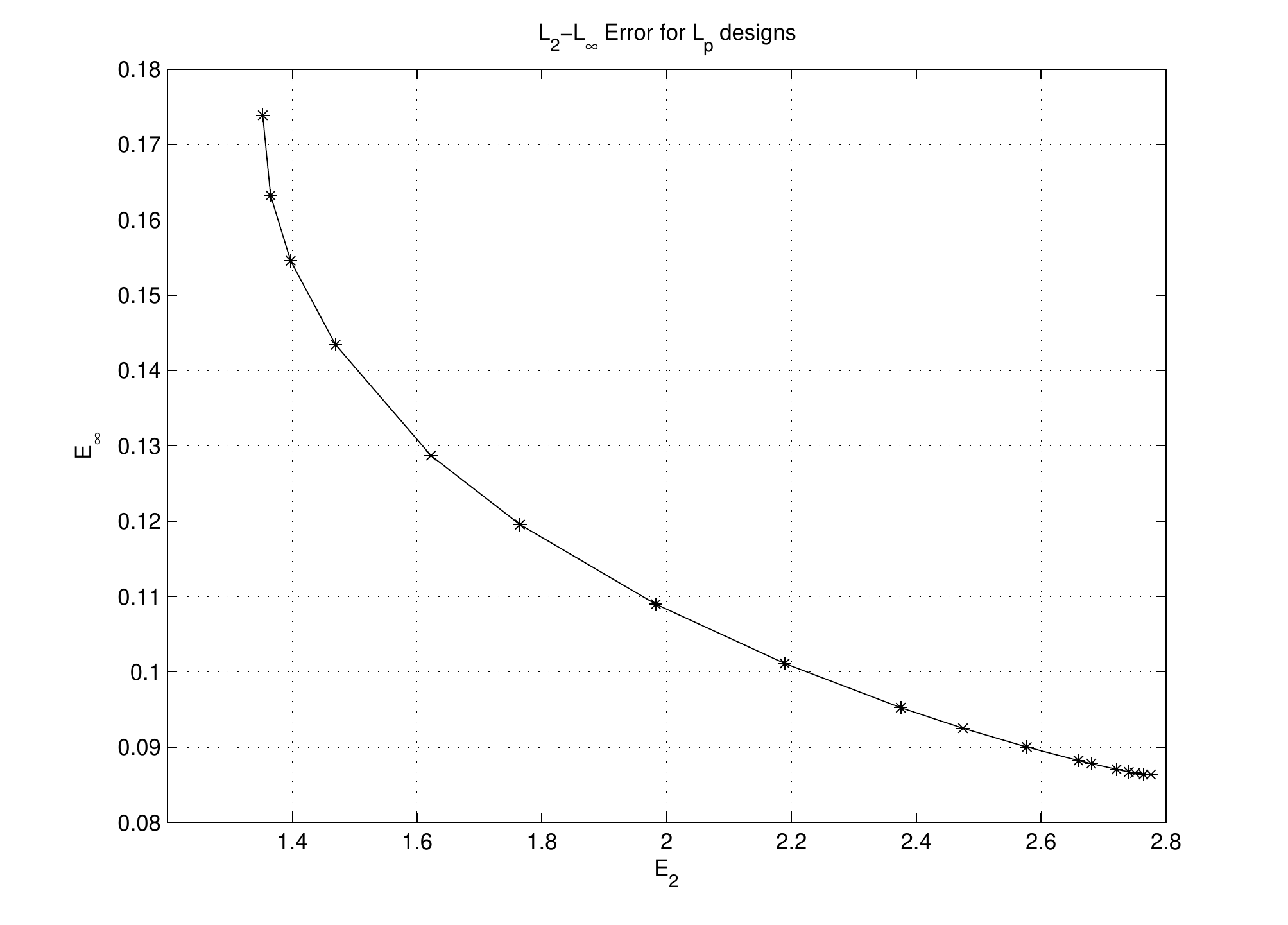,width=3.8in}}
    \caption{Relationship between $l_2$ and $l_{\infty}$ errors for $\lp$ FIR filter design.}
    \label{fir:lp_curve}
\end{figure}

One result worth noting is the relationship between $l_2$ and $l_{\infty}$ solutions and how they compare to $\lp$ designs. Figure \ref{fir:lp_curve} shows a comparison of designs for a length-21 Type-I linear phase low pass FIR filter with transition band defined by $f=\{0.2,0.24\}$. The curve shows the $l_2$ versus $l_{\infty}$ errors (namely $\vep_2$ and $\vep_{\infty}$); the values of $p$ used to make this curve were $p=\{2,2.2,2.5,3,4,5,7,10,15,20,30,50,60,100,150,200,400,\infty\}$ (Matlab's \texttt{firls} and \texttt{firpm} functions were used to design the $l_2$ and $l_{\infty}$ filters respectively). Note the very small decrease in $\vep_{\infty}$ after $p$ reaches $100$. The curve suggests that a better compromise between $\vep_2$ and $\vep_{\infty}$ can be reached by choosing $2\!<\!p\!<\!\infty$. Furthermore, to get better results one can concentrate on values between $p=5$ and $p=20$; fortunately, for values of $p$ so low no numerical complications arise and convergence is reached in a few iterations.

\subsection{Complex $\lp$ problem}
\label{fir:complex} 

The design of linear phase filters has been intensively discussed in literature. For the two most common error criteria ($l_2$ and $l_{\infty}$), optimal solution algorithms exist. The least squares norm filter can be found by solving an overdetermined system of equations, whereas the Chebishev norm filter is easily found by using either the Remez algorithm or linear programming. For many typical applications, linear phase filters are good enough; however, when arbitrary magnitude and phase constraints are required, a more complicated approach must be taken since such design results in a complex approximation problem.  By replacing $\mC$ in the linear phase algorithm with a complex Fourier kernel matrix, and the real desired frequency vector $\vD$ with a complex one, one can use the same algorithm from Section \ref{fir:modirls} to design complex $l_p$ filters.

\subsection{Magnitude $\lp$ problem}
\label{fir:magn} 

In some applications, the effects of phase are not a necessary factor to consider when designing a filter. For these applications, control of the filter's magnitude response is a priority for the designer. In order to improve the magnitude response of a filter, one must not explicitly include a phase, so that the optimization algorithm can look for the best filter that approximates a specified magnitude, without being constrained about optimizing for a phase response too.

\subsubsection{Power approximation formulation}

The magnitude approximation problem can be formulated as follows:
\be \label{fir:power1} \min_{\vh} \; \normp{D(\om) - \abs{H(\om;\vh)}}^p \ee
Unfortunately, the second term inside the norm (namely the absolute value function) is not differentiable when its argument is zero. Although one could propose ways to work around this problem, we propose the use of a different design criterion, namely the approximation of a desired magnitude squared. The resulting problem is
\[ \min_{\vh} \; \normp{D(\om)^2 -\abs{H(\om;\vh)\;}^2}^p \]
The autocorrelation $r(n)$ of a causal length-$N$ FIR filter $h(n)$ is given by 
\be \label{fir:corr} r(n) = h(n)\ast h(-n) = \sum_{k=-(N-1)}^{N-1} h(k)h(n+k) \ee 
The Fourier transform of the autocorrelation $r(n)$ is known as the {\it Power Spectral Density} function \cite{randsig} $R(\om)$ (or simply the SPD), and is defined as follows,
\bean R(\om) & = & \sum_{n=-(N-1)}^{N-1} r(n)e^{-j\om n} \\ & = & \sum_{n=-(N-1)}^{N-1} \sum_{k=-(N-1)}^{N-1} h(n)h(n+k) e^{-j\om n} \eean
From the properties of the Fourier Transform \cite[\S 3.3]{proman} one can show that there exists a frequency domain relationship between $h(n)$ and $r(n)$ given by 
\[ R(\om) = H(\om)\cdot H^{\ast}(-\om) = \abs{H(\om)}^2 \]
This relationship suggests a way to design magnitude-squared filters, namely by using the filter's autocorrelation coefficients instead of the filter coefficients themselves. In this way, one can avoid the use of the non-differentiable magnitude response.

An important property to note at this point is the fact that since the filter coefficients are real, one can see from (\ref{fir:corr}) that the autocorrelation function $r(n)$ is symmetric; thus it is sufficient to consider its last $N$ values. As a result, the PSD can be written as 
\[ R(\om) = \sum_n r(n) e^{-j\om n} = r(0) + \sum_{n=1}^{N-1} 2r(n) \cos \om n \]
in a similar way to the linear phase problem.

The symmetry property introduced above allows for the use of the $\lp$ linear phase algorithm of section (\ref{fir:linph}) to obtain the autocorrelation coefficients of $h(n)$. However, there is an important step missing in this discussion: how to obtain the filter coefficients from its autocorrelation. To achieve this goal, one can follow a procedure known as {\it Spectral Factorization}. The objective is to use the autocorrelation coefficients $\vr\in \real^N$ instead of the filter coefficients $\vh\in \real^N$ as the optimization variables. The variable transformation is done using (\ref{fir:spe3}), which is not a one-to-one transformation. Because of the last result, there is a necessary condition for a vector $\vr\in \real^N$ to be a valid autocorrelation vector of a filter. This is summarized \cite{specfact} in the {\it spectral factorization theorem}, which states that $\vr\in \real^N$ is the autocorrelation function of a filter $h(n)$ if and only if $R(\om)\!\geq\! 0$ for all $\om\in [0,\pi]$. This turns out to be a necessary and sufficient condition \cite{specfact} for the existence of $r(n)$. Once the autocorrelation vector $\vr$ is found using existing robust interior-point algorithms, the filter coefficients can be calculated via spectral factorization techniques.

Assuming a valid vector $\vr\in \real^N$ can be found for a particular filter $\vh$, the problem presented in (\ref{fir:power1}) can be rewritten as 
\begin{equation} \label{fir:spe3} L(\om)^2\!\leq\! R(\om)\!\leq\! U(\om)^2 \; \; \; \; \forall \; \om\in [0,\pi] \end{equation}
In (\ref{fir:spe3}) the existence condition $R(\om)\!\geq\! 0$ is redundant since $0\!\leq\! L(\om)^2$ and, thus, is not included in the problem definition. For each $\om$, the constraints of (\ref{fir:spe3}) constitute a pair of linear inequalities in the vector $\vr$; therefore the constraint is convex in $\vr$. Thus the change of variable transforms a nonconvex optimization problem in $\vh$ into a convex problem in $\vr$.

\subsection{$\lp$ error as a function of frequency}
\label{fir:freqp}

Previous sections have discussed the importance of complex least-square and Chebishev error criteria in the context of filter design. In many applications any of these two approaches would provide adequate results. However, a case could be made where one might want to minimize the error energy in a range of frequencies while keeping control of the maximum error in a different band. This idea results particularly interesting when one considers the use of different $\lp$ norms in different frequency bands. In principle one would be interested in solving
\be \label{fir:lpfreq1} \min_{\vh} \; \norm{D(\om_{pb})-H(\om_{pb};\vh)}_p + \norm{D(\om_{sb})-H(\om_{sb};\vh)}_q \ee
where $\{\om_{pb}\in\Omega_{pb},\om_{sb}\in\Omega_{sb}\}$ represent the pass and stopband frequencies respectively. In principle one would want $\Omega_{pb} \cap \Omega_{sb} = \{\emptyset\}$. Therefore problem (\ref{fir:lpfreq1}) can be written as
\be \begin{split} \label{fir:lpfreq2} \min_{\vh} \; & \sqrt[p]{\sum_{\om_{pb}}\; \abs{D(\om_{pb})-H(\om_{pb};\vh)}^p} \\ & \quad + \sqrt[q]{\sum_{\om_{sb}}\; \abs{D(\om_{sb})-H(\om_{sb};\vh)}^q} \end{split} \ee
One major obstacle in (\ref{fir:lpfreq2}) is the presence of the roots around the summation terms. These roots prevent us from writing (\ref{fir:lpfreq2}) in a simple vector form. Instead, one can consider the use of a similar {\it metric} function as follows
\be \label{fir:lpfreq3} \begin{split} \min_{\vh} \; & \sum_{\om_{pb}}\; \abs{D(\om_{pb})-H(\om_{pb};\vh)}^p + \\ & \quad \sum_{\om_{sb}}\; \abs{D(\om_{sb})-H(\om_{sb};\vh)}^q \end{split} \ee
This expression is similar to (\ref{fir:lpfreq2}) but does not include the root terms. An advantage of using the IRLS approach on (\ref{fir:lpfreq3}) is that one can formulate this problem in the frequency domain and properly separate residual terms from different bands into different vectors. In this manner, the $\lp$ modified measure given by (\ref{fir:lpfreq3}) can be made into a frequency-dependent function of $p(\om)$ as follows,
\[ \min_{\vh} \; \norm{D(\om)-H(\om;\vh)}_{p(\om)}^{p(\om)}= \sum_{\om}\; \abs{D(\om)-H(\om;\vh)}^{p(\om)}\]
 Therefore this {\it frequency-varying} $\lp$ problem can be solved following the modified IRLS algorithm outlined in Section \ref{fir:modirls} with the following modification: at the $i$-th iteration the weights are updated according to
\[ \vw_i = \abs{\vD-\mC\va_{i}}^{p(\om)-2} \]

It is fundamental to note that the proposed method does not indeed solve a linear combination of $\lp$ norms. In fact, it can be shown that the expression (\ref{fir:lpfreq3}) is not a norm but a metric. While from a theoretical perspective this fact might make (\ref{fir:lpfreq3}) a less interesting distance, as it turns out one can use (\ref{fir:lpfreq3}) to solve the far more interesting CLS problem, as discussed below in Section \ref{fir:cls}.
 
\subsection{Constrained Least Squares (CLS) problem}
\label{fir:cls}

One of the common obstacles to innovation occurs when knowledge settles on a particular way of dealing with problems. While new ideas keep appearing suggesting innovative approaches to design digital filters, it is all too common in practice that $l_2$ and $l_{\infty}$ dominate error criteria specifications. This section is devoted to exploring a different way of thinking about digital filters. It is important to note that up to this point we are not discussing an algorithm yet. The main concern being brought into play here is the specification (or description) of the design problem. Once the {\it Constrained Least Squares} (CLS) problem formulation is introduced, we will present an IRLS implementation to solve it, and will justify our approach over other existing approaches. It is the author's belief that under general conditions one should always use our IRLS implementation over other methods, especially when considering the associated management of transition regions.

The CLS problem was introduced in Section \ref{int:fir} and is repeated here for clarity,
\be \label{fir:clsbasic} \begin{array}{ll}
\underset{\vh}{\min}  & \normsq{D(\om)-H(\om;\vh)} \\
\mbox{subject to} & \abs{D(\om)-H(\om;\vh)} \!\leq\! \tau \end{array} \ee
To the best of our knowledge this problem was first introduced in the context of filter design by John Adams \cite{adamsfircls} in 1991. The main idea consists in approximating iteratively a desired frequency response in a least squares sense except in the event that any frequency exhibits an error larger than a specified tolerance $\tau$. At each iteration the problem is adjusted in order to reduce the error on offending frequencies (i.e. those which do not meet the constraint specifications). Ideally, convergence is reached when the {\it altered} least squares problem has a frequency response whose error does not exceed constraint specifications. As will be shown below, this goal might not be attained depending on how the problem is posed.

Adams and some collaborators have worked in this problem and several variations \cite{adamspcsl}. However his main (and original) problem was illustrated in \cite{adamsfircls} with the following important assumption: {\it the definition of a desired frequency response must include a fixed non-zero width transition band}. His method uses Lagrange multiplier theory and alternation methods to find frequencies that exceed constraints and minimize the error at such locations, with an overall least squares error criterion.

Burrus, Selesnick and Lang \cite{constls} looked at this problem from a similar perspective, but relaxed the design specifications so that only a {\it transition frequency} needs to be specified. The actual transition band does indeed exist, and it centers itself around the specified transition frequency; its width adjusts as the algorithm iterates (constraint tolerances are still specified). Their solution method is similar to Adams' approach, and explicitly uses the Karush-Kuhn-Tucker (KKT) conditions together with an alternation method to minimize the least squares error while constraining the maximum error to meet specifications.

C. S. Burrus and the author of this work have been working on the CLS problem using IRLS methods with positive results. This document is the first thorough presentation of the method, contributions, results and code for this approach, and constitutes one of the main contributions of this work. It is crucial to note that there are two separate issues in this problem: on one hand there is the matter of the actual problem formulation, mainly depending on whether a transition band is specified or not; on the other hand there is the question of how the selected problem description is actually met (what algorithm is used). Our approach follows the problem description by Burrus et al. shown in \cite{constls} with an IRLS implementation.
 
\subsubsection{Two problem formulations}
\label{fir:cls2prob}

As mentioned in Section \ref{fir:cls}, one can address problem (\ref{fir:clsbasic}) in two ways depending on how one views the role of the transition band in a CLS problem. The original problem posed by Adams in \cite{adamsfircls} can be written as follows,
\be \label{fir:cls2} \begin{array}{ll}
\underset{\vh}{\min} & \normsq{D(\om)-H(\om;\vh)} \\
\mbox{s.t.} & \abs{D(\om)-H(\om;\vh)} \!\leq\! \tau \quad \forall \; \om\in[0,\om_{pb}]\cup[\om_{sb},\pi] \end{array} \ee
where $0\!<\!\om_{pb}\!<\!\om_{sb}\!<\!\pi$. From a traditional standpoint this formulation feels familiar. It assigns {\bf fixed frequencies} to the transition band edges as a number of filter design techniques do. As it turns out, however, one might not want to do this in CLS design. 

An alternate formulation to (\ref{fir:cls2}) could implicitly introduce a {\it transition} frequency $\om_{tb}$ (where $\om_{pb}\!<\!\om_{tb}\!<\!\om_{sb}$); the user only specifies $\om_{tb}$. Consider
\be \label{fir:cls3} \begin{array}{lll}
\underset{\vh}{\min} & \normsq{D(\om)-H(\om;\vh)} & \forall \; \om\in[0,\pi] \\
\mbox{subject to} & \abs{D(\om)-H(\om;\vh)} \!\leq\! \tau & \forall \; \om\in[0,\om_{pb}]\cup[\om_{sb},\pi] \end{array} \ee
The algorithm at each iteration generates an {\it induced transition band} in order to satisfy the constraints in (\ref{fir:cls3}). Therefore $\{\om_{pb},\om_{sb}\}$ vary at each iteration.

\begin{figure}[ht]
    \centerline{\psfig{figure=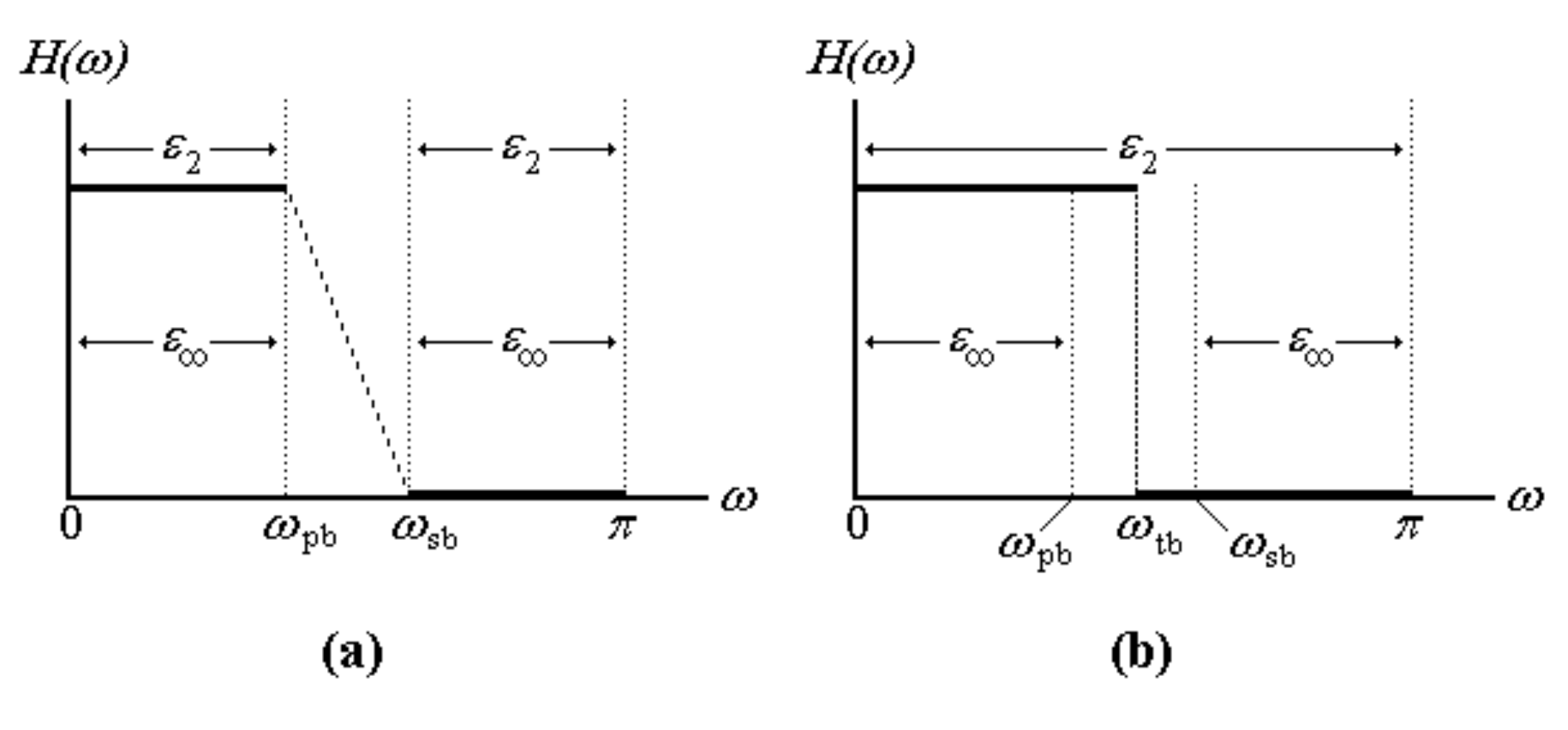,width=3.2in}}
    \caption{Two formulations for Constrained Least Squares problems.}
    \label{fir:clsfilt1}
\end{figure}

It is critical to point out the differences between (\ref{fir:cls2}) and (\ref{fir:cls3}). Figure \ref{fir:clsfilt1}.a explains Adams' CLS formulation, where the desired filter response is only specified at the fixed pass and stop bands. At any iteration, Adam's method attempts to minimize the least squares error ($\vep_2$) at both bands while trying to satisfy the constraint $\tau$. Note that one could think of the constraint requirements in terms of the Chebishev error $\vep_{\infty}$ by writing (\ref{fir:cls2}) as follows,
\[ \begin{array}{ll}
\underset{\vh}{\min} & \normsq{D(\om)-H(\om;\vh)} \\
\mbox{s.t.} & \norm{D(\om)-H(\om;\vh)}_{\infty} \!\leq\! \tau \quad \forall \; \om\in[0,\om_{pb}]\cup[\om_{sb},\pi] \end{array} \]
In contrast, Figure \ref{fir:clsfilt1}.b illustrates our proposed problem (\ref{fir:cls3}). The idea is to minimize the least squared error $\vep_2$ across {\bf all} frequencies while 
ensuring that constraints are met in an intelligent manner. At this point one can think of the interval $(\om_{pb},\om_{sb})$ as an {\it induced} transition band, useful for the purposes of constraining the filter. Section \ref{fir:cls2soln} presents the actual algorithms that solve (\ref{fir:cls3}), including the process of finding $\{\om_{pb},\om_{sb}\}$.

\begin{figure}[ht]
    \centerline{\psfig{figure=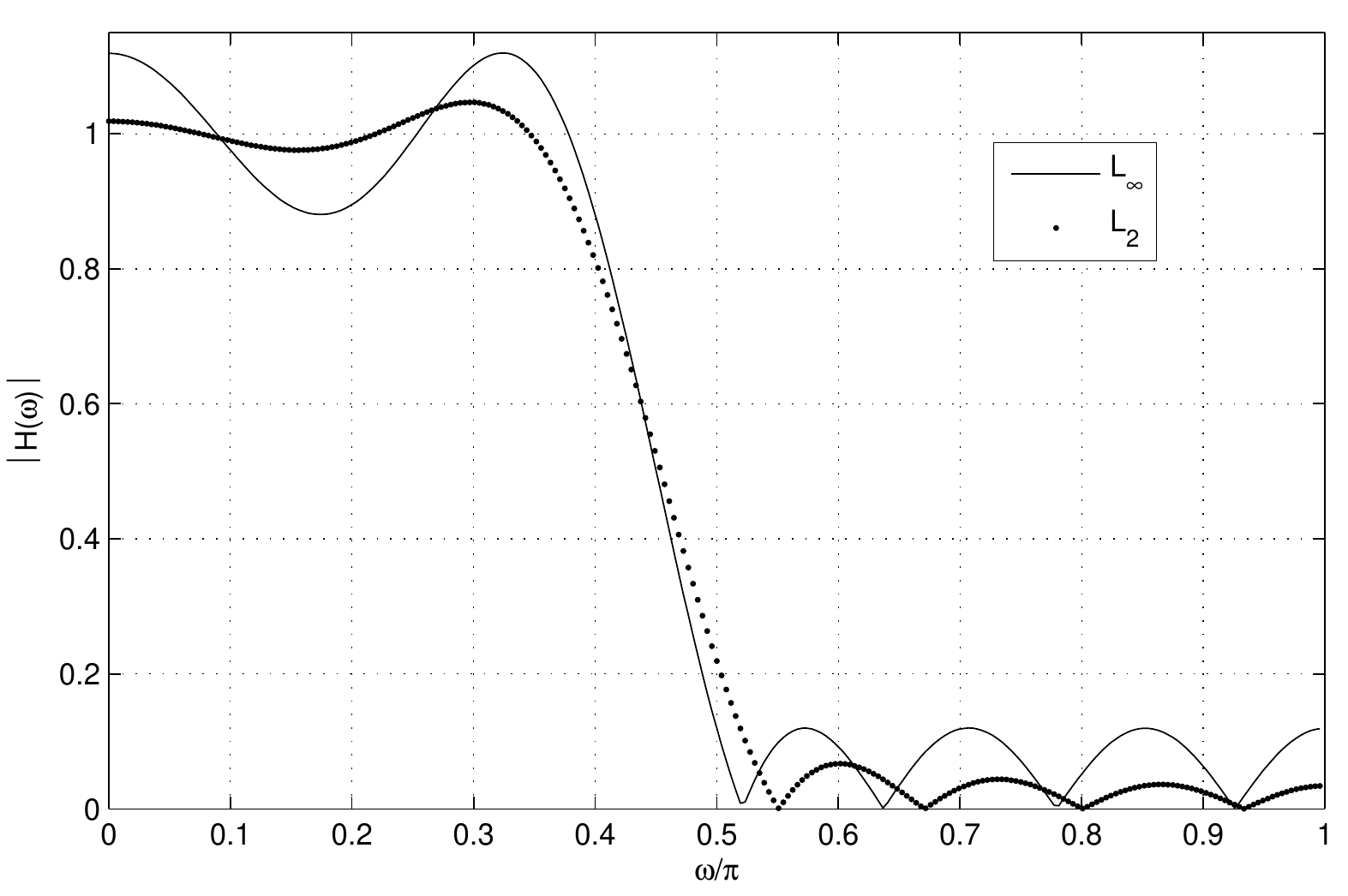,width=3.5in}}
    \caption{Comparison of $l_2$ and $l_{\infty}$ filters.}
    \label{fir:clsl2linf}
\end{figure}

It is important to note an interesting behavior of transition bands and extrema points in $l_2$ and $l_{\infty}$ filters.  Figure \ref{fir:clsl2linf} shows $l_2$ and $l_{\infty}$ length-15 linear phase filters (designed using Matlab's \texttt{firls} and \texttt{firpm} functions); the transition band was specified at $\{\om_{pb}=0.4/\pi,\om_{sb}=0.5/\pi\}$. The dotted $l_2$ filter illustrates an important behavior of least squares filters: typically the maximum error of an $l_2$ filter is located at the transition band. The solid $l_{\infty}$ filter shows why minimax filters are important: despite their larger error across most of the bands, the filter shows the same maximum error at all extrema points, including the transition band edge frequencies. In a CLS problem then, typically an algorithm will attempt to reduce iteratively the maximum error (usually located around the transition band) of a series of least squares filters.

\begin{figure}[ht]
    \centerline{\psfig{figure=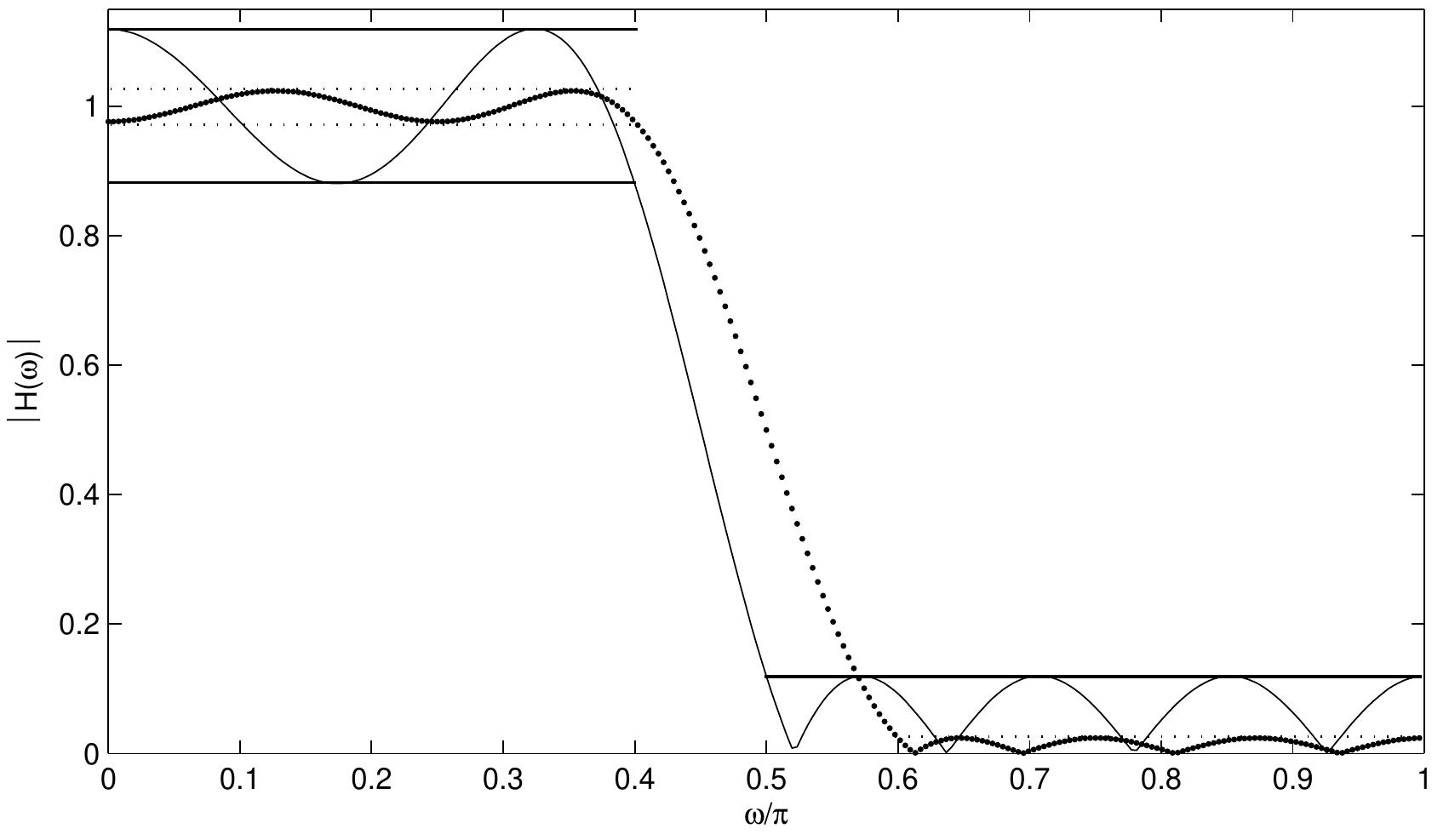,width=3.5in}}
    \caption{Effects of transition bands in $l_{\infty}$ filters.}
    \label{fir:clslinflinf}
\end{figure}

Another important fact results from the relationship between the transition band width and the resulting error amplitude in $l_{\infty}$ filters. Figure \ref{fir:clslinflinf} shows two $l_{\infty}$ designs; the transition bands were set at $\{0.4/\pi,0.5/\pi\}$ for the solid line design, and at $\{0.4/\pi,0.6/\pi\}$ for the dotted line one. One can see that by widening the transition band a decrease in error ripple amplitude is induced. 

These two results together illustrate the importance of the transition bandwidth for a CLS design. Clearly one can decrease maximum error tolerances by widening the transition band. Yet finding the perfect balance between a transition bandwidth and a given tolerance can prove a difficult task, as will be shown in Section \ref{fir:cls2soln}. Hence the relevance of a CLS method that is not restricted by two types of specifications competing against each other. In principle, one should just determine how much error one can live with, and allow an algorithm to find the optimal transition band that meets such tolerance.

\subsubsection{Two problem solutions}
\label{fir:cls2soln}
Section \ref{fir:cls2prob} introduced some important remarks regarding the behavior of extrema points and transition bands in $l_2$ and $l_{\infty}$ filters. As one increases the constraints on an $l_2$ filter, the result is a filter whose frequency response looks more and more like an $l_{\infty}$ filter. 

Section \ref{fir:freqp} introduced the frequency-varying problem and an IRLS-based method to solve it. It was also mentioned that, while the method does not solve the intended problem (but a similar one), it could prove to be useful for the CLS problem. As it turns out, in CLS design one is merely interested in solving an unweighted, constrained least squares problem. In this work, we achieve this by solving a sequence of weighted, unconstrained least squares problems, where the sole role of the weights is to "constraint" the maximum error of the frequency response at each iteration. In other words, one would like to find weights $w$ such that
\[ \begin{array}{ll}
\underset{\vh}{\min} & \normsq{D(\om)-H(\om;\vh)} \\
\mbox{s.t.} & \norm{D(\om)-H(\om;\vh)}_{\infty} \!\leq\! \tau \quad \forall \; \om\in[0,\om_{pb}]\cup[\om_{sb},\pi] \end{array} \]
is equivalent to 
\[ \min_{\vh} \; \normsq{w(\om)\cdot(D(\om)-H(\om;\vh))}  \]
Hence one can revisit the frequency-varying design method and use it to solve the CLS problem. Assuming that one can reasonably approximate $l_{\infty}$ by using high values of $p$, at each iteration the main idea is to use an $\lp$ weighting function only at frequencies where the constraints are exceeded. A formal formulation of this statement is
\[ w\left(\ep(\om)\right) = \left\{ \begin{array}{ll} \abs{\ep(\om)}^{\frac{p-2}{2}} & \mbox{if $\abs{\ep(\om)}\!>\!\tau$} \\ 1 & \mbox{otherwise} \end{array} \right. \]

Assuming a suitable weighting function existed such that the specified tolerances are related to the frequency response constraints, the IRLS method would iterate and assign rather large weights to frequencies exceeding the constraints, while inactive frequencies get a weight of one. As the method iterates, frequencies with large errors move the response closer to the desired tolerance. Ideally, all the active constraint frequencies would eventually meet the constraints. Therefore the task becomes to find a suitable weighting function that {\it penalizes} large errors in order to have all the frequencies satisfying the constraints; once this condition is met, we have reached the desired solution.
\begin{figure}[ht]
    \centerline{\psfig{figure=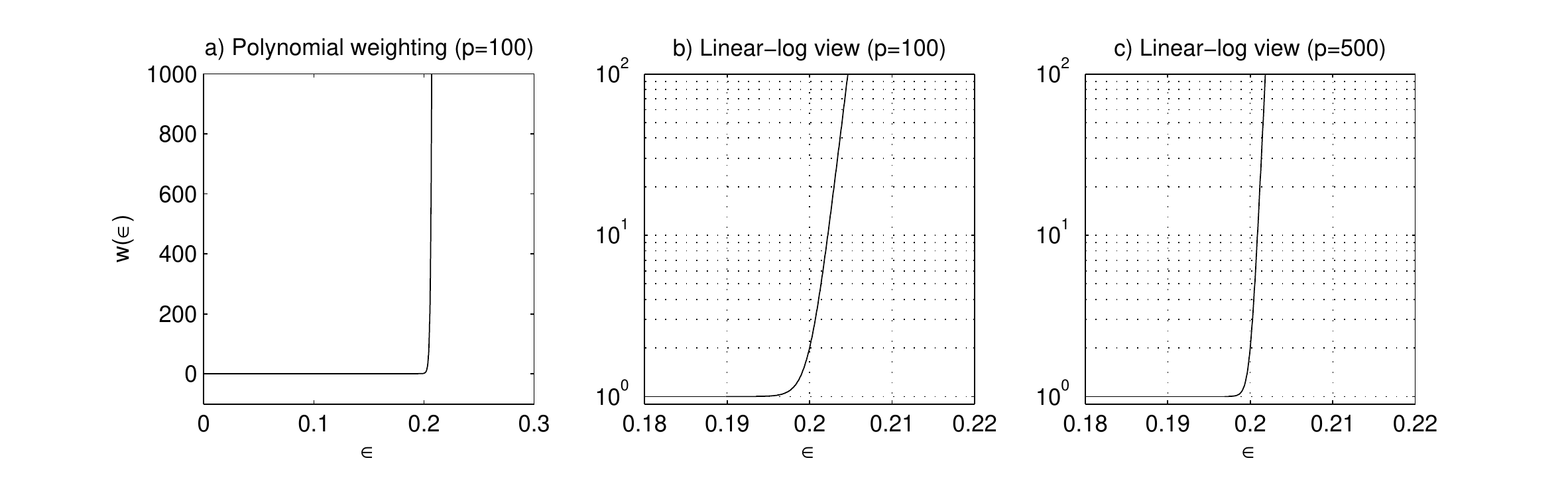,width=3.9in}}
    \caption{CLS polynomial weighting function.}
    \label{fir:poly_weight}
\end{figure}

One proposed way to find adequate weights to meet constraints is given by a {\it polynomial weighting function} of the form
\[ w(\om) = 1 + \left|\frac{\ep(\om)}{\tau}\right|^{\frac{p-2}{2}} \]
where $\tau$ effectively serves as a threshold to determine whether a weight is dominated by either unity or the familiar $\lp$ weighting term. Figure \ref{fir:poly_weight} illustrates the behavior of such a curve. 
\begin{figure}[h]
    \centerline{\psfig{figure=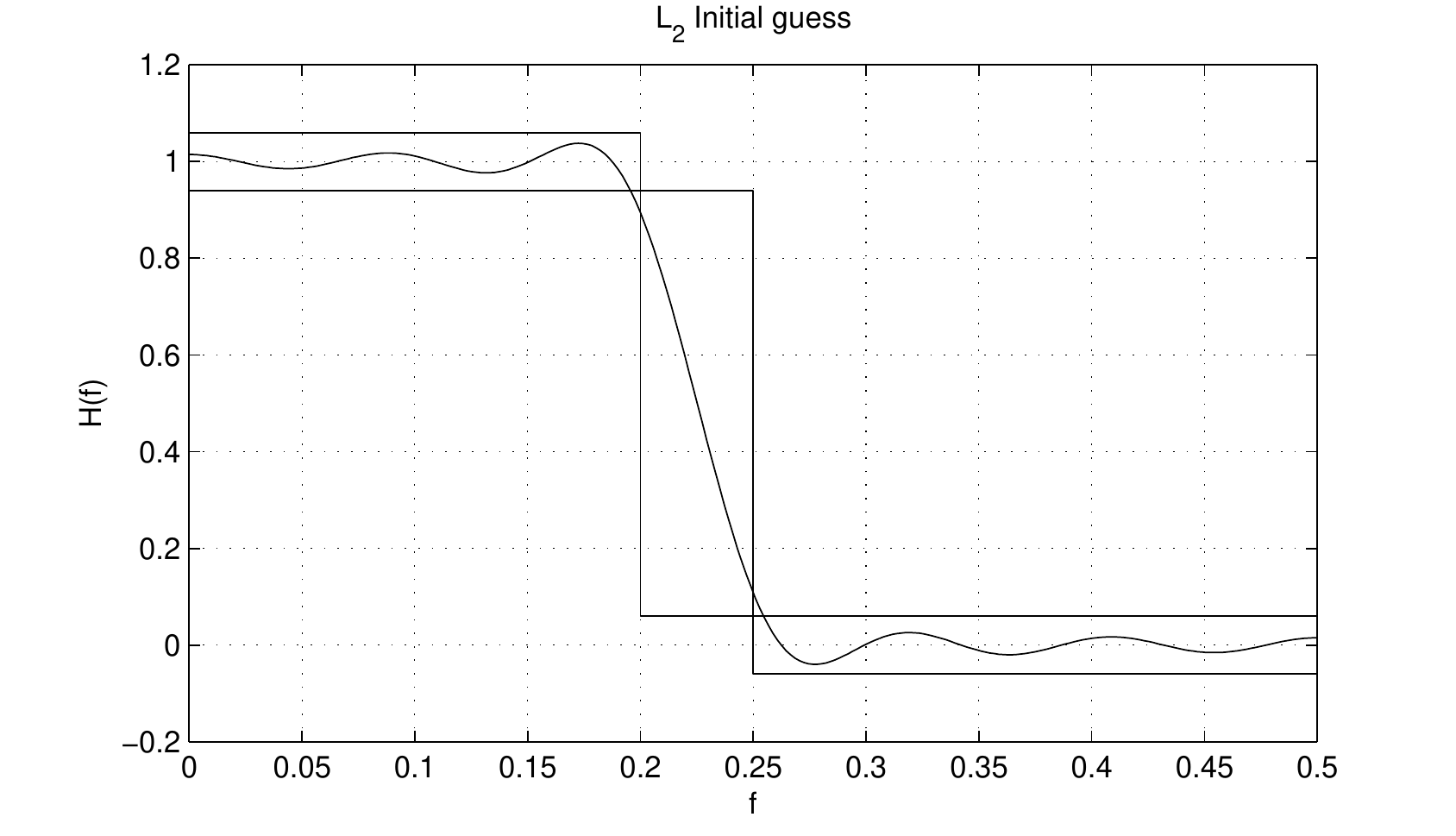,width=3.7in}}
    \caption{Original $l_2$ guess for CLS algorithm.}
    \label{fir:cls_orig_L2}
\end{figure}
\begin{figure}[p]
    \centerline{\psfig{figure=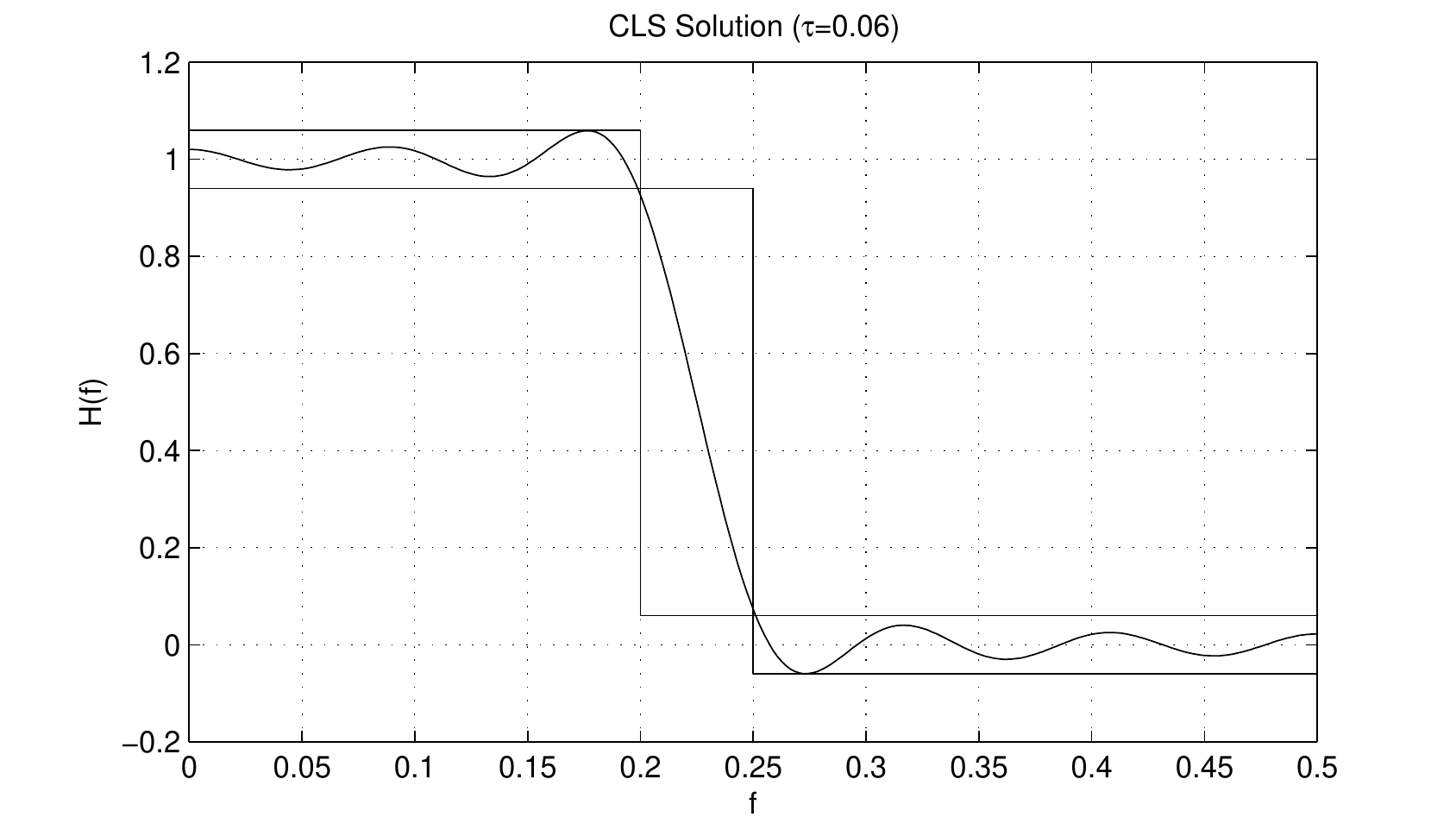,width=3.7in}}
    \caption{CLS design example using mild constraints.}
    \label{fir:cls_mild_stb}
\end{figure}
\begin{figure}[p]
    \centerline{\psfig{figure=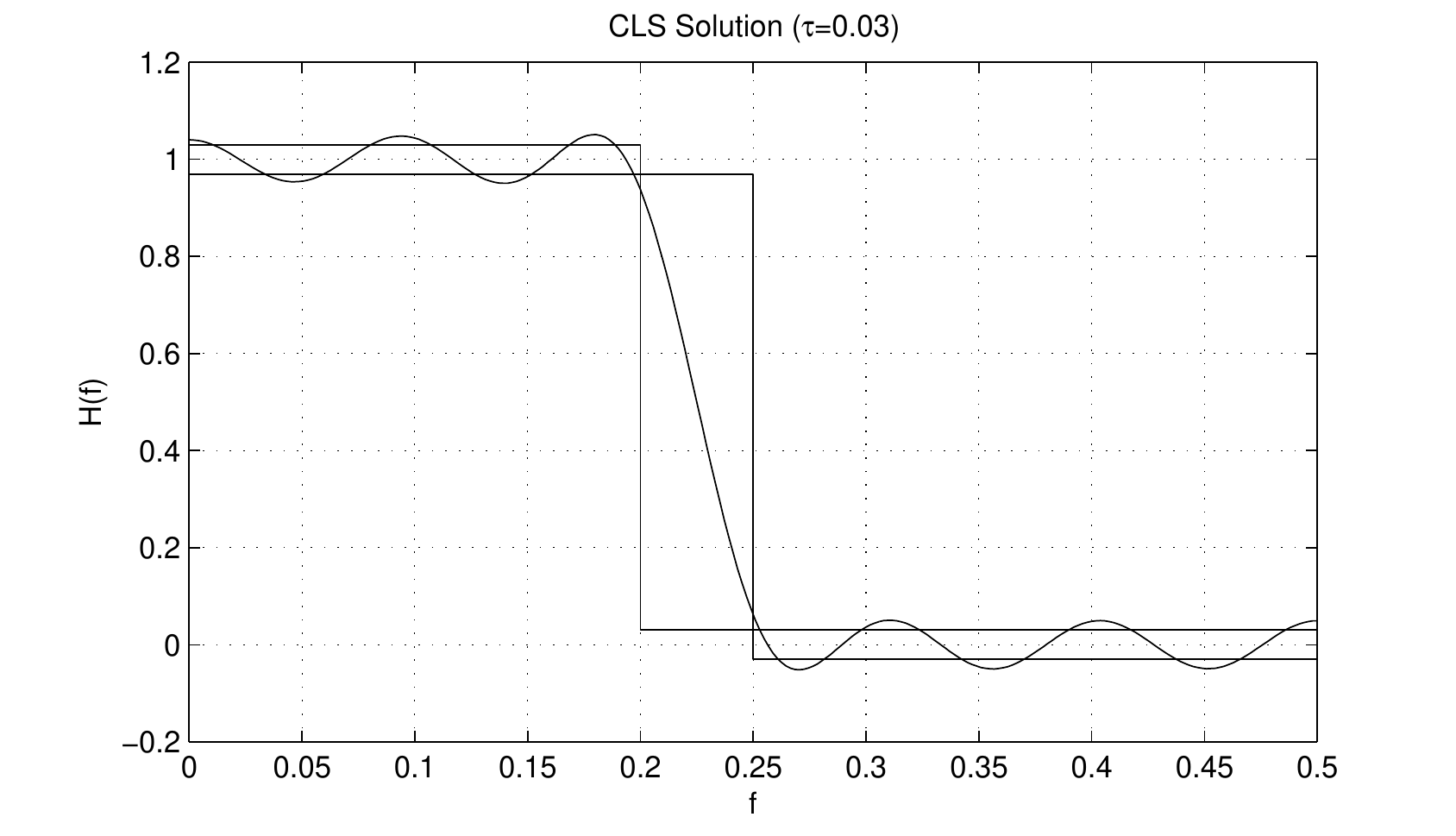,width=3.7in}}
    \caption{CLS design example using tight constraints.}
    \label{fir:cls_hard_stb}
\end{figure}
\begin{figure}[p]
    \centerline{\psfig{figure=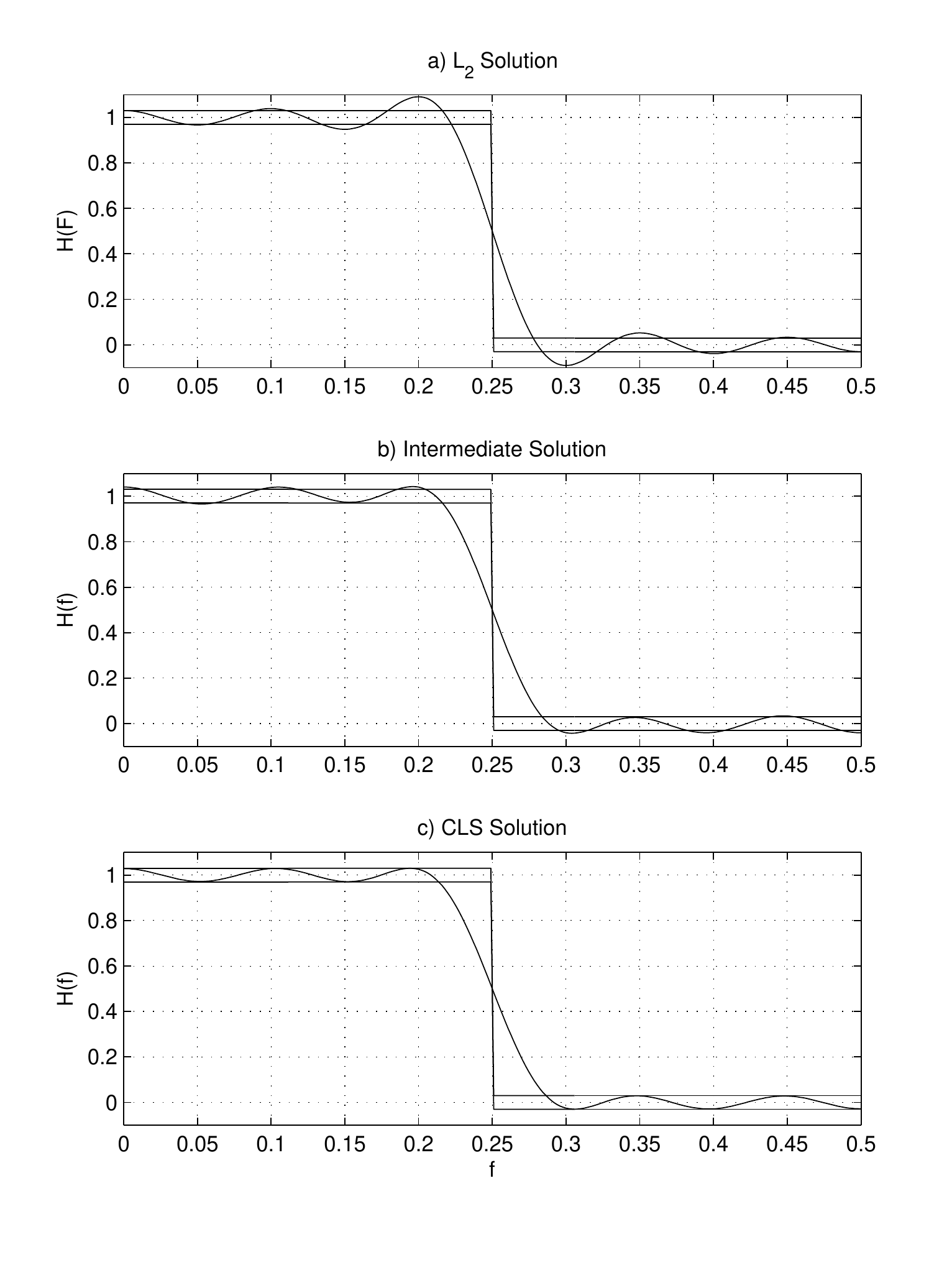,width=3.3in}}
    \caption{CLS design example without transition bands.}
    \label{fir:cls_ntb}
\end{figure}

In practice the method outlined above has proven robust particularly in connection with the specified transition band design. Consider the least squares design in Figure \ref{fir:cls_orig_L2} (using a length-21 Type-I linear phase low-pass FIR filter with linear transition frequencies $\{0.2,0.25\}$). This example illustrates the typical effect of CLS methods over $l_2$ designs; the largest error (in an $l_{\infty}$ sense) can be located at the edges of the transition band. Figures \ref{fir:cls_mild_stb} and \ref{fir:cls_hard_stb} illustrate design examples using the proposed approach. Figure \ref{fir:cls_mild_stb} shows an example of a mild constraint ($\tau=0.6$), whereas \ref{fir:cls_hard_stb} illustrates an advantage of this method, associated to a hard constraint ($\tau=0.3$). The method is trying iteratively to reduce the maximum error towards the constraint; however the specified constraint in Figure \ref{fir:cls_hard_stb} is such that even at the point where an equiripple response is reached for the specified transition bands the constraint is not met. At this point the method converges to an optimal $\lp$ solution that approximates equiripple as $p$ increases (the examples provided use $p=50$).

A different behavior occurs when no transition bands are defined. Departing from an initial $l_2$ guess (as shown in Figure \ref{fir:cls_ntb}.a) the proposed IRLS-based CLS algorithm begins weighting frequencies selectively in order to reduce the $l_{\infty}$ error towards the constraints $\tau$ at each iteration. Eventually an equiripple behavior can be observed if the constraints are too harsh (as in Figure \ref{fir:cls_ntb}.b). The algorithm will keep weighting until all frequencies meet the constraints (as in Figure \ref{fir:cls_ntb}.c).  The absence of a specified transition band presents some ambiguity in defining valid frequencies for weighting. One cannot (or rather should not) apply weights too close to the transition frequency specified as this would result in an effort by the algorithm to create a steep transition region (which as mentioned previously is counterintuitive to finding an equiripple solution). In a sense, this would mean having two opposite effects working at the same time and the algorithm cannot accommodate both, usually leading to numerical problems.
\begin{figure}[ht]
    \centerline{\psfig{figure=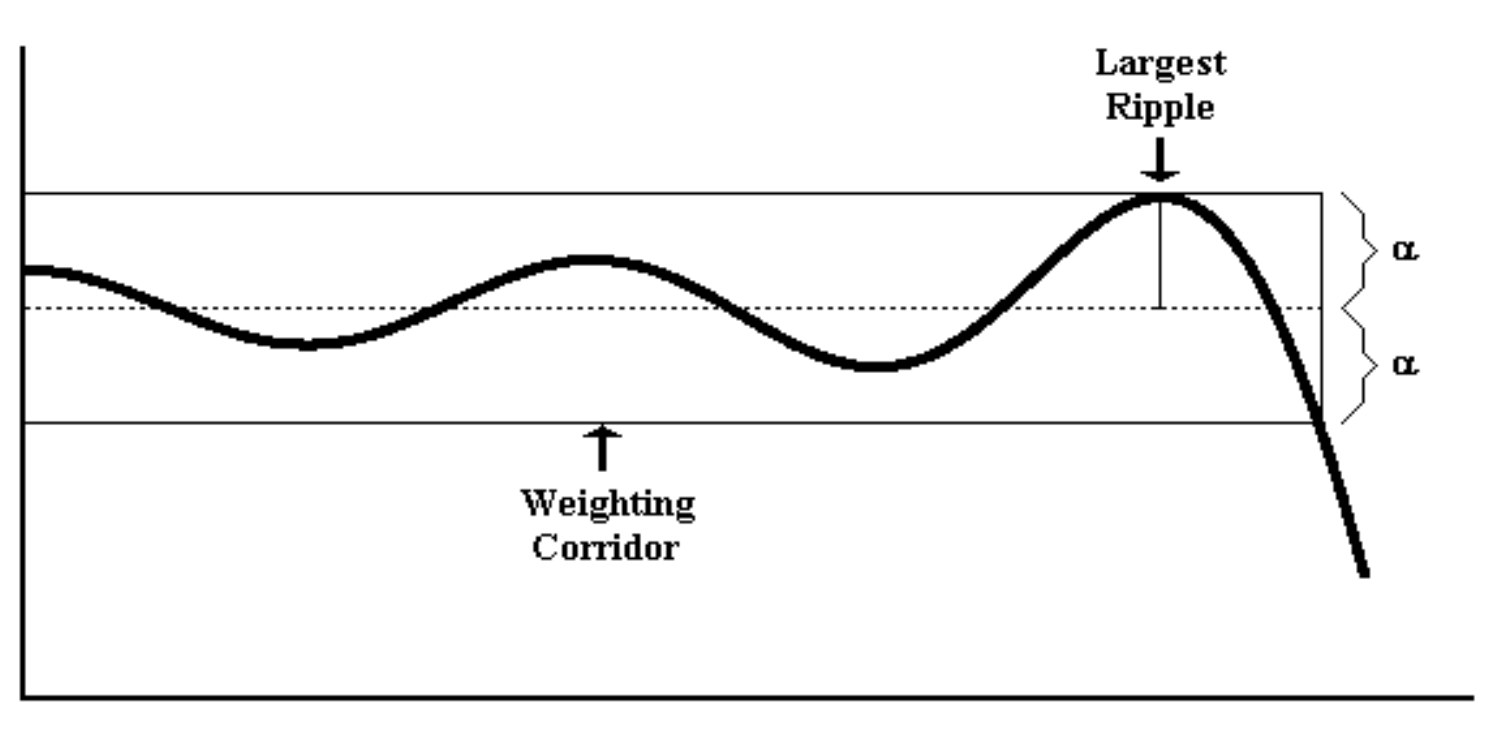,width=3.2in}}
    \caption{Definition of {\it induced} transition band.}
    \label{fir:induced_band1}
\end{figure}

In order to avoid these issues, an algorithm can be devised that selects a subset of the sampled frequencies for weighting purposes at each iteration. The idea is to identify the largest ripple per band at each iteration (the ripple associated with the largest error for a given band) and select the frequencies within that band with errors equal or smaller than such ripple error. In this way one avoids weighting frequencies around the transition frequency. This idea is illustrated in Figure \ref{fir:induced_band1}.
\begin{figure}[ht]
    \centerline{\psfig{figure=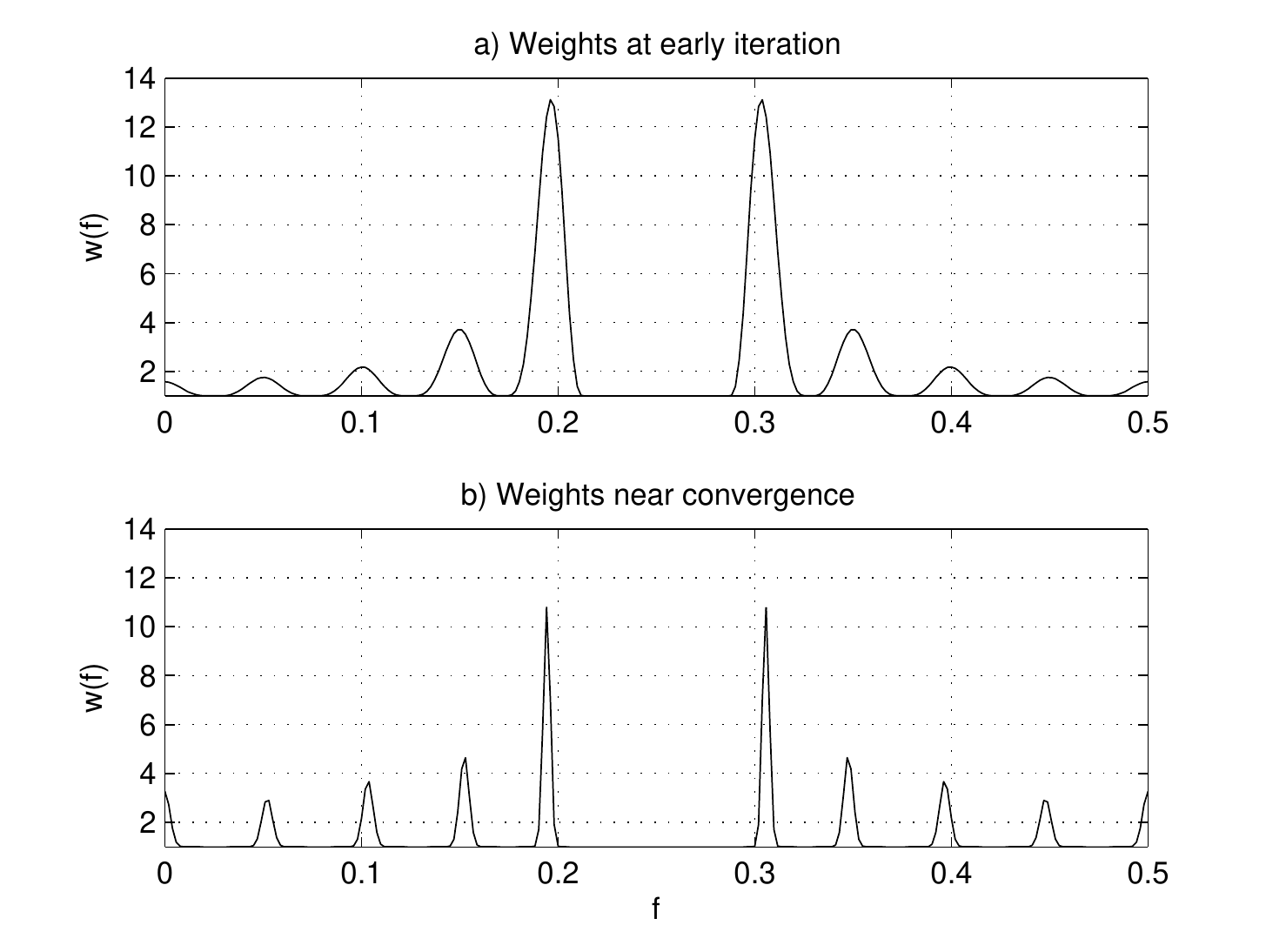,width=3.5in}}
    \caption{CLS weights.}
    \label{fir:cls_weights}
\end{figure}

The previous example is fundamental since it illustrates the relevance of this method: since for a particular transition band the tightest constraint that one can get is given by the equiripple (or minimax) design (as shown in Section \ref{fir:cls2prob}), a problem might arise when specifications are tighter than what the minimax design can meet. Adams found this problem (as reported in \cite{adamsfircls}); his method breaks under these conditions. The method proposed here overcomes an inadequate constraint and relaxes the transition band to meet the constraint.

It is worth noting that the polynomial weighting form works even when no transition bands are specified (this must become evident from Figure \ref{fir:cls_ntb}.c above).  However, the user must be aware of some practical issues related to this approach. Figure \ref{fir:cls_weights} shows a typical CLS polynomial weighting function. Its "spiky" character becomes more dramatic as $p$ increases (the method still follows the homotopy and partial updating ideas from previous sections) as shown in Figure \ref{fir:cls_weights}.b.  It must be evident that the algorithm will assign heavy weights to frequencies with large errors, but at $p$ increases the difference in weighting exaggerates. At some point the user must make sure that proper sampling is done to ensure that frequencies with large weights (from a theoretical perspective) are being included in the problem, without compromising conputational efficiency (by means of massive oversampling, which can lead to ill-conditioning in numerical least squares methods). Also as $p$ increases, the range of frequencies with signifficantly large weights becomes narrower, thus reducing the overall weighting effect and affecting convergence speed.

\begin{figure}[ht]
    \centerline{\psfig{figure=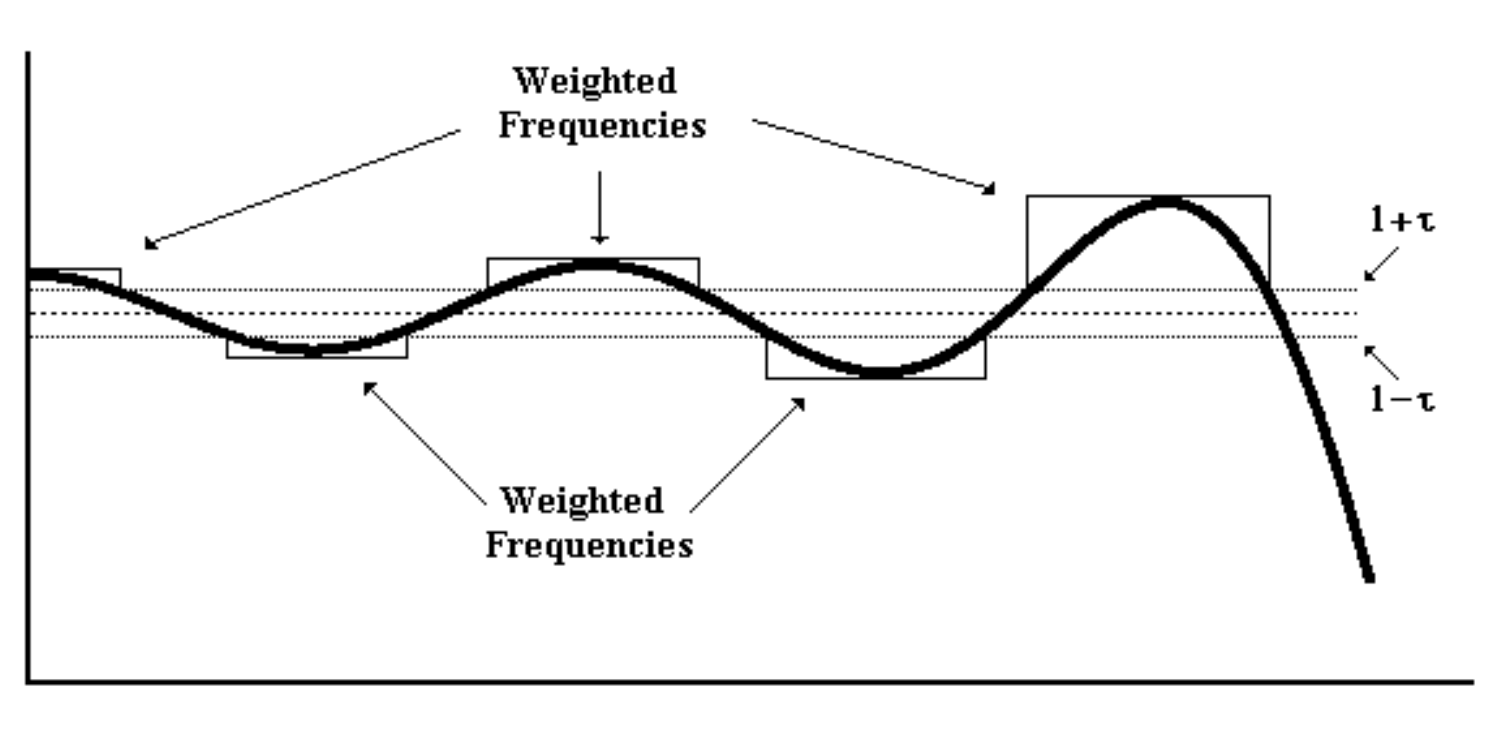,width=3.2in}}
    \caption{CLS envelope weighting function.}
    \label{fir:env_weights}
\end{figure}

A second weighting form can be defined where envelopes are used. The {\it envelope weighting function} approach works by assigning a weight to all frequencies not meeting a constraint. The value of such weights are assigned as flat intervals as illustrated in Figure \ref{fir:env_weights}.  Intervals are determined by the edge frequencies within neighborhoods around peak error frequencies for which constraints are not met. Clearly these neighborhoods could change at each iteration. The weight of the $k$-th interval is still determined by our typical expression,
\[ w_k(\om) = \abs{\ep(\om_k^{+})}^{\frac{p-2}{2}} \]
where $\om_k^{+}$ is the frequency with largest error within the $k$-th interval.

\begin{figure}[ht]
    \centerline{\psfig{figure=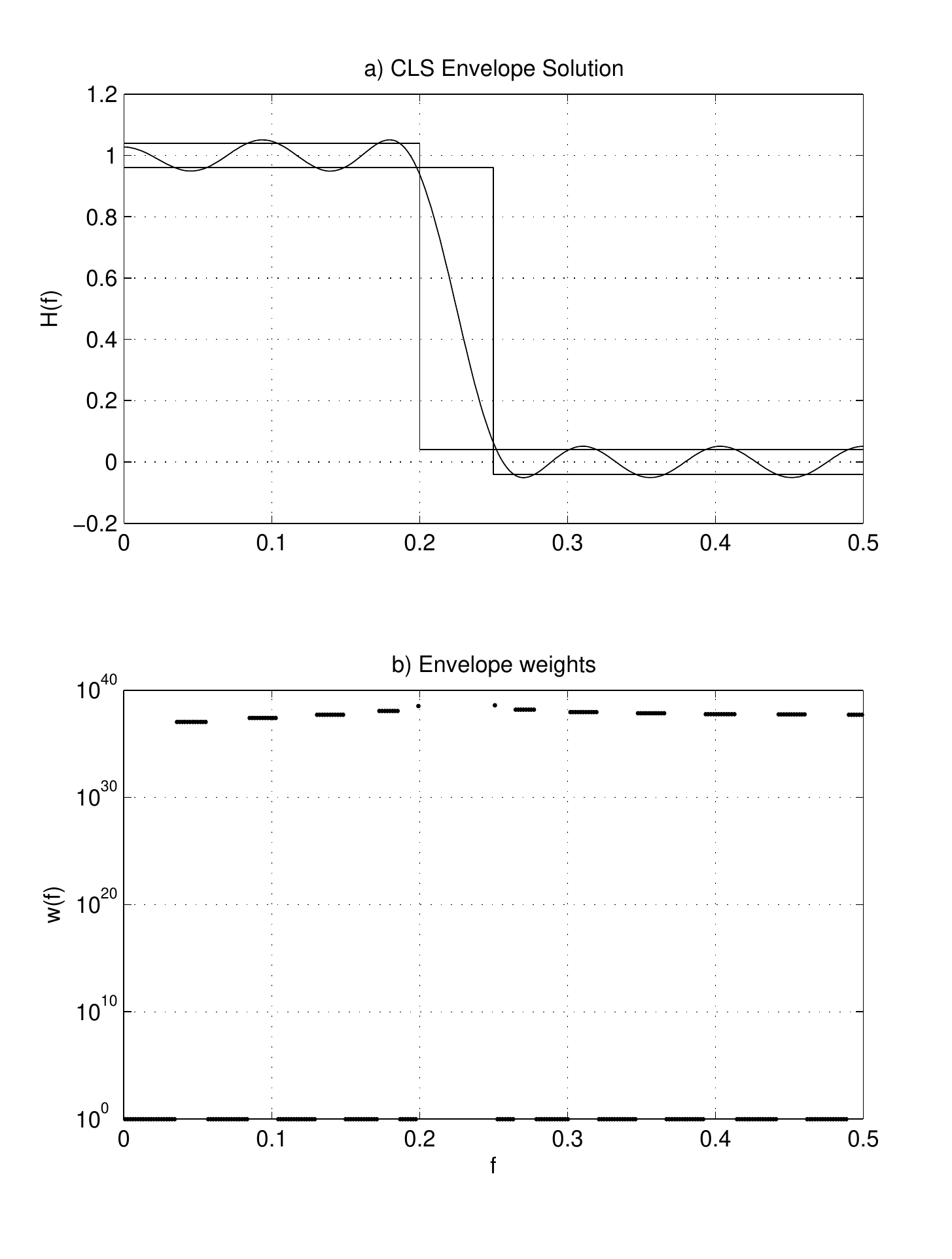,width=3.5in}}
    \caption{CLS design example using envelope weights.}
    \label{fir:cls_env_stb}
\end{figure}

Envelope weighting has been applied in practice with good results. It is particularly effective at reaching high values of $p$ without ill-conditioning, allowing for a true alternative to minimax design. Figure \ref{fir:cls_env_stb} shows an example using $\tau=0.4$; the algorithm managed to find a solution for $p=500$.  By specifying transition bands and unachievable constraints one can produce an almost equiripple solution in an efficient manner, with the added flexibility that milder constraints will result in CLS designs.

\subsubsection{Comparison with $l_p$ problem}
\label{fir:clslpcomp}

This chapter presented two problems with similar effects. On one hand, Section \ref{fir:modirls} illustrated the fact (see Figure \ref{fir:lp_curve}) that as $p$ increases towards infinity, an $l_p$ filter will approximate an $l_{\infty}$ one. On the other hand, Section \ref{fir:cls} presented the constrained least squared problem, and introduced IRLS-based algorithms that produce filters that approximate equiripple behavior as the constraint specifications tighten. 

\begin{figure}[h]
    \centerline{\psfig{figure=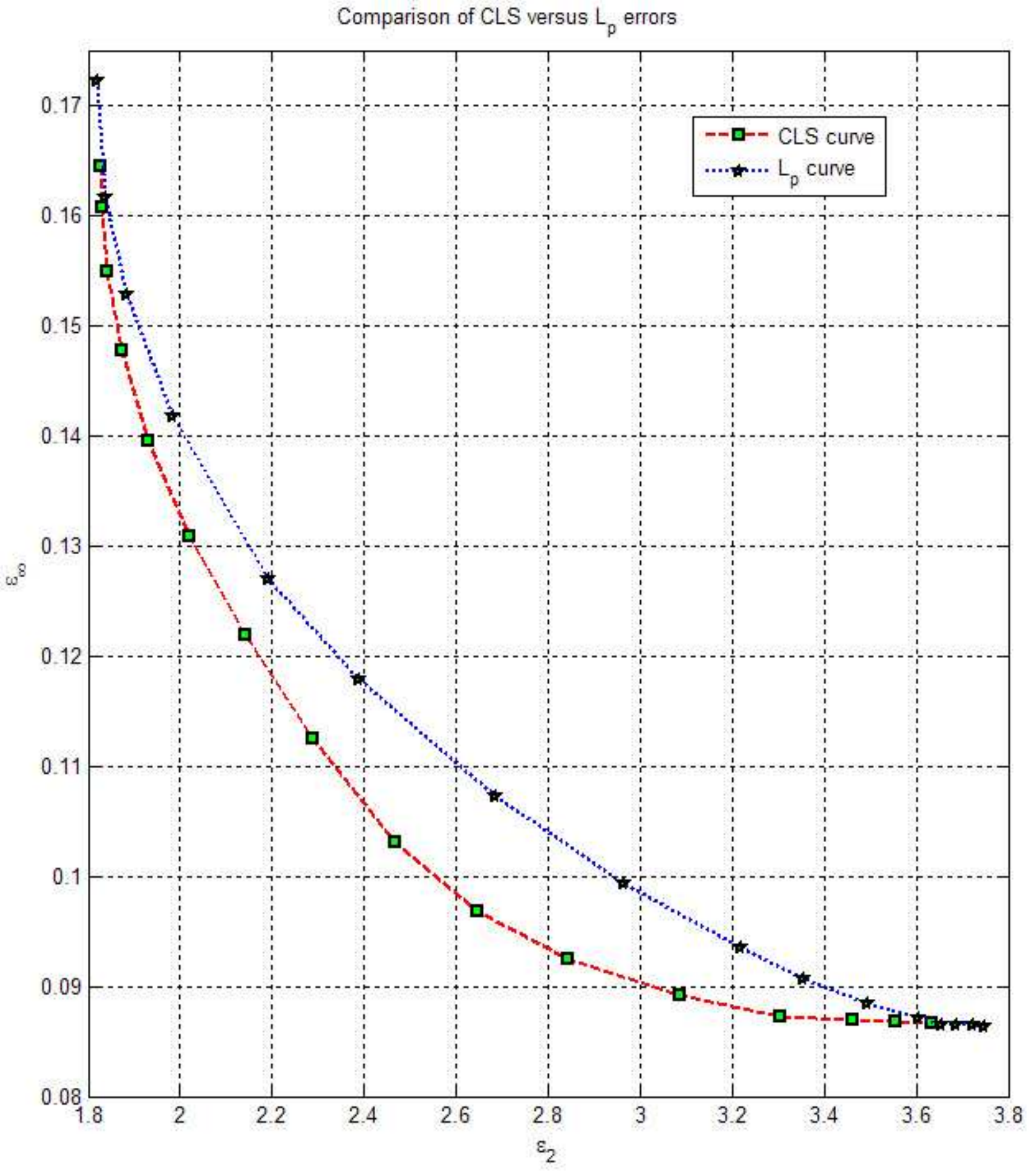,width=3.0in}}
    \caption{Comparison between CLS and $l_p$ problems.}
    \label{fir:clslp_comp}
\end{figure}

A natural question arises: how do these methods compare with each other?  In principle it should be possible to compare their performances, as long as the necessary assumptions about the problem to be solved are compatible in both methods. Figure \ref{fir:clslp_comp} shows a comparison of these algorithms with the following specifications:
\begin{itemize}
\item Both methods designed length-21 Type-I lowpass linear phase digital filters with fixed transition bands defined by $f=\{0.2,0.24\}$ (in normalized linear frequency).
\item The $l_p$ experiment used the following values of $p=$ \{2, 2.2, 2.5, 3,4, 5, 7, 10, 15, 20, 30, 50, 70, 100, 170, 400\}
\item The CLS experiment used the polynomial weighting method with fixed transition bands and a value of $p=60$. The error tolerances were $\tau=$ \{.06, .077, .078, .8, .084, .088, .093, .1, .11, .12, .13, .14, .15, .16, .17, .18\}
\end{itemize}
Some conclusions can be derived from Figure \ref{fir:clslp_comp}. Even though at the extremes of the curves they both seem to meet, the CLS curve lies just below the $l_p$ curve for most values of $p$ and $\tau$. These two facts should be expected: on one hand, in principle the CLS algorithm gives an $l_2$ filter if the constraints are so mild that they are not active for any frequency after the first iteration (hence the two curves should match around $p=2$). On the other hand, once the constraints become too harsh, the fixed transition band CLS method basically should design an equiripple filter, as only the active constraint frequencies are $l_p$-weighted (this effects is more noticeable with higher values of $p$). Therefore for tight constraints the CLS filter should approximate an $l_{\infty}$ filter.

The reason why the CLS curve lies under the $l_p$ curve is because for a given error tolerance (which could be interpreted as {\it for a given minimax error} $\vep_{\infty}$) the CLS method finds the optimal $l_2$ filter. An $l_p$ filter is optimal in an $l_p$ sense; it is not meant to be optimal in either the $l_2$ or $l_{\infty}$ senses. Hence for a given $\tau$ it cannot beat the CLS filter in an $l_2$ sense (it can only match it, which happens around $p=2$ or $p=\infty$).

It is important to note that both curves are not drastically different. While the CLS curve represents optimality in an $L_2-l_{\infty}$ sense, not all the problems mentioned in this work can be solved using CLS filters (for example, the {\it magnitude} IIR problem presented in Section \ref{iir:magsect}). Also, one of the objectives of this work is to motivate the use of $l_p$ norms for filter design problems, and the proposed CLS implementations (which absolutely depends on IRLS-based $l_p$ formulations) are good examples of the flexibility and value of $l_p$ IRLS methods discussed in this work.

\section{Infinite Impulse Response Filters}
\label{ch:iir}

Chapter \ref{ch:fir} introduced the problem of designing $\lp$ FIR filters, along with several design scenarios and their corresponding design algorithms. This chapter considers the design of $\lp$ IIR filters and examines the similarities and differences compared to $\lp$ FIR filter design.  It was mentioned in Section \ref{int:fir} that $\lp$ FIR design involves a polynomial approximation.  The problem becomes more complicated in the case of IIR filters as the approximation problem is a ratio of two polynomials. In fact, the case of FIR polynomial approximation is a special form of IIR rational approximation where the denominator is equal to 1. 

{\it Infinite Impulse Response} (or {\it recursive}) digital filters constitute an important analysis tool in many areas of science (such as signal processing, statistics and biology). The problem of designing IIR filters has been the object of extensive study. Several approaches are typically used in designing IIR filters, but a general procedure follows: given a desired filter specification (which may consist of an impulse response or a frequency specification), a predetermined approximation error criterion is optimized. Although one of the most widely used error criteria in Finite Impulse Response (FIR) filters is the {\it least-squares} criterion (which in most scenarios merely requires the solution of a linear system), least-squares ($l_2$) approximation for IIR filters requires an optimization over an infinite number of filter coefficients (in the time domain approximation case). Furthermore, optimizing for an IIR frequency response leads to a rational (nonlinear) approximation problem rather than the polynomial problem of FIR design.

As discussed in the previous chapter, a successful IRLS-based $\lp$ algorithm depends to a large extent in the solution of a weighted $l_2$ problem. One could argue that one of the most important aspects contrasting FIR and IIR $\lp$ filter design lies in the $l_2$ optimization step. This chapter presents the theoretical and computational issues involved in the design of both $l_2$ and $\lp$ IIR filters and explores several approaches taken to handle the resulting nonlinear $l_2$ optimization problem. Section \ref{iir:gen} introduces the IIR filter formulation and the nonlinear least-squares design problem. Section \ref{iir:ls} presents the $l_2$ problem more formally, covering relevant methods as a manner of background and to lay down a framework for the approach proposed in this work.  Some of the methods covered here date back to the 1960's, yet others are the result of current active work by a number of research groups; the approach employed in this work is described in section \ref{iir:soe}. Finally, Section \ref{iir:lp} considers different design problems concerning IIR filters in an $\lp$ sense, including IIR versions of the complex, frequency-varying and magnitude filter design problems as well as the proposed algorithms and their corresponding results. 

\subsection{IIR filters}
\label{iir:gen}

An IIR filter describes a system with input $x(n)$ and output $y(n)$, related by the following expression 
\[ y(n)=\sum_{k=0}^M b(k)x(n-k)-\sum_{k=1}^N a(k)y(n-k) \]
Since the current output $y(n)$ depends on the input as well as on $N$ previous output values, the output of an IIR filter might not be zero well after $x(n)$ becomes zero (hence the name ``Infinite''). Typically IIR filters are described by a rational transfer function of the form 
\begin{equation} \label{iir:transf} H(z)=\frac{B(z)}{A(z)}=\frac{b_0+b_1z^{-1}+\cdots+b_Mz^{-M}}{1+a_1z^{-1}+\cdots +a_Nz^{-N}} \end{equation} 
where
\begin{equation} \label{iir:impresp} H(z)= \sum_{n=0}^\infty h(n) z^{-n} \end{equation}
and $h(n)$ is the {\it infinite impulse response} of the filter. Its {\it frequency response} is given by 
\begin{equation} \label{iir:ztofreq} H(\om)=H(z)|_{z=e^{j\om}}\end{equation} 
Substituting (\ref{iir:transf}) into (\ref{iir:ztofreq}) we obtain 
\begin{equation} \label{iir:freqprob} H(\om)=\frac{B(\om)}{A(\om)}=\frac{\sum\limits_{n=0}^M b_n e^{-j\om n}}{1+\sum\limits_{n=1}^N a_n e^{-j\om n}} \end{equation}
Given a {\it desired} frequency response $D(\om)$, the $l_2$ IIR design problem consists of solving the following problem
\begin{equation} \label{iir:opt_prob} \min_{a_n,b_n} \; \left|\frac{B(\om)}{A(\om)} - D(\om) \right|_2^2 \end{equation} 
for the $M+N+1$ real filter coefficients $a_n, b_n$ with $\om\in\Om$ (where $\Om$ is the set of frequencies for which the approximation is done). A discrete version of (\ref{iir:opt_prob}) is given by
\begin{equation} \label{iir:opt_prob_disc} \min_{a_n,b_n} \; \sum_{\om_k} \left| \frac{\sum\limits_{n=0}^M b_n e^{-j\om_k n}}{1+\sum\limits_{n=1}^N a_n e^{-j\om_k n}} - D(\om_k) \right|^2
\end{equation}
where $\om_k$ are the $L$ frequency samples over which the approximation is made. Clearly, (\ref{iir:opt_prob_disc}) is a nonlinear least squares optimization problem with respect to the filter coefficients.

\subsection{Least squares design of IIR filters}
\label{iir:ls}

Section \ref{iir:gen} introduced the IIR least squares design problem, as illustrated in (\ref{iir:opt_prob_disc}). Such problem cannot be solved in the same manner as in the FIR case; therefore more sophisticated methods must be employed. As will be discussed later in Section \ref{iir:lp}, some tradeoffs are desirable for $\lp$ optimization. As in the case of FIR design, when designing $\lp$ IIR filters one must use $l_2$ methods as internal steps over which one iterates while moving between diferent values of $p$. Clearly this internal iteration must not be too demanding computationally since an outer $\lp$ loop will invoke it repeatedly (this process will be further illustrated in Section \ref{iir:cplxfreq}). With this issue in mind, one needs to select an $l_2$ algorithm that remains accurate within reasonable error bounds while remaining computationally efficient.

This section begins by summarizing some of the traditional approaches that have been employed for $l_2$ rational approximation, both within and outside filter design applications. Amongst the several existing traditional nonlinear optimization approaches, the Davidon-Fletcher-Powell (DFP) and the Gauss-Newton methods have been often used and remain relatively well understood in the filter design community. A brief introduction to both methods is presented in Section \ref{iir:ls_trad}, and their caveats briefly explored.

An alternative to attacking a complex nonlinear problem like (\ref{iir:opt_prob_disc}) with general nonlinear optimization tools consists in {\it linearization}, an attempt to "linearize" a nonlinear problem and to solve it by using linear optimization tools. Multiple efforts have been applied to similar problems in different areas of statistics and systems analysis and design. Section \ref{iir:ls_lin} introduces the notion of an {\it Equation Error}, a linear expression related to the actual {\it Solution Error} that one is interested in minimizing in $l_2$ design. The equation error formulation is nonetheles important for a number of filter design methods (including the ones presented in this work) such as Levy's method, one of the earliest and most relevant frequency domain linearization approaches. Section \ref{iir:pronypade} presents a frequency domain equation error algorithm based on the methods by Prony and Pad{\'e}. This algorithm illustrates the usefulness of the equation error formulation as it is fundamental to the implementation of the methods proposed later in this work (in Section \ref{iir:lp}).

An important class of linearization methods fall under the name of {\it iterative prefiltering} algorithms, presented in Section \ref{iir:iterfilt}. The {\it Sanathanan-Koerner} (\textbf{SK}) algorithm and the {\it Steiglitz-McBride} (\textbf{SMB}) methods are well known and commonly used examples in this category, and their strengths and weaknesses are explored. Another recent development in this area is the method by Jackson, also presented in this section. Finally, Soewito's {\it quasilinearization} (the method of choice for least squares IIR approximation in this work) is presented in Section \ref{iir:soe}.

\subsubsection{Traditional optimization methods}
\label{iir:ls_trad}

One way to adress (\ref{iir:opt_prob_disc}) is to attempt to solve it with general nonlinear optimization tools.  One of the most typical approach in nonlinear optimization is to apply either Newton's method or a Newton-based algorithm.  One assumption of Newton's method is that the optimization function resembles a quadratic function near the solution. In order to update a current estimate, Newton's method requires first and second order information through the use of gradient and Hessian matrices.  A {\it quasi-Newton method} is one that estimates in a certain way the second order information based on gradients (by generalizing the secant method to multiple dimensions).

One of the most commonly used quasi-Newton methods in IIR filter design is the {\it Davidon-Fletcher-Powell} ({\bf DFP}) method \cite{dfp}. In 1970 K. Steiglitz \cite{CADiir} used the DFP method to solve an IIR magnitude approximation to a desired real frequency response. For stability concerns he used a cascade form of the IIR filter given in (\ref{iir:transf}) through
\be \label{iir:dfp0} H(z)=\al \prod_{r=1}^M \frac{1+a_rz^{-1}+b_rz^{-2}}{1+c_rz^{-1}+d_rz^{-2}} \ee
Therefore he considered the following problem,
\[ \min_{a_n,b_n,c_n,d_n} \; \sum_{\om_k} \left( \left| \al \prod_{r=1}^M \frac{1+a_re^{-j\om_k}+b_re^{-2j\om_k}}{1+c_re^{-j\om_k}+d_re^{-2j\om_k}} \right| - D(\om_k) \right)^2
\]
His method is a direct implementation of the DFP algorithm in the problem described above.

In 1972 Andrew Deczky \cite{deczky} employed the DFP algorithm to solve a complex IIR least-$p$ approximation to a desired frequency response. Like Steiglitz, Deczky chose to employ the cascaded IIR structure of (\ref{iir:dfp0}), mainly for stability reasons but also because he claims that for this structure it is simpler to derive the first order information required for the DFP method.

The MATLAB Signal Processing Toolbox includes a function called INVFREQZ, originally written by J. Smith and J. Little \cite{invfreqz}. Invfreqz uses the algorithm by Levy (see \S \ref{iir:levy}) as an initial step and then begins an iterative algorithm based on the damped Gauss-Newton \cite{num_met_unc} to minimize the solution error $\vep_s$ according to the least-squared error criteria. This method performs a line search after every iteration to find the optimal direction for the next step. Invfreqz evaluates the roots of $A(z)$ after each iteration to verify that the poles of $H(z)$ lie inside the unit circle; otherwise it will convert the pole into its reciprocal. This approach guarantees a stable filter.

Among other Newton-based approaches, Spanos and Mingori\cite{newt_freq} use a Newton algorithm combined with the Levenberg-Marquardt technique to improve the algorithm's convergence properties. Their idea is to express the denominator function $A(\om)$ as a sum of second-order rational polynomials. Thus $H(\om)$ can be written as
\[ H(\om)=\sum_{r=1}^{L-1} \frac{b_r+j\om\beta_r}{a_r+j\om\beta_r-\om^2} + d \]
Their global descent approach is similar to the one presented in \cite{soren_newt}. As any Newton-based method, this approach suffers under a poor initial guess, and does not guarantee to converge (if convergence occurs) to a local minimum. However, in such case, convergence is quadratic.

Kumaresan's method \cite{kumar_burrus} considers a three-step approach. It is not clear whether his method attempts to minimize the equation error or the solution error. He uses divided differences \cite{hilde} to reformulate the solution error in terms of the coefficients $a_k$. Using Lagrange multiplier theory, he defines
\begin{equation} \label{iir:kumar1} \e=\vy^T\mC^T[\mC\mC^T]^{-1}\mC\vy \end{equation}
where $\vy=[H_0 H_1 \cdots H_{L-1}]^T$ contains the frequency samples and $\mC$ is a composition matrix containing the frequency divided differences and the coefficients $a_k$ (a more detailed derivation can be found in \cite{kumar_ident}). Equation (\ref{iir:kumar1}) is iterated until convergence of the coefficient vector $\hat{a}$ is reached. This vector is used as initial guess in the second step, involving a Newton-Raphson search of the optimal $\hat{a}$ that minimizes $\normsq{\e}$. Finally the vector $\hat{b}$ is found by solving a linear system of equations.

\subsubsection{Equation error linearization methods}
\label{iir:ls_lin}

Typically general use optimization tools prove effective in finding a solution. However in the context of IIR filter design, they often tend to take a rather large number of iterations, generate large matrices or require complicated steps like solving or estimating (and often inverting) vectors and matrices of first and second order information \cite{soe}.  Using gradient-based tools for nonlinear problems like (\ref{iir:opt_prob_disc}) certainly seems like a suboptimal approach. Also, typical Newton-based methods tend to converge quick (quadratically), yet they make assumptions about radii of convergence and initial proximity to the solution (otherwise performance is suboptimal). In the context of filter design one should wonder if better performance could be achieved by exploiting characteristics from the problem. This section introduces the concept of linearization, an alternative to general optimization methods that has proven successful in the context of rational approximation. The main idea behind linearization approaches consists in transforming a complex nonlinear problem into a sequence of linear ones, an idea that is parallel to the approach followed in our development of IRLS $\lp$ optimization.

A common notion used in this work (as well as some of the literature related to linearization and filter design) is that there are two different error measures that authors often refer to. It is important to recognize the differences between them as one browses through literature. Typically one would be interested in minimizing the $l_2$ error given by:
\be \label{iir:e_sol} \vep = \| \e(\om) \|_2^2 = \left\| D(\om) - \frac{B(\om)}{A(\om)}\right\|_2^2 \ee
This quantity is often referred to as the {\it solution error} (denoted by $\vep_s$); we refer to the function $\e(\om)$ in (\ref{iir:e_sol}) as the {\it solution error function}, denoted by $\e_s(\om)$. Also, in linearization algorithms the following measure often arises,
\be \label{iir:e_eqn} \vep = \| \e(\om) \|_2^2 = \left\| A(\om)D(\om) - B(\om)\right\|_2^2 \ee
This measure is often referred to as the {\it equation error} $\vep_e$; we denote the function $\e(\om)$ in (\ref{iir:e_eqn}) as the {\it equation error function} $\e_e(\om)$. Keeping the notation previously introduced, it can be seen that the two errors relate by one being a weighted version of the other,
\[ \e_e(\om) = A(\om)\e_s(\om) \]

\subsubsection{Levy's method}
\label{iir:levy}

E. C. Levy \cite{levy} considered in 1959 the following problem in the context of analog systems (electrical networks to be more precise): define\footnote{For consistency with the rest of this document, notation has been modified from the author's original paper whenever deemed necessary.} 
\be \label{iir:levy0} H(j\om) = \frac{B_0+B_1(j\om)+B_2(j\om)^2+\cdots}{A_0+A_1(j\om)+A_2(j\om)^2+\cdots} = \frac{B(\om)}{A(\om)} \ee
Given $L$ samples of a desired complex-valued function $D(j\om_k)=R(\om_k)+jI(\om_k)$ (where $R,I$ are both real funtions of $\om$), Levy defines 
\[ \e(\om) = D(j\om) - H(j\om) = D(j\om) - \frac{B(\om)}{A(\om)} \]
or
\be \label{iir:levy1} \vep = \sum_{k=0}^L |\e(\om_k)|^2 = \sum_{k=0}^L |A(\om_k)D(j\om_k) - B(\om_k)|^2 \ee
Observing the linear structure (in the coefficients $A_k,B_k$) of equation (\ref{iir:levy1}), Levy proposed minimizing the quantity $\vep$. He actually realized that this measure (what we would denote as the equation error) was indeed a {\bf weighted} version of the actual solution error that one might be interested in; in fact, the denominator function $A(\om)$ became the weighting function.

Levy's proposed method for minimizing (\ref{iir:levy1}) begins by writing $\vep$ as follows,
\be \label{iir:levy2} \vep = \sum_{k=0}^L \left[(R_k\sig_k-\om_k\tau_kI_k-\al_k)^2 + (\om_k\tau_kR_k+\sig_kI_k-\om_k\beta_k)^2\right] \ee
by recognizing that (\ref{iir:levy0}) can be reformulated in terms of its real and imaginary parts, 
\[ H(j\om) = \frac{\al+j\om\beta}{\sig+j\om\tau} \]
with
\bean \begin{split} \al+j\om\beta = & (B_0-B_2\om^2+B_4\om^4\cdots) \\ & \quad + j\om(B_1-B_3\om^2+B_5\om^4\cdots) \\
\sig+j\om\tau = & (A_0-A_2\om^2+A_4\om^4\cdots) \\ & \quad + j\om(A_1-A_3\om^2+A_5\om^4\cdots) \end{split} \eean
and performing appropriate manipulations\footnote{For further details on the algebraic manipulations involved, the reader should refer to \cite{levy}.}. Note that the optimal set of coefficients $A_k$, $B_k$ must satisfy
\[ \frac{\pa \vep}{\pa A_0} = \frac{\pa \vep}{A_1} = \ldots =\frac{\pa \vep}{\pa B_0} = \ldots = 0 \]
The conditions introduced above generate a linear system in the filter coefficients.  Levy derives the system 
\be \label{iir:levy3} \mC\vx=\vy \ee
where $\mC = \left\{ \mC_1 \quad \mC_2 \right\}$ with
\ben \mC_1 = \left\{ \begin{array}{cccccc}
\lambda_0 & 0 & -\lambda_2 & 0 & \lambda_4 & \cdots \\
0 & \lambda_2 & 0 & -\lambda_4 & 0 & \cdots \\
\lambda_2 & 0 & -\lambda_4 & 0 & \lambda_6 & \cdots \\
\vdots & \vdots & \vdots & \vdots & \vdots & \\
T_1 & -S_2 & -T_3 & S_4 & T_5 & \cdots \\
S_2 & T_3 & -S_4 & -T_5 & S_6 & \cdots \\
T_3 & -S_4 & -T_5 & S_6 & T_7 & \cdots \\
\vdots & \vdots & \vdots & \vdots & \vdots & 
\end{array} \right\} \een
\ben \mC_2 = \left\{ \begin{array}{cccccc}
T_1 & S_2 & -T_3 & -S_4 & T_5 & \cdots \\
-S_2 & T_3 & S_4 & -T_5 & -S_6 &  \\
T_3 & S_4 & -T_5 & -S_6 & T_7 & \\
\vdots & \vdots & \vdots & \vdots & \vdots & \\
U_2 & 0 & -U_4 & 0 & U_6 & \cdots \\
0 & U_4 & 0 & -U_6 & 0 & \\
U_4 & 0 & -U_6 & 0 & U_8 & \cdots \\
\vdots & \vdots & \vdots & \vdots & \vdots &
\end{array} \right\} \een
and
\be \label{iir:levy4} \vx=\left\{ \begin{array}{c} B_0 \\ B_1 \\ B_2 \\ \vdots \\ A_1 \\ A_2 \\ \vdots \end{array} \right\} 
\hspace{1.5 cm}  \vy= \left\{ \begin{array}{c} S_0 \\ T_1 \\ S_2 \\ T_3 \\ \vdots \\ 0 \\ U_2 \\ 0 \\ U_4 \\ \vdots  \end{array} \right\} \ee
with
\begin{eqnarray*} \lambda_h & = & \sum_{l=0}^{L-1} \om_l^h \\ S_h & = & \sum_{l=0}^{L-1} \om_l^h R_l \\ T_h & = & \sum_{l=0}^{L-1} \om_l^hI_l \\ U_h
& = & \sum_{l=0}^{L-1} \om_l^h (R_l^2+ I_l^2) \end{eqnarray*}
Solving for the vector $\vx$ from (\ref{iir:levy3}) gives the desired coefficients (note the trivial assumption that $A_0=1$). It is important to remember that although Levy's algorithm leads to a linear system of equations in the coefficients, his approach is indeed an equation error method. Matlab's \verb'invfreqz' function uses an adaptation of Levy's algorithm for its least-squares equation error solution.

\subsubsection{Prony-based equation error linearization}
\label{iir:pronypade}

A number of algorithms that consider the approximation of functions in a least-squared sense using rational functions relate to Prony's method. This section summarizes these methods especially in the context of filter design. 

\subsubsection{Prony's method}
The first method considered in this section is due to Gaspard Riche Baron de Prony, a Lyonnais mathematician and physicist which, in 1795, proposed to model the expansion properties of different gases by sums of damped exponentials. His method \cite{prony} approximates a sampled function $f(n)$ (where $f(n)=0$ for $n\!<\!0$) with a sum of $N$ exponentials,
\begin{equation} \label{iir:prony_orig}  f(n)= \sum_{k=1}^N c_k e^{s_k n} = \sum_{k=1}^N c_k \lambda_k^n \end{equation} 
where $\lambda_k=e^{s_k}$. The objective is to determine the $N$ parameters $c_k$ and the $N$ parameters $s_k$ in (\ref{iir:prony_orig}) given $2N$ samples of $f(n)$. 

It is possible to express (\ref{iir:prony_orig}) in matrix form as follows,
\begin{equation} \label{iir:vander} \left[ \begin{array}{cccc} 1& 1& \cdots& 1 \\ \lambda_1& \lambda_2& \cdots& \lambda_N \\ \vdots& \vdots& \ddots& \vdots \\
\lambda_1^{N-1}& \lambda_2^{N-1}& \cdots& \lambda_N^{N-1} \end{array}\right] \left[ \begin{array}{c} c_1\\ c_2\\
\vdots\\ c_N \end{array} \right] = \left[ \begin{array}{c} f(0)\\ f(1)\\ \vdots\\ f(N-1)\end{array}\right] \end{equation} 
System (\ref{iir:vander}) has a Vandermonde structure with $N$ equations, but $2N$ unknowns (both $c_k$ and $\lambda_k$ are unknown) and thus it cannot be solved directly. Yet the major contribution of Prony's work is to recognize that $f(n)$ as given in (\ref{iir:prony_orig}) is indeed the solution of a homogeneous order-$N$ {\it Linear Constant Coefficient Difference Equation} (LCCDE) \cite[ch. 4]{diffeqn} given by 
\begin{equation} \label{iir:LCCDE} \sum_{p=0}^N a_p f(m-p)=0 \end{equation}
with $a_0=1$. Since $f(n)$ is known for $0\!\leq\! n \!\leq\! 2N-1$, we can extend (\ref{iir:LCCDE}) into an ($N\times N$) system of the form
\begin{equation} \label{iir:prony0} \left[ \begin{array}{cccc} f(N-1)& f(N-2)& \cdots& f(0) \\ f(N)& f(N-1)& \cdots& f(1) \\ \vdots& \vdots& \ddots& \vdots \\ f(2N-2)& f(2N-3)& \cdots& f(N-1) \end{array} \right] \left[ \begin{array}{c} a_1 \\ a_2 \\ \vdots \\ a_N \end{array} \right] = \hat{f} \end{equation}
where
\[ \hat{f} = \left[ \begin{array}{c} -f(N) \\ -f(N+1) \\ \vdots \\ -f(2N-1) \end{array} \right] \]
which we can solve for the coefficients $a_p$. Such coefficients are then used in the {\it characteristic equation} \cite[\S 2.3]{intro_diffeqn} of (\ref{iir:LCCDE}), 
\begin{equation} \lambda^N+ a_1\lambda^{N-1}+\cdots+a_{N-1}\lambda+a_N=0 \end{equation}
The $N$ roots $\lambda_k$ of (\ref{iir:prony0}) are called the {\it characteristic roots} of (\ref{iir:LCCDE}). From the $\lambda_k$ we can find the parameters $s_k$ using $s_k=\ln \lambda_k$. Finally, it is now possible to solve (\ref{iir:vander}) for the parameters $c_k$.

The method described above is an adequate representation of Prony's original method \cite{prony}. More detailed analysis is presented in \cite{onfluids,rev_prony,pronyZpade,exp_prony} and \cite[\S 11.4]{marpledsa}. Prony's method is an adequate algorithm for interpolating $2N$ data samples with $N$ exponentials. Yet it is not a filter design algorithm as it stands. Its connection with IIR filter design, however, exists and will be discussed in the following sections.

\subsubsection{Pad{\'e}'s method}
The work by Prony served as inspiration to Henry Pad{\'e}, a French mathematician which in 1892 published a work \cite{pade} discussing the problem of rational approximation. His objective was to approximate a function that could be represented by a power series expansion using a rational function of two polynomials.

Assume that a function $f(x)$ can be represented with a power series expansion of the form
\begin{equation} f(x)=\sum_{k=0}^\infty c_k x^k \end{equation}
Pad{\'e}'s idea was to approximate $f(x)$ using the function
\begin{equation} \hat{f}(x) = \frac{B(x)}{A(x)} \end{equation}
where \[ B(x)= \sum_{k=0}^M b_k x^k \] and \[ A(x)= 1+ \sum_{k=1}^N a_k x^k \] 
The objective is to determine the coefficients $a_k$ and $b_k$ so that the first $M+N+1$ terms of the residual \[ r(x)= A(x)f(x) - B(x) \] dissappear (i.e. the first $N+M$ derivatives of $f(x)$ and $\hat{f}(x)$ are equal \cite{appnuman}). That is, \cite{pade_mech}, 
\[r(x) = A(x) \sum_{k=0}^\infty c_k x^k -B(x) = x^{M+N+1} \sum_{k=0}^\infty d_k x^k \]
To do this, consider $A(x)f(x)=B(x)$ \cite{pronyZpade}
\[ \begin{split} (1+a_1x+\cdots+a_Nx^N)\cdot(c_0 + c_1x & +\cdots+c_{i}x^{i}+\cdots) = \\ & b_0+b_1x+\cdots+b_Mx^M \end{split} \]
By equating the terms with same exponent up to order $M+N+1$, we obtain two sets of equations,
\begin{equation} \label{iir:pade1}\left\{ \begin{array}{rcl} c_0& =& b_0 \\ a_1c_0+c_1& =& b_1 \\ a_2c_0+a_1c_1+c_2& =& b_2 \\ a_3c_0+a_2c_1+a_1c_2+c_3& =& b_3 \\ &\vdots& \\ a_Nc_{M-N}+a_{N-1}c_{M-N+1}+\cdots + c_M& =& b_M \end{array} \right. \end{equation}
\begin{equation} \label{iir:pade2}\left\{ \begin{array}{rcl} a_Nc_{M-N+1}+a_{N-1}c_{M-N+2}+\cdots+c_{M+1}& =& 0\\ a_Nc_{M-N+2}+a_{N-1}c_{M-N+3}+\cdots+c_{M+2}& =& 0\\ &\vdots& \\ a_Nc_{M}+a_{N-1}c_{M+1}+\cdots+c_{M+N}& =& 0 \end{array} \right. \end{equation} 
Equation (\ref{iir:pade2}) represents an $N\times N$ system that can be solved for the coefficients $a_k$ given $c(n)$ for $0\!\leq\! n\!\leq\! N+M$. These values can then be used in (\ref{iir:pade1}) to solve for the coefficients $b_k$.  The result is a system whose impulse response matches the first $N+M+1$ values of $f(n)$.

\subsubsection{Prony-based filter design methods}

Both the original methods by Prony and Pade were meant to interpolate data from applications that have little in common with filter design. What is relevant to this work is their use of rational functions of polynomials as models for data, and the linearization process they both employ.

When designing FIR filters, a common approach is to take $L$ samples of the desired frequency response $D(\om)$ and calculate the inverse DFT of the samples. This design approach is known as {\it frequency sampling}. It has been shown \cite{dfd} that by designing a length-$L$ filter $h(n)$ via the frequency sampling method and symmetrically truncating $h(n)$ to $N$ values ($N\ll L$) it is possible to obtain a least-squares optimal length-$N$ filter $h_N(n)$. It is not possible however to extend completely this method to the IIR problem. This section presents an extension based on the methods by Prony and Pade, and illustrates the shortcomings of its application.

Consider the frequency response defined in (\ref{iir:freqprob}). One can choose $L$ {\bf equally spaced} samples of $H(\om)$ to obtain 
\begin{equation} \label{iir:freqsamp} H(\om_k)=H_k=\frac{B_k}{A_k}\,\qquad \text{for }k=0,1,\ldots,L-1 \end{equation} 
where $A_k$ and $B_k$ represent the length-$L$ DFTs of the filter coefficients $a_n$ and $b_n$ respectively. The division in (\ref{iir:freqsamp}) is done point-by-point over the $L$ values of $A_k$ and $B_k$. The objective is to use the relationship in described in (\ref{iir:freqsamp}) to calculate $a_n$ and $b_n$.

One can express (\ref{iir:freqsamp}) as $B_k=H_kA_k$. This operation represents the length-$L$ circular convolution $b(n)=h(n)\circonv a(n)$ defined as follows \cite[\S 8.7.5]{dtsp} 
\begin{equation} \label{iir:circ_conv} b(n)=h(n)\circonv a(n)=\sum_{m=0}^{L-1} h \left[((n-m))_L\right]a(m)\, ,\qquad 0\!\leq\! n\!\leq\! L-1 \end{equation} 
where $h(n)$ is the length-$L$ inverse DFT of $H_k$ and the operator $((\cdot))_L$ represents modulo $L$. Let
\begin{equation} \label{iir:matr1} \vha =\left[ \begin{array}{c} 1 \\ a_1 \\ \vdots \\ a_N \\ 0 \\ \vdots \\ 0 \end{array} \right] \quad \mbox{and} \quad \vhb = \left[ \begin{array}{c} b_0 \\ b_1 \\ \vdots \\ b_M \\ 0 \\ \vdots \\ 0 \end{array} \right] \end{equation} 
Therefore (\ref{iir:circ_conv}) can be posed as a matrix operation \cite[\S 7.4.1]{dfd} of the form 
\be \label{iir:hhat} \mhH\vha=\vhb \ee
where $\mhH = [\mhH_1 \mhH_2]$ with 
\begin{equation*} \mhH_1=\left[ \begin{tabular}{cccccccc}
$h_0$ & $h_{L-1}$ & $\cdots$ & $h_{L-N}$ \\ $h_1$ & $h_0$ & $\cdots$ & $h_{L-N+1}$ \\
$\vdots$ & $\vdots$ & & $\vdots$ \\ $h_M$ & $h_{M-1}$ & $\cdots$ & $h_{((L-N+M))_L}$ \\
$h_{M+1}$ & $h_M$ & $\cdots$ & $h_{((L-N+M+1))_L}$ \\ $\vdots$ & $\vdots$ & & $\vdots$ \\
$h_{L-2}$ & $h_{L-3}$ & $\cdots$ & $h_{L-N-2}$ \\ $h_{L-1}$ &$h_{L-2}$ & $\cdots$ & $h_{L-N-1}$
\end{tabular} \right] \end{equation*}
\begin{equation*} \mhH_2=\left[ \begin{array}{cccccccc}
h_{L-N-1} & \cdots & h_2 & h_1 \\ h_{L-N} & \cdots & h_3 & h_2 \\
\vdots & & \vdots & \vdots \\ h_{((L-N+M-1))_L} & \cdots & h_{M+2} & h_{M+1} \\
h_{((L-N+M))_L} & \cdots & h_{M+3} & h_{M+2} \\ \vdots & & \vdots & \vdots \\
h_{L-N-3} & \cdots & h_0 & h_{L-1} \\ h_{L-N-2} & \cdots & h_1 &h_0
\end{array} \right] \end{equation*}
Hence $\mhH$ is an $L\times L$ matrix. From (\ref{iir:matr1}) it is clear that the $L-(N+1)$ rightmost columns of $\mhH$ can be discarded (since the last $L-(N+1)$ values of $\vha$ in (\ref{iir:matr1}) are equal to 0). Therefore equation (\ref{iir:hhat}) can be rewritten as
\begin{equation} \label{iir:matr1b} \left[ \begin{array}{cccc} h_0 & h_{L-1} & \cdots & h_{L-N} \\ h_1 & h_0 & \cdots & h_{L-N+1} \\ \vdots & \vdots & & \vdots \\ h_M & h_{M-1} & \cdots & h_{((L-N+M))_L} \\ h_{M+1} & h_M & \cdots & h_{((L-N+M+1))_L} \\ \vdots & \vdots & & \vdots \\ h_{L-2} & h_{L-3} & \cdots & h_{L-N-2} \\ h_{L-1} & h_{L-2} & \cdots & h_{L-N-1} \end{array} \right] \left[ \begin{array}{c} 1 \\ a_1 \\ \vdots \\ a_N \end{array} \right] = \left[ \begin{array}{c} b_0 \\ b_1 \\ \vdots \\ b_M \\ 0 \\ \vdots \\ 0 \\ 0 \end{array} \right] \end{equation}
or in matrix notation
\be \label{iir:justleft} \mH\left[ \begin{array}{c} 1\\ \va \end{array} \right] = \left[ \begin{array}{c} \vb \\ \bf{0} \end{array} \right] \qquad \text{or} \qquad \mH\vta = \vhb \ee
where $\va$ and $\vb$ correspond to the length-$N$ and $(M+1)$ filter coefficient vectors respectively and $\mH$ contains the first $N+1$ columns of $\mhH$. It is possible to uncouple the calculation of $\va$ and $\vb$ from (\ref{iir:justleft}) by breaking $\mH$ furthermore as follows,
\[ \mH= \left[ \begin{array}{c} \colorbox{dark}{\begin{tabular}{cccc} $h_0$ & $h_{L-1}$ & $\cdots$ & $h_{L-N}$ \\ $h_1$ & $h_0$ & $\cdots$ & $h_{L-N+1}$ \\
$\vdots$ & $\vdots$ & & $\vdots$ \\ $h_M$ & $h_{M-1}$ & $\cdots$ & $h_{((L-N+M))_L}$ \end{tabular}} \\ \colorbox{light}{\begin{tabular}{cccc}
$h_{M+1}$ & $h_M$ & $\cdots$ & $h_{((L-N+M+1))_L}$ \\ $\vdots$ & $\vdots$ & & $\vdots$ \\
$h_{L-2}$ & $h_{L-3}$ & $\cdots$ & $h_{L-N-2}$ \\ $h_{L-1}$ & $h_{L-2}$ & $\cdots$ & $h_{L-N-1}$
\end{tabular}} \end{array} \right] \] 
Therefore
\be \label{iir:sysreduced} \mH = \left[ \begin{array}{c} \colorbox{dark}{$\mH_1$} \\ \colorbox{light}{$\mH_2$} \end{array} \right]\vta= \left[ \begin{array}{c} \vb \\ {\bf 0} \end{array} \right] \ee
with 
\[ \vta=\left[\begin{array}{c} 1 \\ \va \end{array} \right] \]
as defined in (\ref{iir:justleft}). This formulation allows to uncouple the calculations for $\va$ and $\vb$ using two systems, 
\begin{eqnarray*} \mH_1 \vta & = & \vb\\ \mH_2 \vta & = & {\bf 0} \end{eqnarray*} 
Note that the last equation can be expressed as
\begin{equation} \label{iir:find_a} \mhH_2 \va = -\vhh_2 \end{equation}
where $\mH_2=[\vhh_2 \hspace{.2cm} \mhH_2]$ (that is, $\vhh_2$ and $\mhH_2$ contain the first and second through $N$-th columns of $\mhH_2$ respectively).

From (\ref{iir:find_a}) one can conclude that if $L=N+M+1$ and if $\mhH_2$ and $\mH_1$ are nonsingular, then they can be inverted \footnote{In practice one should not invert the matrices $\mH_1$ and $\mhH_2$ but use a more robust and efficient algorithm. See \cite{matcomp} for details.} to solve for the filter coefficient vectors $\va$ in (\ref{iir:find_a}) and solve for $\vb$ using $\mH_1 \vta=\vb$.

The algorithm described above is an {\bf interpolation} method rather than an {\bf approximation} one. If $L \!>\! N+M+1$ and $\mhH_2$ is full column rank then (\ref{iir:find_a}) is an overdetermined linear system for which no exact solution exists; therefore an approximation must be found. From (\ref{iir:freqsamp}) we can define the {\it solution error} function $\e_s(\om_k)$ as
\begin{equation} \label{iir:sol_error} \e_s(\om_k) = \frac{B(\om_k)}{A(\om_k)} - H(\om_k) \end{equation} 
Using this notation, the design objective is to solve the nonlinear problem 
\[ \min_{\va,\vb} \; \normsq{\e_s(\om_k)}^2 \]

Consider the system in equation (\ref{iir:justleft}). If $\mH_2$ is overdetermined, one can define an approximation problem by introducing an error vector $\ve$,
\begin{equation} \label{iir:overdet} \vhb = \mH\vta - \ve \end{equation}
where
\[ \ve=\left[\begin{array}{c} \ve_1 \\ \ve_2 \end{array} \right] \] 
Again, it is possible to uncouple (\ref{iir:overdet}) as follows,
\begin{eqnarray} \label{iir:error_b} \vb & = & \mH_1 \vta-\ve_1 \\ \label{iir:error_a} \ve_2 & = & \vhh_2 + \mhH_2\va \end{eqnarray} 
One can minimize the least-squared error norm $\| \ve_2\|_2$ of the overdetermined system (\ref{iir:error_a}) by solving the normal equations \cite{num_met_unc}
\[ \mhH_2^T\vhh_2 = -\mhH_2^T\mhH_2\va \] 
so that 
\[ \va = -[\mhH_2^T\mhH_2]^{-1}\mhH_2^T\vhh_2 \] 
and use this result in (\ref{iir:error_b})
\begin{equation} \label{iir:eq_error_t}  \vb = \mH_1 \vta \end{equation}

Equation (\ref{iir:error_b}) represents the following time-domain operation, 
\[ \vep(n) = b(n) - h(n) \circonv a(n)\, , \; \; 0\!\leq\! n \!\leq\! M  \]
(where $\circonv$ denotes {\it circular convolution}) and can be interpreted in the frequency domain as follows,
\begin{equation} \label{iir:eq_error_f} \e_e(\om_k) = B(\om_k) - H(\om_k)A(\om_k) \end{equation} 
Equation (\ref{iir:eq_error_f}) is a weighted version of (\ref{iir:sol_error}), as follows
\[ \e_e(\om_k) = A(\om_k)\e_s(\om_k) \]
Therefore the algorithm presented above will find the filter coefficient vectors $\va$ and $\vb$ that minimize the {\it equation error} $\e_e$ in (\ref{iir:eq_error_f}) in the least-squares sense. Unfortunately, this error is not what one may want to optimize, since it is a weighted version of the {\it solution error} $\e_s$. 

\subsubsection{Iterative prefiltering linearization methods}
\label{iir:iterfilt}

Section \ref{iir:ls_lin} introduced the equation error formulation and several algorithms that minimize it. In a general sense however one is more interested in minimizing the solution error problem from (\ref{iir:opt_prob_disc}). This section presents several algorithms that attempt to minimize the solution error formulation from (\ref{iir:opt_prob}) by {\it prefiltering} the desired response $D(\om)$ in (\ref{iir:e_eqn}) with $A(\om)$. Then a new set of coefficients $\{a_n,b_n\}$ are found with an equation error formulation and the prefiltering step is repeated, hence defining an iterative procedure.

\subsubsection{Sanathanan-Koerner ({\bf SK}) method}

The method by Levy presented in Section \ref{iir:levy} suggests a relatively easy-to-implement approach to the problem of rational approximation. While interesting in itself, the equation error $\vep_e$ does not really represent what in principle one would like to minimize. A natural extension to Levy's method is the one proposed \cite{trans_ratio} by C. K. Sanathanan and J. Koerner in 1963. The algorithm iteratively {\it prefilters} the equation error formulation of Levy with an estimate of $A(\om)$.  The \textbf{SK} method considers the solution error function $\e_s$ defined by 
\begin{equation} \label{iir:sk1} \begin{split} \e_s(\om)= D(\om)- \frac{B(\om)}{A(\om)} & =\frac{1}{A(\om)} \left[A(\om) D(\om)- B(\om)\right] \\ & = \frac{1}{A(\om)} \e_e(\om) \end{split} \end{equation}
Then the solution error problem can be written as
\be \label{iir:sk3} \min_{a_k,b_k} \; \vep_s \ee
where
\bea \vep_s & = & \sum_{k=0}^L |\e_s(\om_k)|^2 \nonumber \\ & = & \sum_{k=0}^L \frac{1}{|A(\om)|^2} |\e_e(\om_k)|^2 \nonumber \\ & = & W(\om)|\e_e(\om_k)|^2 \label{iir:sk2} \eea
Note that given $A(\om)$, one can obtain an estimate for $B(\om)$ by minimizing $\e_e$ as Levy did.  This approach provides an estimate, though, because one would need to know the optimal value of $A(\om)$ to truly optimize for $B(\om)$. The idea behind this method is that by solving iteratively for $A(\om)$ and $B(\om)$ the algorithm would eventually converge to the solution of the desired solution error problem defined by (\ref{iir:sk3}). Since $A(\om)$ is not known from the beginning, it must be initialized with a reasonable value (such as $A(\om_k)=1$). 

To solve (\ref{iir:sk3}) Sanathanan and Koerner defined the same linear system from (\ref{iir:levy3}) with the same matrix and vector definitions. However the scalar terms used in the matrix and vectors reflect the presence of the weighting function $W(\om)$ in $\vep_s$ as follows,
\bean \lambda_h & = & \sum_{l=0}^{L-1} \om_l^hW(\om_l) \\ S_h & = & \sum_{l=0}^{L-1} \om_l^h R_l W(\om_l) \\ T_h & = & \sum_{l=0}^{L-1} \om_l^hI_l W(\om_l) \\ U_h & = & \sum_{l=0}^{L-1} \om_l^h (R_l^2+ I_l^2) W(\om_l) \eean
Then, given an initial definition of $A(\om)$, at the $p$-th iteration one sets 
\be \label{iir:sk4} W(\om)=\frac{1}{\left| A_{p-1}(\om_k)\right|^2} \ee
and solves (\ref{iir:levy3}) using $\{\lam,S,T,U\}$ as defined above until a convergence criterion is reached. Clearly, solving (\ref{iir:sk3}) using (\ref{iir:sk2}) is equivalent to solving a series of weighted least squares problems where the weighting function consists of the estimated values of $A(\om)$ from the previous iteration. This method is similar to a time-domain method proposed by Steiglitz and McBride \cite{smb}, presented later in this chapter.

\subsubsection{Method of Sid-Ahmed, Chottera and Jullien}

The methods by Levy and Sanathanan and Koerner did arise from an analog analysis problem formulation, and cannot therefore be used directly to design digital filters. However these two methods present important ideas that can be translated to the context of filter design. In 1978 M. Sid-Ahmed, A. Chottera and G. Jullien followed on these two important works and adapted \cite{sidahmed} the matrix and vectors used by Levy to account for the design of IIR digital filters, given samples of a desired frequency response. Consider the frequency response $H(\om)$ defined in (\ref{iir:freqprob}). In parallel with Levy's development, the corresponding equation error can be written as
\be \label{iir:ahmed1} \vep_e=\sum_{k=0}^L \left| F_k(\om) \right|^2 \ee
with
\[ F_k(\om) = \left\{(R_k+jI_k) \left(1+\sum_{c=1}^N a_i e^{-j\om_k c}\right)- \left(\sum_{c=0}^M b_i e^{-j\om_k c}\right) \right\} \]
One can follow a similar differentiation step as Levy by setting
\[ \frac{\pa \vep_e}{\pa a_1} = \frac{\pa \vep_e}{a_2} = \ldots =\frac{\pa \vep_e}{\pa b_0} = \ldots = 0 \]
with as defined in (\ref{iir:ahmed1}). Doing so results in a linear system of the form
\[ \mC\vx=\vy \]
where the vectors $\vx$ and $\vy$ are given by
\be \label{iir:ahmed2} \vx=\left\{ \begin{array}{c} b_0 \\ \vdots \\ b_M \\a_1 \\ \vdots \\ a_N \end{array} \right\} 
\hspace{1.5 cm}  \vy= \left\{ \begin{array}{c} \phi_0-r_0 \\ \vdots \\ \phi_M-r_M \\ -\beta_1 \\ \vdots \\ -\beta_N \end{array} \right\} \ee
The matrix $\mC$ has a special structure given by
\ben \mC=\left[ \begin{array}{lc} \mathbf{\Psi} & \mathbf{\Phi} \\ \mathbf{\Phi}^T & \mathbf{\Upsilon} \end{array} \right] \een
where $\mathbf{\Psi}$ and $\mathbf{\Upsilon}$ are symmetric Toeplitz matrices of order $M+1$ and $N$ respectively, and their first row is given by
\bean \begin{array}{ll} \mathbf{\Psi}_{1m} = \eta_{m-1} & \hspace{1.0cm}\mbox{for } \; m=1,\ldots,M+1 \\ \mathbf{\Upsilon}_{1m} = \beta_{m-1} & \hspace{1.0cm}\mbox{for } \; m=1,\ldots,N \end{array} \eean
Matrix $\mathbf{\Phi}$ has order $M+1\times N$ and has the property that elements on a given diagonal are identical (i.e. $\mathbf{\Phi}_{i,j}=\mathbf{\Phi}_{i+1,j+1}$). Its entries are given by
\bean \begin{array}{ll} \mathbf{\Phi}_{1m} = \phi_m + r_m & \hspace{1.0cm}\mbox{for } \; m=1,\ldots,N \\ \mathbf{\Phi}_{m1} = \phi_{m-2} - r_{m-2} & \hspace{1.0cm}\mbox{for } \; m=2,\ldots,M+1 \\ \end{array} \eean
The parameters $\{\eta,\phi,r,\beta\}$ are given by
\begin{alignat*}{2} 
\eta_i &= \sum_{k=0}^L \cos i\om_k & & \qquad \text{for } 0\!\leq\! i\!\leq\! M \\
\beta_i &= \sum_{k=0}^L |D(\om_k)|^2\cos i\om_k & & \qquad \text{for } 0\!\leq \!i\!\leq\! N-1 \\
\phi_i &= \sum_{k=0}^L R_k\cos i\om_k & & \qquad \text{for } 0\!\leq\! i\!\leq\! \max(N,M-1)  \\
r_i &= \sum_{k=0}^L I_k\sin i\om_k & & \qquad \text{for } 0\!\leq\! i\!\leq\! \max(N,M-1) 
\end{alignat*}
The rest of the algorithm works the same way as Levy's. For a solution error approach, one must weight each of the parameters mentioned above with the factor from (\ref{iir:sk4}) as in the SK method.

There are two important details worth mentioning at this point: on one hand the methods discussed up to this point (Levy, SK and Sid-Ahmed et al.) do not put any limitation on the spacing of the frequency samples; one can sample as fine or as coarse as desired in the frequency domain. On the other hand there is no way to decouple the solution of both numerator and denominator vectors. In other words, from (\ref{iir:levy4}) and (\ref{iir:ahmed2}) one can see that the linear systems that solve for vector $\vx$ solve for all the variables in it. This is more of an issue for the iterative methods (SK \& Sid-Ahmed), since at each iteration one solves for all the variables, but for the purposes of updating one needs only to keep the denominator variables (they get used in the weighting function); the numerator variables are never used within an iteration (in contrast to Burrus' Prony-based method presented in Section \ref{iir:pronypade}). This approach {\it decouples} the numerator and denominator computation into two separate linear systems. One only needs to compute the denominator variables until convergence is reached, and only then it becomes necessary to compute the numerator variables. Therefore most of the iterations solve a smaller linear system than the methods involved up to this point.

\subsubsection{Steiglitz-McBride iterative algorithm}

In 1965 K. Steiglitz and L. McBride presented an algorithm \cite{smb,time_iter} that has become quite popular in statistics and engineering applications. The {\it Steiglitz-McBride} method ({\bf SMB}) considers the problem of deriving a transfer function for either an analog or digital system from their input and output data; in essence it is a time-domain method. Therefore it is mentioned in this work for completeness as it closely relates to the methods by Levy, SK and Sid-Ahmed, yet it is far better known and understood.

The derivation of the SMB method follows closely that of SK. In the Z-domain, the transfer function of a digital system is defined by
\[ H(z) = \frac{B(z)}{A(z)} = \frac{b_0+b_1z^{-1}+\ldots+b_Nz^{-N}}{1+a_1z_{-1}+\ldots+a_Nz^{-N}} \]
Furthermore
\[ Y(z) = H(z)X(z) = \frac{B(z)}{A(z)}X(z) \]
Steiglitz and McBride define the following problem,
\be \label{iir:smb1} \min \; \vep_s = \sum_i \e_i(z)^2=\frac{1}{2\pi j}\oint \left| X(z)\frac{B(z)}{A(z)}-D(z)  \right|^2 \frac{dz}{z} \ee
where $X(z)=\sum_j x_jz^{-j}$ and $D(z)=\sum_j d_jz^{-j}$ represent the z-transforms of the input and desired signals respectively. Equation (\ref{iir:smb1}) is the familiar nonlinear solution error function expressed in the Z-domain. Steiglitz and McBride realized the complexity of such function and proposed the iterative solution (\ref{iir:smb1}) using a simpler problem defined by
\be \label{iir:smb2} \min \; \vep_e = \sum_i \e_i(z)^2=\frac{1}{2\pi j}\oint \left| X(z)B(z)-D(z)A(z)  \right|^2 \frac{dz}{z} \ee
This linearized error function is the familiar equation error in the Z-domain. Steiglitz and McBride proposed a two-mode iterative approach. The {\bf SMB Mode 1} iteration is similar to the SK method, in that at the $k$-th iteration a {\it linearized} error criterion based on (\ref{iir:smb2}) is used,
\be \begin{split} \e_k(z) & = \frac{B_k(z)}{A_{k-1}(z)}X(z) - \frac{A_k(z)}{A_{k-1}(z)}D(z) \\ & = W_k(z)\left[B_k(z)X(z)-A_k(z)D(z)\right] \end{split} \ee
where
\[ W_k(z) = \frac{1}{A_{k-1}(z)} \]
Their derivation\footnote{For more details the reader should refer to \cite{smb,time_iter}.} leads to the familiar linear system
\[ \mC\vx=\vy \]
with the following vector definitions
\[ \vx=\left\{ \begin{array}{c} b_0 \\ \vdots \\ b_N \\a_1 \\ \vdots \\ a_N \end{array} \right\} 
\hspace{1.5 cm}  \vq_j= \left\{ \begin{array}{c} x_j \\ \vdots \\ x_{j-N+1} \\ d_{j-1} \\ \vdots \\ d_{j-N} \end{array} \right\} \]
The vector $\vq_j$ is referred to as the {\it input-output} vector. Then
\bean \mC&  = & \sum_j \vq_j \vq_j^T \\ \vy & = & \sum_j d_j \vq_j \eean

{\bf SMB Mode 2} is an attempt at reducing further the error once Mode 1 produces an estimate close enough to the actual solution. The idea behind Mode 2 is to consider the solution error defined by (\ref{iir:smb1}) and equate its partial derivatives with respect to the coefficients to zero. Steiglitz and McBride showed \cite{smb,time_iter} that this could be attained by defining a new vector
\[ \vr_j= \left\{ \begin{array}{c} x_j \\ \vdots \\ x_{j-N+1} \\ y_{j-1} \\ \vdots \\ y_{j-N} \end{array} \right\} \]
Then
\bean \mC&  = & \sum_j \vr_j \vq_j^T \\ \vy & = & \sum_j d_j \vr_j \eean

The main diference between Mode 1 and Mode 2 is the fact that Mode 1 uses the desired values to compute its vectors and matrices, whereas Mode 2 uses the actual output values from the filter. The rationale behind this is that at the beggining, the output function $y(t)$ is not accurate, so the desired function provides better data for computations. On the other hand, Mode 1 does not really solve the desired problem. Once Mode 1 is deemed to have reached the vicinity of the solution, one can use true partial derivatives to compute the gradient and find the actual solution; this is what Mode 2 does. 

It has been claimed that under certain conditions the Steiglitz-McBride algorithm converges. However no guarantee of global convergence exists. A more thorough discussion of the Steiglitz-McBride algorithm and its relationships to other parameter estimation algorithms (such as the Iterative Quadratic Maximum Likelihood algorithm, or IQML) are found in \cite{smb_globconv, smb_rev, smb_iqml}. 

\subsubsection{Jackson's method}

The following is a recent approach (from 2008) by Leland Jackson \cite{jackson} based in the frequency domain. Consider vectors $\va\in\real^N$ and $\vb\in\real^M$ such that
\[ H(\om)=\frac{B(\om)}{A(\om)} \]
where $H(\om),B(\om),A(\om)$ are the Fourier transforms of $\vh,\vb$ and $\va$ respectively. For a discrete frequency set one can describe Fourier transform vectors $\vB=\mW_b \vb$ and $\vA=\mW_a \va$ (where $\mW_b,\mW_a$ correspond to the discrete Fourier kernels for $\vb,\va$ respectively). Define
\[ H_a(\om_k)=\frac{1}{A(\om_k)} \]
In vector notation, let $\mD_a=\diag(\vH_a)=\diag(1/\vA)$. Then
\be \label{zero} H(\om)=\frac{B(\om)}{A(\om)} = H_a(\om)B(\om) \Rightarrow \vH=\mD_a  \vB \ee
Let $H_d(\om)$ be the desired complex frequency response. Define $\mD_d=\diag(\vH_d)$. Then one wants to solve
\[ \min \; \vE^*\vE = \norm{\vE}_2^2 \]
where $\vE=\vH-\vH_d$. From (\ref{zero}) one can write $\vH=\vH_d+\vE$ as
\be \vH=\mD_a\vB=\mD_a  \mW_b  \vb \ee
Therefore
\be \label{one} \vH_d=\vH-\vE=\mD_a \mW_b \vb-\vE \ee
Solving (\ref{one}) for $\vb$ one gets
\be \label{two} \vb=(\mD_a \mW_b) \bs \vH_d \ee
Also,
\[ \vH_d=\mD_d  \hat{\vI} =\mD_d  \mD_a  \vA=\mD_a  \mD_d  \vA=\mD_a  \mD_d  \mW_a  \va \]
where $\hat{\vI}$ is a unit column vector. Therefore 
\[ \vH-\vE=\vH_d=\mD_a  \mD_d  \mW_a  \va \]
From (\ref{one}) we get
\[ \mD_a  \mW_b  \vb-\vE=\mD_a  \mD_d  \mW_a  \va \]
or
\[ \mD_a  \mD_d  \mW_a  \va+\vE=\mD_a \mW_b \vb \]
which in a least squares sense results in
\be \label{three} \va=(\mD_a  \mD_d  \mW_a) \bs (\mD_a  \mW_b  \vb) \ee
From (\ref{two}) one gets
\[ \va=(\mD_a  \mD_d  \mW_a) \bs (\mD_a  \mW_b  [(\mD_a  \mW_b) \bs \vH_d]) \]
As a summary, at the $i$-th iteration one can write (\ref{one}) and (\ref{three}) as follows,
\begin{align}
\vb_i &= (\diag(1/\vA_{i-1})  \mW_b) \bs \vH_d \notag \\
\va_i &= (\diag(1/\vA_{i-1})  \diag(\vH_d)  \mW_a) \bs (\diag(1/\vA_{i-1})  \mW_b  \vb_i) \notag
\end{align}

\subsubsection{Soewito's quasilinearization method}
\label{iir:soe}

Consider the {\it equation error} residual function
\begin{align} \notag \begin{split} e(\om_k) &= B(\om_k)-D(\om_k)\cdot A(\om_k) \\
&= \sum_{n=0}^M b_n e^{-j\om_k n} -D(\om_k)\cdot\left(1+\sum_{n=1}^N a_n e^{-j\om_k n}\right) \\
&= b_0+b_1e^{-j\om_k}+\cdots+b_M e^{-j\om_k M} \cdots \\ & \hspace{1cm} -D_k-D_k a_1e^{-j\om_k}-\cdots-D_k a_N e^{-j\om_k N} \\
&= \left(b_0+\cdots b_M e^{-j\om_kM}\right)- \cdots \\ & \hspace{1cm} D_k\left(a_1e^{-j\om_k}+\cdots a_Ne^{-j\om_kN}\right)-D_k
\end{split} \end{align}
with $D_k=D(\om_k)$. The last equation indicates that one can represent the equation error in matrix form as follows,
\[ \ve=\mF\vh-\vD \]
where $\mF=[\mF_1\quad\mF_2]$ and
\ben \mF_1 =\left[ \begin{array}{ccccccc} 1 & e^{-j\om_0} & \cdots & e^{-j\om_0 M} \\ 
\vdots & \vdots & & \vdots \\ 
1 & e^{-j\om_{L-1}} & \cdots & e^{-j\om_{L-1} M} \end{array} \right] \een
\ben \mF_2 =\left[ \begin{array}{ccccccc} -D_0e^{-j\om_0} & \cdots & - D_0e^{-j\om_0 N} \\ 
\vdots & & \vdots \\ 
-D_{L-1}e^{-j\om_{L-1}} & \cdots & - D_{L-1}e^{-j\om_{L-1} N} \end{array} \right] \een
and
\ben \vh =\left[ \begin{array}{c} b_0 \\ b_1 \\ \vdots \\ b_M \\ a_1 \\ \vdots \\ a_N \end{array} \right] \quad \mbox{and} \quad \vD =\left[ \begin{array}{c} D_0 \\ \vdots \\ D_{L-1} \end{array} \right] \een
Consider now the {\it solution error} residual function
\begin{align*} s(\om_k) &= H(\om_k)-D(\om_k) = \frac{B(\om_k)}{A(\om_k)}-D(\om_k) \\ &= \frac{1}{A(\om_k)}\left[B(\om_k)-D(\om_k)\cdot A(\om_k)\right] \\ &= W(\om_k) e(\om_k) \end{align*}
Therefore one can write the solution error in matrix form as follows
\be \label{serr} \vs=\mW(\mF\vh-\vD) \ee
where $\mW$ is a diagonal matrix with $\frac{1}{A(\om)}$ in its diagonal. From (\ref{serr}) the least-squared solution error $\vep_s=\vs^*\vs$ can be minimized by
\be \label{smb1} \vh=(\mF^*\mW^2\mF)^{-1}\mF^*\mW^2\vD \ee
From (\ref{smb1}) an iteration\footnote{Soewito refers to this expression as the Steiglitz-McBride Mode-1 in frequency domain.} could be defined as follows
\[ \vh_{i+1}=(\mF^*\mW_i^2\mF)^{-1}\mF^*\mW_i^2\vD \]
by setting the weights $\mW$ in (\ref{serr}) equal to $A_k(\om)$, the Fourier transform of the current solution for $\va$.

A more formal approach to minimizing $\vep_s$ consists in using a gradient method (these approaches are often referred to as {\it Newton-like} methods). First one needs to compute the {\it Jacobian} matrix $\mJ$ of $\vs$, where the $pq$-th term of $\mJ$ is given by $\mJ_{pq}=\frac{\pa \vs_p}{\pa \vh_q}$ with $\vs$ as defined in (\ref{serr}). Note that the $p$-th element of $\vs$ is given by
\[ s_p= H_p-D_p=\frac{B_p}{A_p}-D_p \]
For simplicity one can consider these reduced form expressions for the independent components of $\vh$,
\begin{align*} 
\frac{\pa s_p}{\pa b_q} &= \frac{1}{A_p} \frac{\pa}{\pa b_q} \sum_{n=0}^M b_n e^{-j\om_p n} = W_p e^{-j\om_p q} \\ 
\frac{\pa s_p}{\pa a_q} &= B_p \frac{\pa}{\pa a_q} \frac{1}{A_p} = \frac{-B_p}{A_p^2}\frac{\pa}{\pa a_q}\left(1+\sum_{n=1}^N a_n e^{-j\om_p n}\right) \\
& = \frac{-1}{A_p}\cdot \frac{B_p}{A_p}\cdot e^{-j\om_p q} = -W_pH_p e^{-j\om_p q} \end{align*}
Therefore on can express the Jacobian $\mJ$ as follows,
\be \label{jac} \mJ = \mW\mG \ee
where $\mG=[\mG_1\quad\mG_2]$ and
\ben \mG_1 =\left[ \begin{array}{cccc} 1 & e^{-j\om_0} & \cdots & e^{-j\om_0 M} \\ \vdots & \vdots & & \vdots \\ 1 & e^{-j\om_{L-1}} & \cdots & e^{-j\om_{L-1} M} \end{array} \right] \een
\ben \mG_2 =\left[ \begin{array}{ccc} -H_0e^{-j\om_0} & \cdots & - H_0e^{-j\om_0 N} \\ \vdots & & \vdots \\ -H_{L-1}e^{-j\om_{L-1}} & \cdots & - H_{L-1}e^{-j\om_{L-1} N} \end{array} \right] \een
Consider the {\it solution error} least-squares problem given by
\[ \min_{\vh} f(\vh) = \vs^T\vs \]
where $\vs$ is the solution error residual vector as defined in (\ref{serr}) and depends on $\vh$. It can be shown \cite[pp. 219]{num_met_unc} that the gradient of the squared error $f(\vh)$ (namely $\nabla\vf$) is given by
\be \label{grad} \nabla\vf = \mJ^*\vs \ee
A necessary condition for a vector $\vh$ to be a local minimizer of $f(\vh)$ is that the gradient $\nabla \vf$ be zero at such vector. With this in mind and combining (\ref{serr}) and (\ref{jac}) in (\ref{grad}) one gets
\be \label{grad2} \nabla\vf=\mG^*\mW^2(\mF\vh-\vD)=\mathbf{0} \ee
Solving the system (\ref{grad2}) gives
\[ \vh=(\mG^*\mW^2\mF)^{-1}\mG^*\mW^2\vD \]
An iteration can be defined as follows\footnote{Soewito refers to this expression as the Steiglitz-McBride Mode-2 in frequency domain. Compare to the Mode-1 expression and the use of $G_i$ instead of $F$.} 
\be \label{smb2} \vh_{i+1}=(\mG_i^*\mW_i^2\mF)^{-1}\mG_i^*\mW_i^2\vD \ee
where matrices $\mW$ and $\mG$ reflect their dependency on current values of $\va$ and $\vb$.

Atmadji Soewito \cite{soe} expanded the method of {\it quasilinearization} of Bellman and Kalaba \cite{quasilin} to the design of IIR filters. To understand his method consider the first order of Taylor's expansion near $H_i(z)$, given by
\begin{align*} H_{i+1}(z) &= H_i(z) + \\ & \frac{[B_{i+1}(z)-B_i(z)]A_i(z)-[A_{i+1}(z)-A_i(z)]B_i(z)}{A_i^2(z)} \\
&= H_i(z) + \\ & \frac{B_{i+1}(z)-B_i(z)}{A_i(z)} - \frac{B_i(z)[A_{i+1}(z)-A_i(z)]}{A_i^2(z)}
\end{align*}
Using the last result in the solution error residual function $s(\om)$ and applying simplification leads to
\be \begin{split} s(\om) & = \frac{B_{i+1}(\om)}{A_i(\om)}-\frac{H_i(\om)A_{i+1}(\om)}{A_i(\om)}+\frac{B_i(\om)}{A_i(\om)}-D(\om) \\ &= \frac{1}{A_i(\om)}[B_{i+1}(\om)-H_i(\om)A_{i+1}(\om)+B_i(\om) \ldots \\
 & \hspace{2cm} -A_i(\om)D(\om)] \label{solnsoe} \end{split} \ee
Equation (\ref{solnsoe}) can be expressed (dropping the use of $\om$ for simplicity) as 
\be \label{serr2} s=W\Big(\big( \left[B_{i+1}-H_i(A_{i+1}-1)\right]-H_i \big) + \big(\left[B_i-D(A_i-1)\right]-D \big)\Big) \ee
One can recognize the two terms in brackets as $\mG\vh_{i+1}$ and $\mF\vh_i$ respectively. Therefore (\ref{serr2}) can be represented in matrix notation as follows,
\be \label{soevec} \vs=\mW[\mG\vh_{i+1}-(\vD+\vH_i-\mF\vh_i)] \ee
with $\vH=[H_0,H_1,\cdots,H_{L-1}]^T$. Therefore one can minimize $\vs^T\vs$ from (\ref{soevec}) with
\be \label{soe3} \vh_{i+1}=(\mG_i^*\mW_i^2\mG_i)^{-1}\mG_i^*\mW_i^2(\vD+\vH_i-\mF\vh_i) \ee
since all the terms inside the parenthesis in (\ref{soe3}) are constant at the $(i+1)$-th iteration. In a sense, (\ref{soe3}) is similar to (\ref{smb2}), where the desired function is updated from iteration to iteration as in (\ref{soe3}).

It is important to note that any of the three algorithms can be modified to solve a {\it weighted} $l_2$ IIR approximation using a weighting function $W(\om)$ by defining 
\be \label{weights} V(\om)=\frac{W(\om)}{A(\om)} \ee
Taking (\ref{weights}) into account, the following is a summary of the three different updates discussed so far:
\begin{align*}
\mbox{SMB Frequency Mode-1:} &\hspace{.3cm} \vh_{i+1} = (\mF^*\mV_i^2\mF)^{-1}\mF^*\mV_i^2\vD \\
\mbox{SMB Frequency Mode-2:} &\hspace{.3cm} \vh_{i+1} = (\mG_i^*\mV_i^2\mF)^{-1}\mG_i^*\mV_i^2\vD \\
\mbox{Soewito's quasilinearization:} &\hspace{.3cm} \vh_{i+1} = (\mG_i^*\mV_i^2\mG_i)^{-1}\ast \ldots \\
 &\hspace{1.8cm} \mG_i^*\mV_i^2(\vD+\vH_i-\mF\vh_i) 
\end{align*}

\subsection{$\lp$ approximation} 
\label{iir:lp}

Infinite Impulse Response (IIR) filters are important tools in signal processing. The flexibility they offer with the use of poles and zeros allows for relatively small filters meeting specifications that would require somewhat larger FIR filters. Therefore designing IIR filters in an efficient and robust manner is an inportant problem.

This section covers the design of a number of important $\lp$ IIR problems. The methods proposed are consistent with the methods presented for FIR filters, allowing one to build up on the lessons learned from FIR design problems. The complex $\lp$ IIR problem is first presented in Section \ref{iir:cplxfreq}, being an essential tool for other relevant problems. The $\lp$ frequency-dependent IIR problem is also introduced in Section \ref{iir:cplxfreq}. While the frequency-dependent formulation might not be practical in itself as a filter design formulation, it is fundamental for the more relevant magnitude $\lp$ IIR filter design problem, presented in Section \ref{iir:magsect}.

Some complications appear when designing IIR filters, among which the intrinsic least squares solving step clearly arises from the rest. Being a nonlinear problem, special handling of this step is required. It was detemined after thorough experimentation that the {\it quasilinearization} method of Soewito presented in Section \ref{iir:soe} can be employed successfully to handle this issue.  

\begin{figure}[h]
    \centerline{\psfig{figure=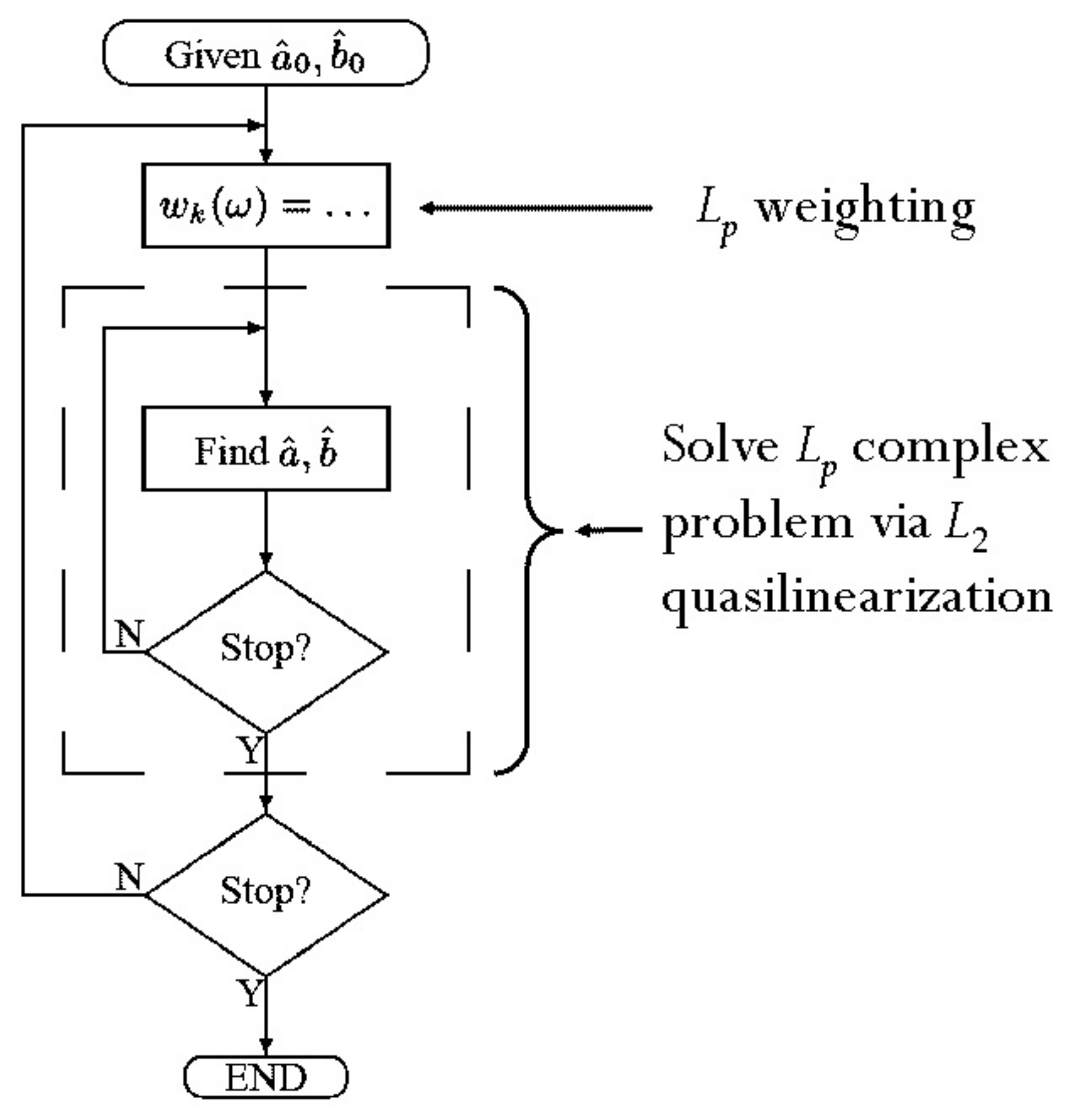,width=2.4in}}
    \caption{Block diagram for complex $\lp$ IIR algorithm.}
    \label{iir:iir_cplx_eps}
\end{figure}

\subsubsection{Complex and frequency-dependent $\lp$ approximation}
\label{iir:cplxfreq}

Chapter \ref{ch:fir} introduced the problem of designing $\lp$ complex FIR filters.  The complex $\lp$ IIR  algorithm builds up on its FIR counterpart by introducing a {\it nested structure} that internally solves for an $l_2$ complex IIR problem. Figure \ref{iir:iir_cplx_eps} illustrates this procedure in more detail. This method was first presented in \cite{iir_icassp01}.

\begin{figure}[h]
    \centerline{\psfig{figure=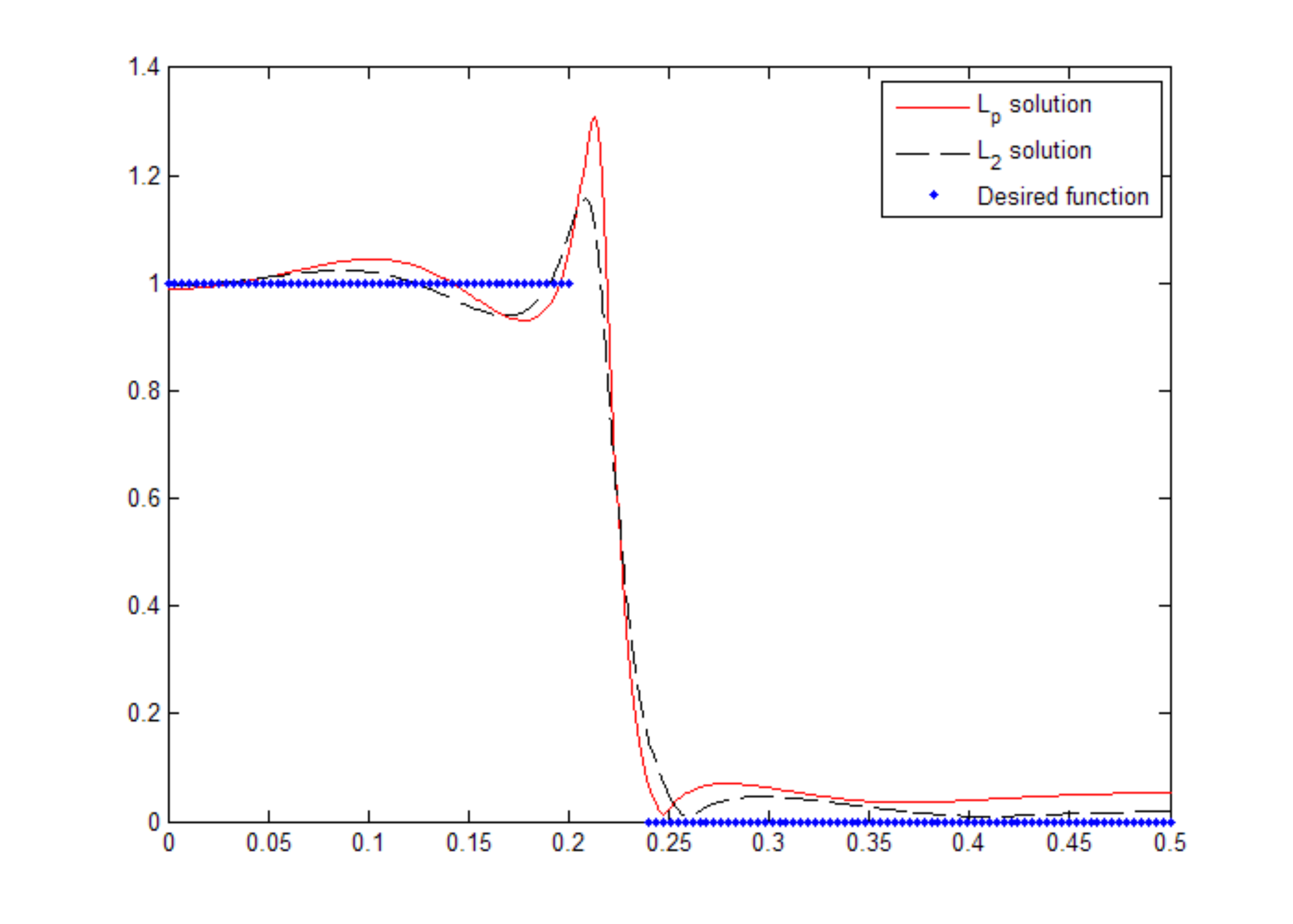,width=3.8in}}
    \caption{Results for complex $l_{100}$ IIR design.}
    \label{iir_lpiircplx1}
\end{figure}

\begin{figure}[h]
    \centerline{\psfig{figure=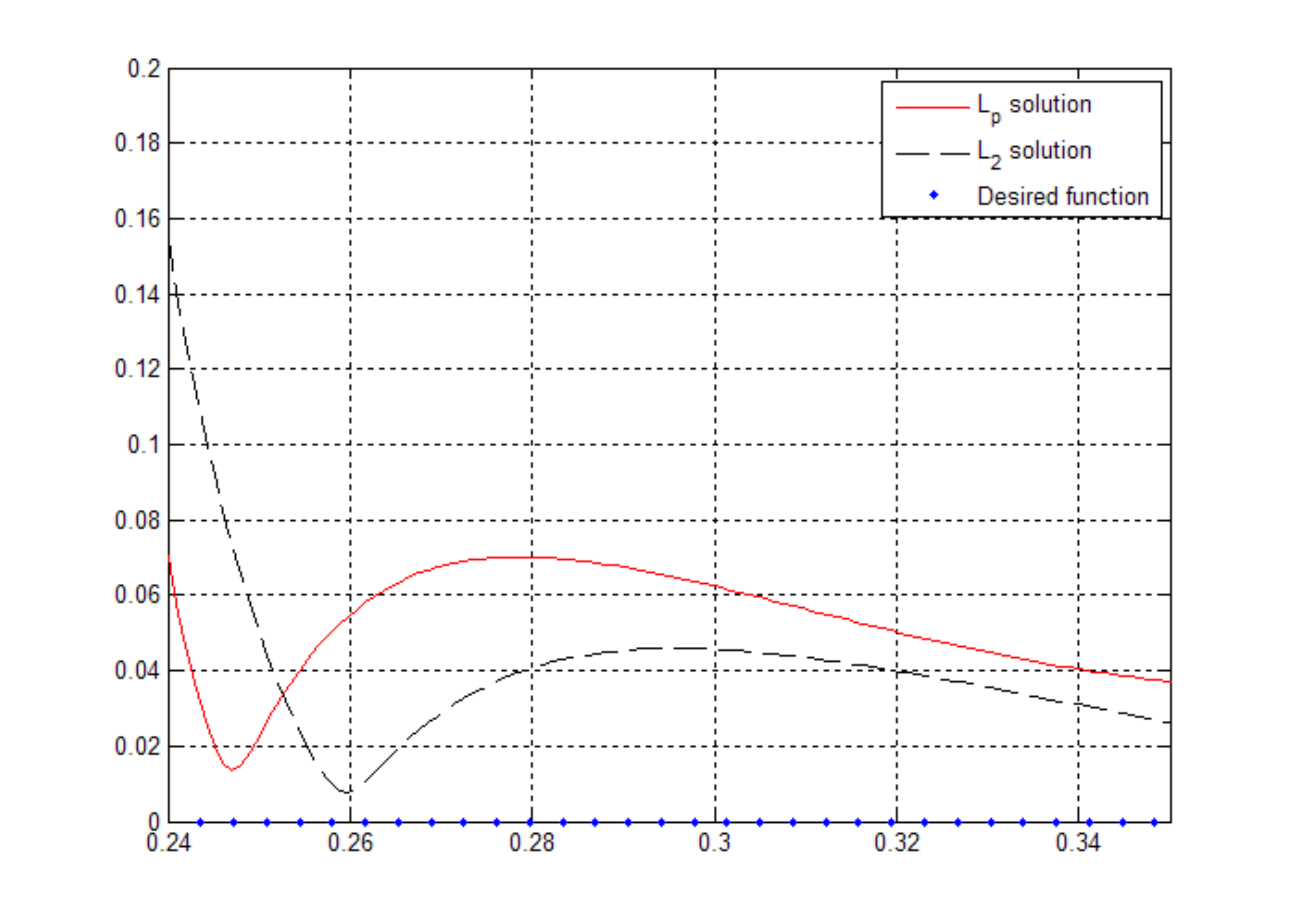,width=3.8in}}
    \caption{Maximum error for $l_2$ and $l_{100}$ complex IIR designs.}
    \label{iir_lpiircplx2}
\end{figure}

\begin{figure}[h]
    \centerline{\psfig{figure=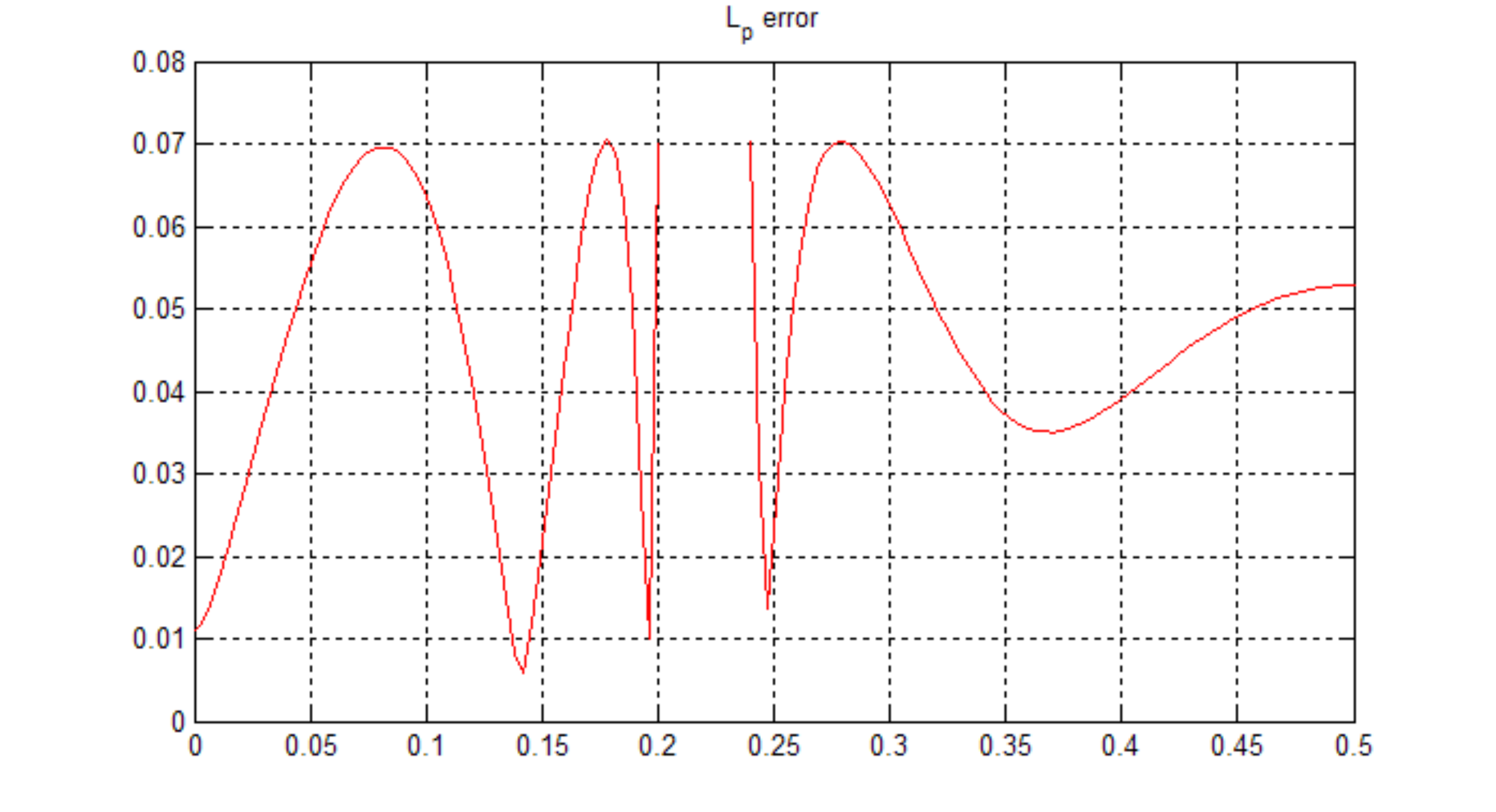,width=4.0in}}
    \caption{Error curve for $l_{100}$ complex IIR design.}
    \label{iir_lpiircplx3}
\end{figure}

Compared to its FIR counterpart, the IIR method only replaces the weighted linear least squares problem for Soewito's quasilinearization algorithm. While this nesting approach might suggest an increase in computational expense, it was found in practice that after the initial $l_2$ iteration, in general the $\lp$ iterations only require from one to only a few internal weighted $l_2$ quasilinearization iterations, thus maintaining the algorithm efficiency. Figures \ref{iir_lpiircplx1} through \ref{iir_lpiircplx3} present results for a design example using a length-5 IIR filter with $p=100$ and transition edge frequencies of 0.2 and 0.24 (in normalized frequency).  

Figure \ref{iir_lpiircplx1} compares the $l_2$ and $\lp$ results and includes the desired frequency samples. Note that no transition band was specified.  Figure \ref{iir_lpiircplx2} illustrates the effect of increasing $p$. The largest error for the $l_2$ solution is located at the transition band edges. As $p$ increases the algorithm weights the larger errors heavier; as a result the largest errors tend to decrease.  In this case the magnitude of the frequency response went from 0.155 at the stopband edge (in the $l_2$ case) to 0.07 (for the $\lp$ design).  Figure \ref{iir_lpiircplx3} shows the error function for the $\lp$ design, illustrating the quasiequiripple behavior for large values of $p$.

Another fact worth noting from Figure \ref{iir_lpiircplx1} is the increase in the peak in the right hand side of the passband edge (around $f=0.22$). The $l_p$ solution increased the amplitude of this peak with respect to the corresponding $l_2$ solution. This is to be expected, since this peak occurs at frequencies not included in the specifications, and since the $l_p$ algorithm will move poles and zeros around in order to meet find the optimal $l_p$ solution (based on the frequencies included for the filter derivation).  The addition of a specified transition band function (such as a spline) would allow for control of this effect, depending on the user's preferences.

The frequency-dependent FIR problem was first introduced in Section \ref{fir:freqp}. Following the FIR approach, one can design IIR frequency-dependent filters by merely replacing the linear weighted least squares step by a nonlinear approach, such as the quasilinearization method presented in Section \ref{iir:soe} (as in the complex $\lp$ IIR case).  This problem illustrates the flexibility in design for $\lp$ IRLS-based methods. 

\begin{figure}[h]
    \centerline{\psfig{figure=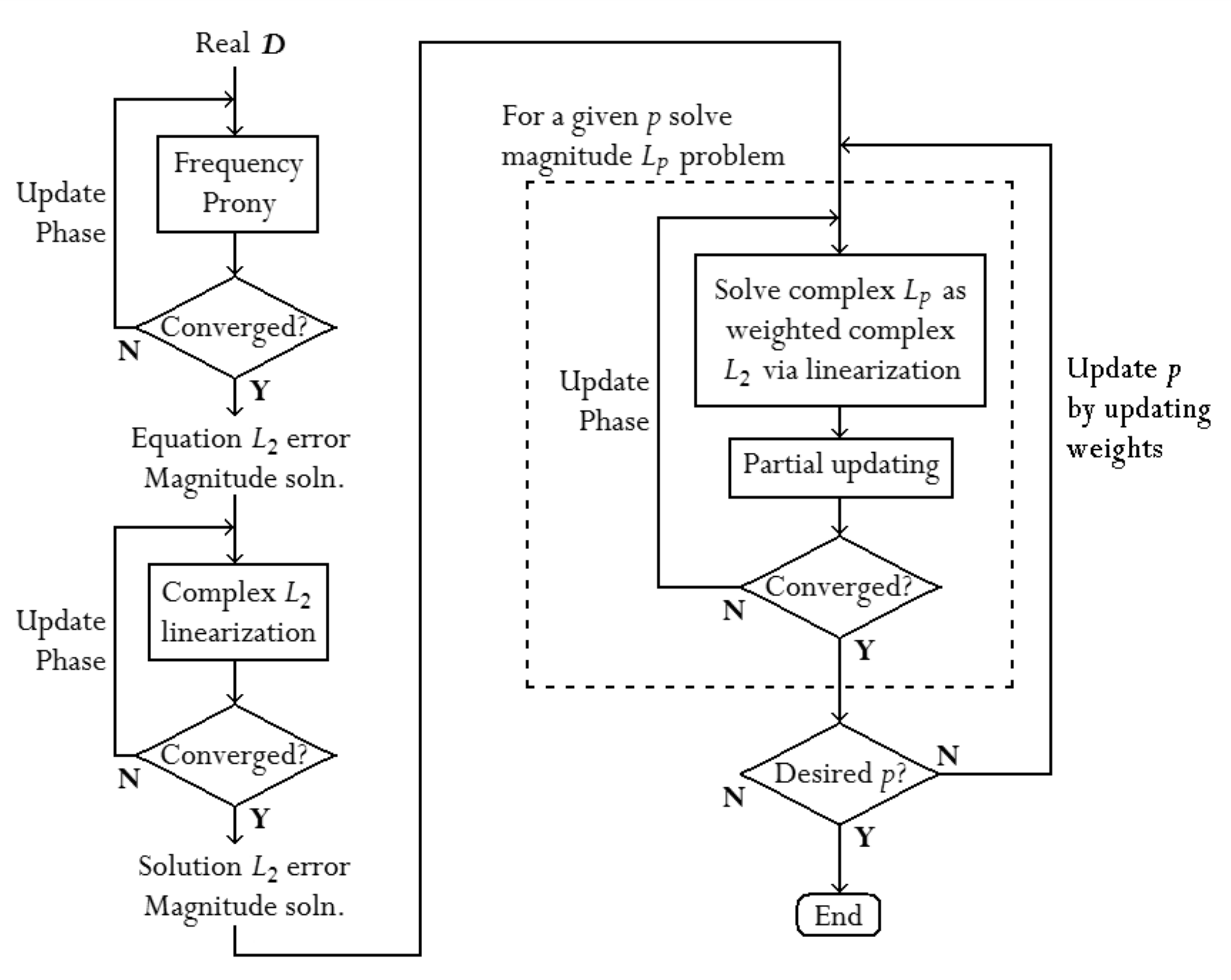,width=3.4in}}
    \caption{Block diagram for magnitude $\lp$ IIR method.}
    \label{iir_lpmag}
\end{figure}

\subsubsection{Magnitude $\lp$ IIR design}
\label{iir:magsect}

The previous sections present algorithms that are based on complex specifications; that is, the user must specify both desired magnitude and phase responses. In some cases it might be better to specify a desired magnitude response only, while allowing an algorihm to select the phase that optimally minimizes the magnitude error. Note that if an algorithm is given a phase in addition to a magnitude function, it must then make a compromise between approximating both functions. The magnitude $\lp$ IIR approximation problem overcomes this dilemma by posing the problem only in terms of a desired magnitude function. The algorithm would then find the optimal phase that provides the optimal magnitude approximation. A mathematical formulation follows,
\be \label{iir:mag1} \min_{\va,\vb} \;\; \left\| \; \left|D(\om)\right| - \left| \frac{B(\om;\vb)}{A(\om;\va)} \right| \; \right\|_p^p \ee

A critical idea behind the magnitude approach is to allow the algorithm to find the optimum phase for a magnitude approximation. It is important to recognize that the optimal magnitude filter indeed has a complex frequency response. Atmadji Soewito \cite{soe} published in 1990 a theorem in the context of $l_2$ IIR design that demonstrated that the phase corresponding to an optimal magnitude approximation could be found iteratively by {\bf updating} the {\it desired phase} in a complex approximation scenario. In other words, given a desired complex response $D_0$ one can solve a complex $l_2$ problem and take the resulting phase to form a new desired response $D^+$ from the original desired magnitude response with the new phase. That is,
\[ D_{i+1}= |D_0|e^{j\phi_i} \]
where $D_0$ represents the original desired magnitude response and $e^{j\phi_i}$ is the resulting phase from the previous iteration. This approach was independently suggested \cite{jackson} by Leland Jackson and Stephen Kay in 2008. 

This work introduces an algorithm to solve the magnitude $\lp$ IIR problem by combining the IRLS-based complex $\lp$ IIR algorithm from Section \ref{iir:cplxfreq} with the phase updating ideas from Soewito, Jackson and Kay. The resulting algorithm is robust, efficient and flexible, allowing for different orders in the numerator and denominator as well as even or uneven sampling in frequency space, plus the optional use of specified transition bands. A block diagram for this method is presented in Figure \ref{iir_lpmag}.

The overall $\lp$ IIR magnitude procedure can be summarized as follows,
\begin{enumerate}
\item Experimental analysis demonstrated that a reasonable initial solution for each of the three main stages would allow for faster convergence. It was found that the frequency domain Prony method by Burrus \cite{dfd} (presented in Section \ref{iir:pronypade}) offered a good {\it initial guess}. In Figure \ref{iir_lpmag} this method is iterated to update the specified phase. The outcome of this step would be an {\bf equation error $l_2$ magnitude} design. 

\item The equation error $l_2$ magnitude solution from the previous step initializes a second stage where one uses quasilinearization to update the desired phase. Quasilinearization solves the true {\it solution error} complex approximation. Therefore by iterating on the phase one finds at convergence a {\bf solution error $l_2$ magnitude} design.

\item The rest of the algorithm follows the same idea as in the previous step, except that the least squared step becomes a {\it weighted} one (to account for the necessary $\lp$ homotopy weighting). It is also crucial to include the partial updating introduced in Section \ref{fir:irlsmeth}. By iterating on the weights one would find a {\bf solution error $\lp$ magnitude} design.
\end{enumerate}

\begin{figure}[h]
    \centerline{\psfig{figure=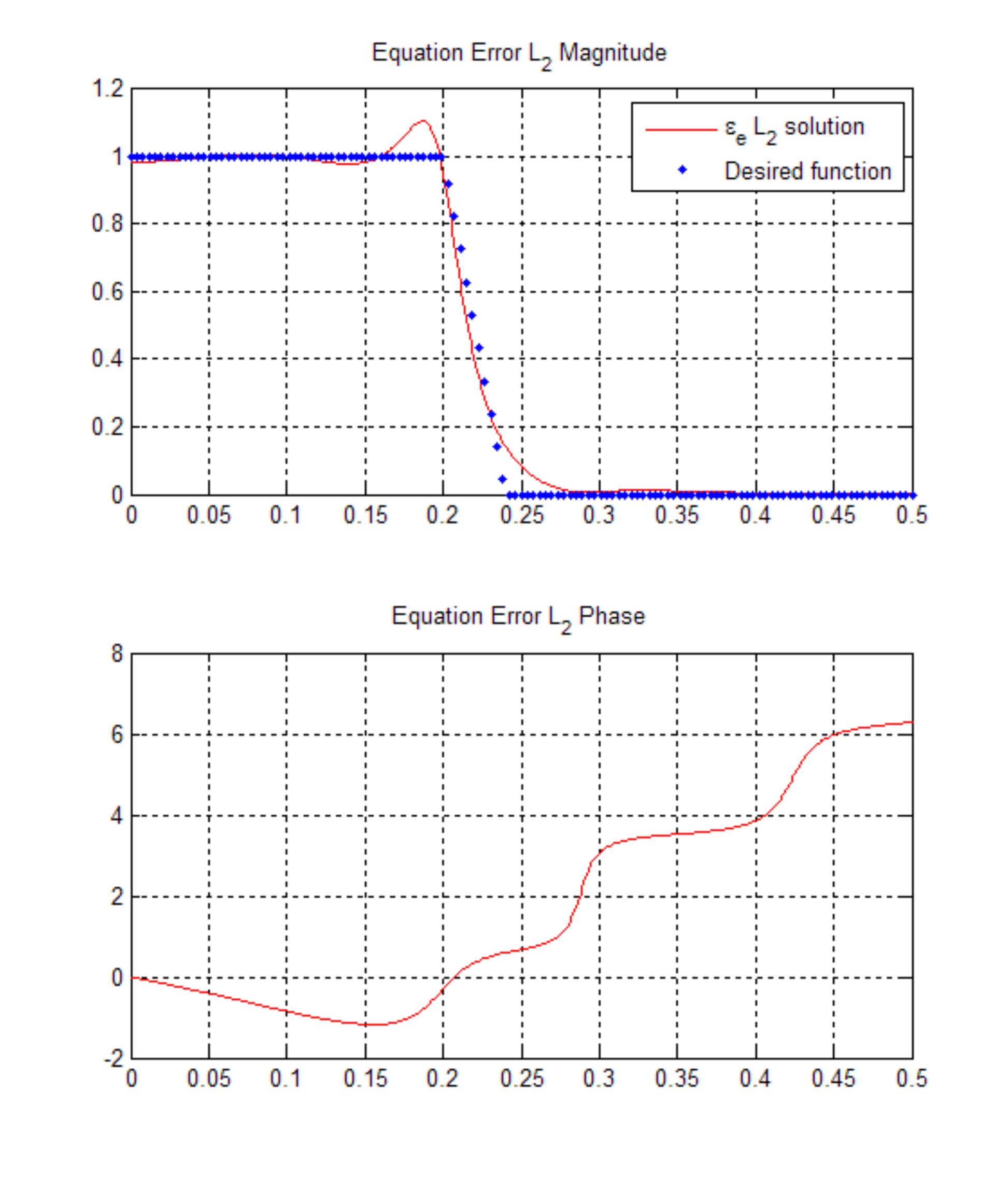,width=3.8in}}
    \caption{Equation error $l_2$ magnitude design.}
    \label{lpiirmag1}
\end{figure}

\begin{figure}[h]
    \centerline{\psfig{figure=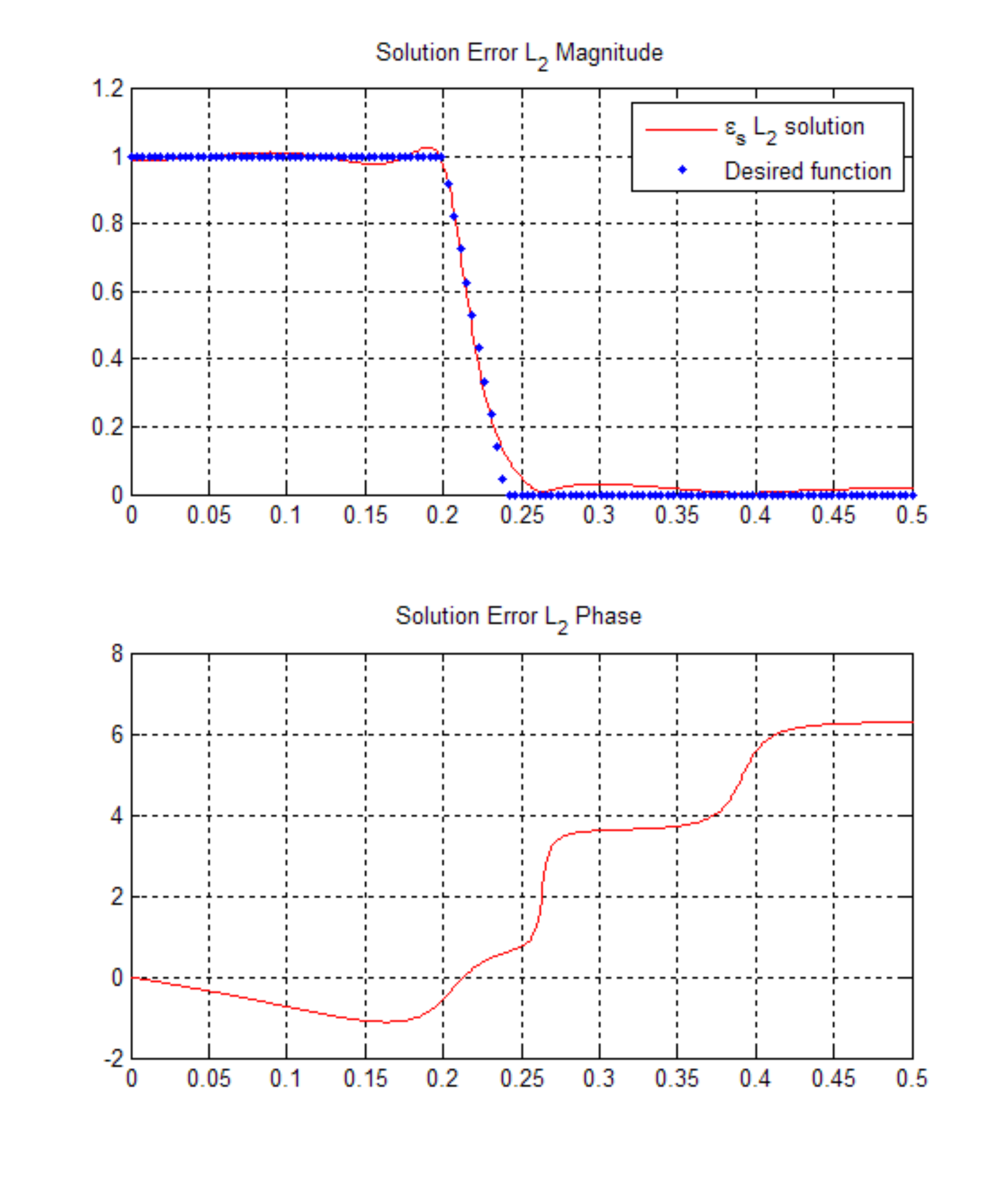,width=3.8in}}
    \caption{Solution error $l_2$ magnitude design.}
    \label{lpiirmag2}
\end{figure}

\begin{figure}[h]
    \centerline{\psfig{figure=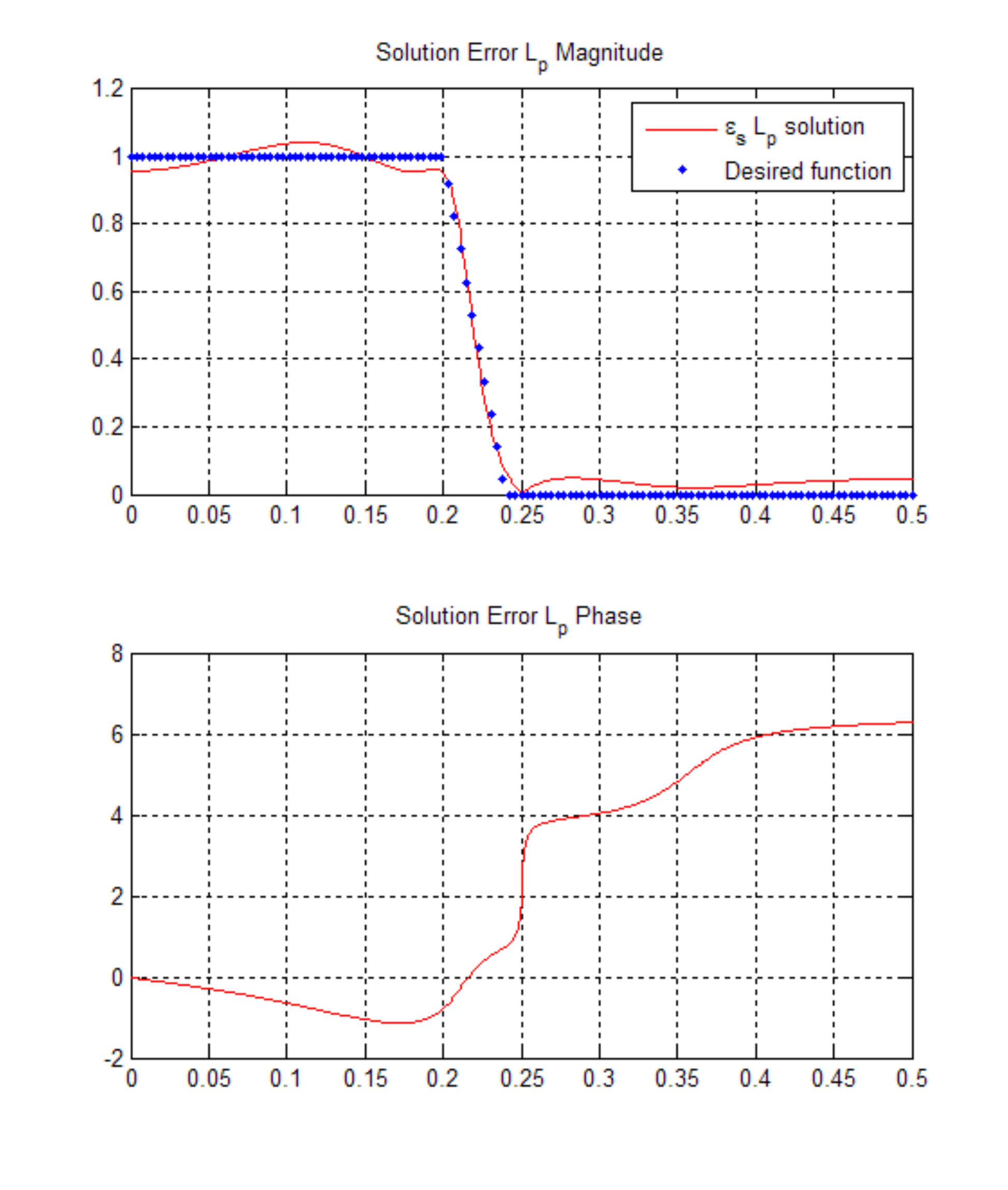,width=3.8in}}
    \caption{Solution error $\lp$ magnitude design.}
    \label{lpiirmag3}
\end{figure}

\begin{figure}[h]
    \centerline{\psfig{figure=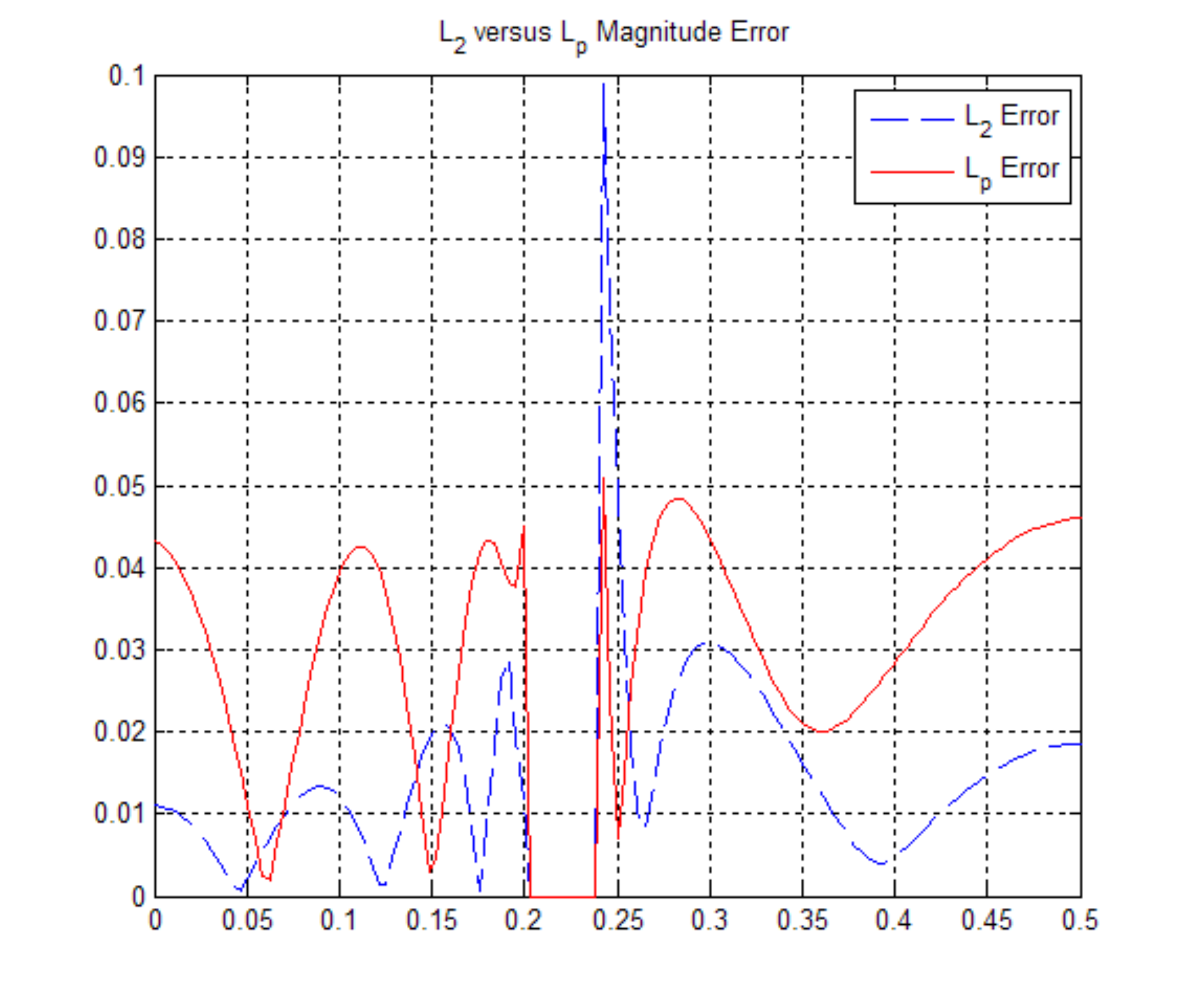,width=3.8in}}
    \caption{Comparison of $l_2$ and $\lp$ IIR magnitude designs}
    \label{lpiirmag4}
\end{figure}

\begin{figure}[h]
    \centerline{\psfig{figure=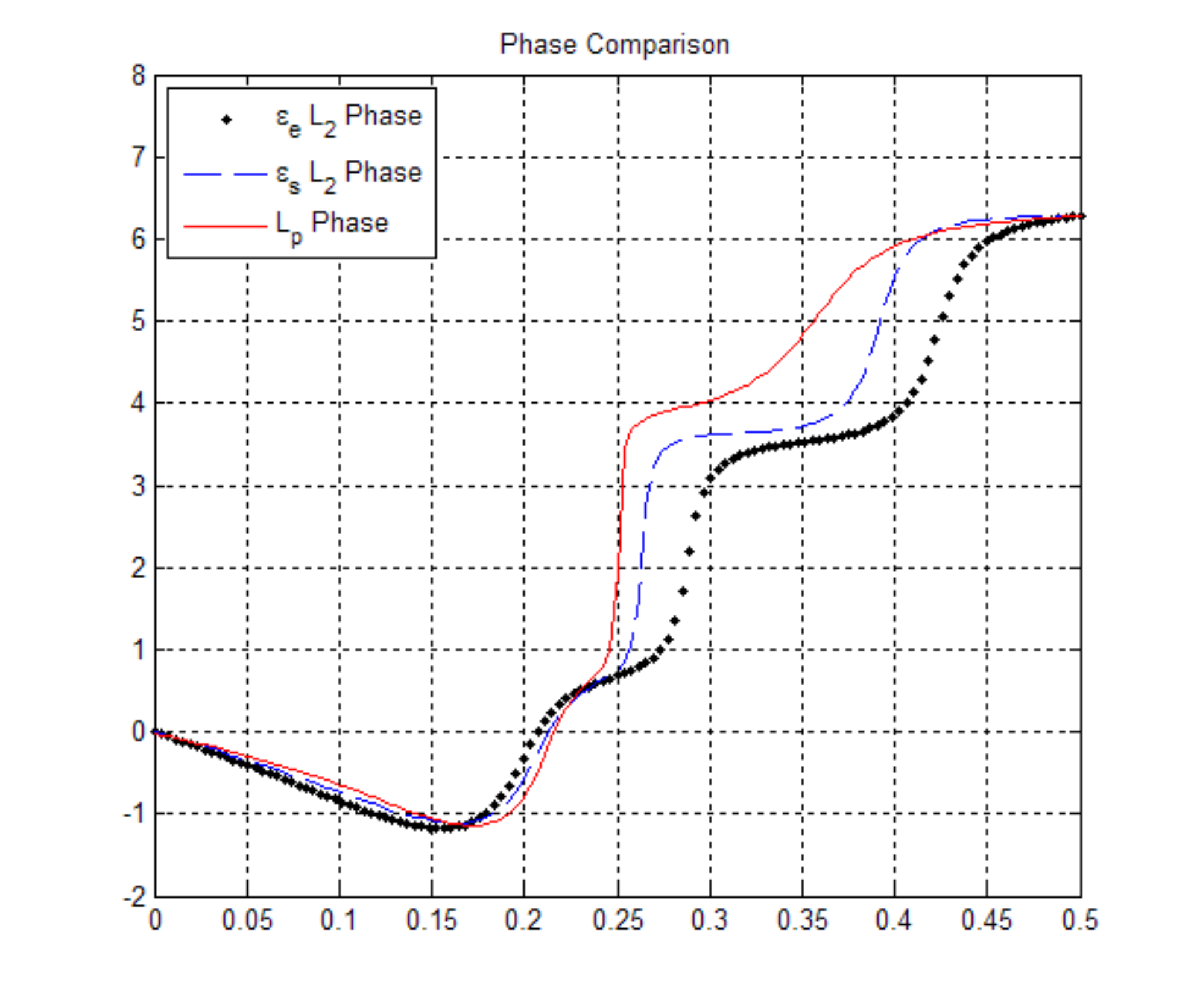,width=3.8in}}
    \caption{Phase responses for $l_2$ and $\lp$ IIR magnitude designs.}
    \label{lpiirmag5}
\end{figure}

Figures \ref{lpiirmag1} through \ref{lpiirmag5} illustrate the effectiveness of this algorithm at each of the three different stages for length-5 filters $\va$ and $\vb$, with transition edge frequencies of 0.2 and 0.24 (in normalized frequency) and $p=30$. A linear transition band was specified. Figures \ref{lpiirmag1}, \ref{lpiirmag1} and \ref{lpiirmag1} show the equation error $l_2$, solution error $l_2$ and solution error $\lp$. Figure \ref{lpiirmag4} shows a comparison of the magnitude error functions for the solution error $l_2$ and $\lp$ designs.  Figure \ref{lpiirmag5} shows the phase responses for the three designs.

From Figures \ref{lpiirmag4} and \ref{lpiirmag5} one can see that the algorithm has changed the phase response in a way that makes the maximum magnitude error (located in the stopband edge frequency) to be reduced by approximately half its value.  Furthermore, Figure \ref{lpiirmag4} demonstrates that one can reach quasiequiripple behavior with relatively low values of $p$ (for the examples shown, $p$ was set to 30).

\section{Conclusions}
\label{ch:concl}

Digital filters are essential building blocks for signal processing applications. One of the main goals of this work is to illustrate the versatility and relevance of $l_p$ norms in the design of digital filters.  While popular and well understood, $l_2$ and $l_{\infty}$ filters do tend to accentuate specific benefits from their respective designs; filters designed using $l_p$ norms as optimality criteria can offer a tradeoff between the benefits of these two commonly used criteria.  This work presented a number of applications of $L_p$ norms in both FIR and IIR filter design, and their corresponding design algorithms and software implementation.

The basic workhorse for the methods presented in this document is the {\it Iterative Reweighted Least Squares} algorithm, a simple yet powerful method that sets itself naturally adept for the design of $l_p$ filters. The notion of converting a mathematically complex problem into a series of significantly easier optimization problems is common in optimization. Nevertheless, the existence results from Theorem \ref{int:wexist} strongly motivate the use of IRLS methods to design $l_p$ filters. Knowing that optimal weights exist that would turn the solution of a weighted least squared problem into the solution of a least-$p$ problem must at the very least captivate the curiosity of the reader. The challenge lies in finding a robust and efficient method to find such weights.  All the methods presented in this work work under this basic framework, updating iteratively the weighting function of a least squares problem in order to find the optimal $l_p$ filter for a given application. Therefore it is possible to develop a suite of computer programs in a modular way, where with few adjustments one can solve a variety of problems.

Throughout this document one can find examples of the versatility of the IRLS approach. One can change the internal linear objective function from a complex exponential kernel to a sinusoidal one to solve complex and linear phase FIR filters respectively using the same algorithm. Further adaptations can be incorporated with ease, such as the proposed {\it adaptive solution} to improve robustness. 

Another important design example permits to make $p$ into a function of frequency to allow for different $p$-norms in different frequency bands. Such design merely requires a few changes in the implementation of the algorithm, yet allows for fancier, more elegant problems to be solved, such as the {\it Constrained Least Squares} (CLS) problem. In the context of FIR filters, this document presents the CLS problem from an $l_p$ prespective. While the work by John Adams \cite{adamsfircls} set a milestone in digital filter design, this work introduces a strong algorithm and a different perspective to the problem from that by Adams and other authors. The IRLS $l_p$-based approach from this work proves to be robust and flexible, allowing for even and uneven sampling. Furthermore, while a user can use fixed transition bands, one would benefit much from using a flexible transition band formulation, where the proposed IRLS-based algorithm literally finds the optimal transition band definition based on the constraint specifications. Such flexibility allows for tight constrains that would otherwise cause other algorithms to fail meeting the constraint specifications, or simply not converging at all. Section \ref{fir:cls} introduced two problem formulations as well as results that illustrate the method's effectiveness at solving the CLS problem.

While previous work exists in the area of FIR design (or in linear $l_p$ approximation for that matter), the problem of designing $l_p$ IIR filters has been far less explored. A natural reason for this is the fact that $l_2$ IIR design is in itself an open research area (and a rather complicated problem as well). Traditional linear optimization approaches cannot be directly used for either of these problems, and nonlinear optimization tools often prove either slow or do not converge. 

This work presents the $l_p$ IIR design problem as a natural extension of the FIR counterpart, where in a modular fashion the linear weigthed $l_2$ section of the algorithms is replaced by a nonlinear weighted $l_2$ version. This problem formulation allows for the IIR implementation of virtually all the IRLS FIR methods presented in Chapter \ref{ch:fir}. Dealing with the weighted nonlinear $l_2$ problem is a different story.

The problem of rational $l_2$ approximation has been studied for some time. However the sources of ideas and results related to this problem are scattered across several areas of study. One of the contributions of this work is an organized summary of efforts in rational $l_2$ optimization, particularly related to the design of IIR digital filters. The work in Section \ref{iir:ls} also lays down a framework for the IIR methods proposed in this work.

As mentioned in Section \ref{iir:lp}, some complications arise when designing IIR $l_p$ filters. Aside from the intrinsic $l_2$ problem, it is necessary to properly combine a number of ideas that allowed for robust and efficient $l_p$  FIR methods. A design algorithm for {\it complex} $l_p$ IIF filters were presented in Section \ref{iir:cplxfreq}; this algorithm combined Soewito's quasilinearization with ideas such as $l_p$ homotopy, partial updating and the adaptive modification. In practice, the combination of these ideas showed to be practical and the resulting algorithm remained robust. It was also found that after a few $p$-steps, the internal $l_2$ algorithm required from one to merely a few iterations on average, thus maintaining the algorithm efficient.

One of the main contributions of this work is the introduction of an IRLS-based method to solve $l_p$ IIR design problems.  By properly combining the principle of magnitude approximation via phase updating (from Soewito, Jackson and Kay) with the complex IIR algorithm one can find optimal magntiude $l_p$ designs. This work also introduced a sequence of steps that improve the efficiency and robustness of this algorithm, by dividing the design process into three stages and by using suitable 
initial guesses for each stage.

Some of the examples in this document were designed using Matlab programs. It is worth to notice the common elements between these programs, alluding to the modularity of the implementations. An added benefit to this setup is that further advances in any of the topics covered in this work can easily be ported to most if not all of the algorithms. 

Digital filter design is and will remain an important topic in digital signal processing. It is the hope of the author to have motivated in the reader some curiosity for the use of $l_p$ norms as design criteria for applications in FIR and IIR filter design. This work is by no means comprehensive, and is meant to inspire the consideration of the flexibility of IRLS algorithms for new $l_p$ related problems.

\bibliographystyle{IEEEtran}
\bibliography{dspref}

\end{document}